\documentclass[iop, apj, natbib]{emulateapj}
\usepackage{graphicx}
\usepackage{amsmath}
\usepackage{amssymb}
\usepackage{color}
\usepackage{multirow}
\usepackage{rotating}
\usepackage{booktabs}
\usepackage{times}
\usepackage{natbib}

\newcommand{\imag}{\ensuremath{i^\prime}}
\newcommand{\zmag}{\ensuremath{z^\prime}}
\newcommand{\Msun}{\ensuremath{M_{\odot}}}
\newcommand{\Zsun}{\ensuremath{Z_{\odot}}}

\newcommand{\ar}{\arcsec}
\newcommand{\wb}{\ensuremath{B_{435}}}
\newcommand{\wv}{\ensuremath{V_{660}}}
\newcommand{\wi}{\ensuremath{i_{775}}}
\newcommand{\wz}{\ensuremath{z_{850}}}
\newcommand{\wh}{\ensuremath{H_{160}}}
\newcommand{\ha}{H$\alpha$}
\newcommand{\mg}{\mu_\gamma}
\newcommand{\s}{\sigma}
\newcommand{\sg}{\sigma_\gamma}

\slugcomment{Accepted for Publication in ApJ: Oct 9, 2012}

\shorttitle{Intrinsic Shape of sBzK in GOODS-S and SXDS}
\shortauthors{Yuma et al.}

\begin{document}

\title{Intrinsic Shape of Star-Forming BzK Galaxies II: 
Rest-Frame UV and Optical Structures in GOODS-South and SXDS}
\author{Suraphong Yuma        \altaffilmark{1,2}, 
        Kouji Ohta        \altaffilmark{1}, and
        Kiyoto Yabe        \altaffilmark{1,3}	
        }

\email{yuma@icrr.u-tokyo.ac.jp}

\begin{abstract}
We study statistical intrinsic shape of star-forming BzK galaxies (sBzK galaxies) 
at $z\sim2$ in both rest-frame UV and rest-frame optical wavelengths.
The sBzK galaxies are selected down to $K_{AB} = 24.0$ mag in the Great Observatories 
Origins Deep Survey-South (GOODS-S) and Subaru-XMM Deep Survey (SXDS) fields, 
where high-resolution images from {\it Hubble Space Telescope} are publicly available. 
57\% (583) of all 1028 galaxies in GOODS-S show a single component 
in the Advanced Camera for Survey (ACS)/F850LP image. 
As Wide Field Camera (WFC3)/F160W images cover only some part of GOODS-S and SXDS, 
724/1028 and 2500/29835 sBzK galaxies in the GOODS-S and SXDS have the WFC3 coverage. 
86\% (626) and 82\% (2044) of the sBzK galaxies in WFC3/F160W images appear as a 
single component in the GOODS-S and SXDS, respectively. 
Larger fraction of single-component objects in F850LP images represents 
multiple star-forming regions in galaxies, while they are not so obvious in the F160W image which appears smoother.
Most of the single-component sBzK galaxies show S\'ersic indices of $n=0.5-2.5$, 
in agreement with those of local disk galaxies. Their effective radii are $1.0-3.0$ kpc 
and $1.5-4.0$ kpc in F850LP and F160W images, respectively, regardless of the 
observed fields. Stellar surface mass density of the sBzK galaxies is also comparable 
to that of the local disk galaxies. 
However, the intrinsic shape of sBzK galaxies is not a round disk as seen in the local 
disk galaxies. By comparing apparent axial ratio ($b/a$) distributions of the sBzK galaxies 
with those by assuming tri-axial model with axes $A>B>C$, we found their intrinsic face-on 
$B/A$ ratios peak at $B/A=0.70$ and $B/A=0.77-0.79$ in the rest-frame UV and optical, 
respectively and are statistically more bar-like than that of the local disk galaxies. 
The intrinsic edge-on $C/A$ ratios in both rest-frame UV and optical wavelengths 
peak at 0.26, which is slightly larger than that of the local disk galaxies. 
Possible origins of this bar-like structure are bar instability, galaxy interaction, or 
continuous minor mergers. Study of galaxy structure evolution in cosmological simulations 
is desirable to examine the most likely scenario. 

\end{abstract}


\altaffiltext{1}{Department of Astronomy, Kyoto University, Sakyo-ku, Kyoto 606-8502, Japan}
\altaffiltext{2}{Institute for Cosmic Ray Research, University of Tokyo, Kashiwa-no-ha, Kashiwa 277-8582, Japan}
\altaffiltext{3}{National Astronomical Observatory of Japan, Mitaka, Tokyo 181-8588, Japan}

\section{Introduction}\label{intro}

Disk galaxy is one type of galaxies commonly seen in the present-day universe. 
Origin of the disk galaxy is, therefore, a crucial key to understand the 
galaxy formation and evolution. Observational studies suggest that disk 
galaxies already exist at $z\sim1$. Despite some luminosity evolution up to $z\sim1$ 
\citep[e.g.,][]{scarlata07}, stellar surface mass density does not show significant 
evolution \citep{barden05}. Likewise, there is almost no significant evolution of disk galaxies 
seen in the stellar mass function \citep[e.g.,][]{pannella09} 
and in the scale length function \citep[e.g.,][]{lilly98, sargent07}. 
These studies imply that the origin of disk galaxies is beyond $z\sim1$. 
However, obvious counterpart of disk galaxies has not been identified at $z\sim3$. 
Studies of Lyman break galaxies (LBGs) at $z\sim3$ reveal that 
most of the LBGs show surface brightness distribution consistent with S\'ersic profile 
with S\'ersic index $n$ around one \citep[e.g,][]{steidel96, akiyama08}. 
Nevertheless, \cite{akiyama08} showed that the stellar surface mass density of 
the LBGs is too high to plausibly evolve into the local disk galaxies, suggesting 
that the LBGs at $z\sim3$ are more likely to be progenitors of the local elliptical galaxies. 
In addition, large clustering amplitude of the bright LBGs indicates that they reside in 
massive dark matter halo and are seemingly not a direct progenitor of the 
present-day disk galaxies \citep[e.g.,][]{gm01, ouchi01}. 

At $z\sim2$, galaxies can be efficiently selected by color selection \citep[e.g.,][]{steidel04, daddi04}. 
The clustering study of star-forming BzK (sBzK) galaxies, which are selected based on 
$B-z$ and $z-K$ color \citep{daddi04}, showed that faint sBzK galaxies 
($K_{AB} < 23.2$ mag) reside in dark matter halos with comparable mass 
to that of the local disk galaxies \citep{hayashi07}. 
\cite{forster09} studied kinematic properties of 80 star-forming galaxies at $1.3 < z < 2.6$ 
through spectroscopic imaging observations of \ha~emission and found that 
about 30\% of their sample is rotationally supported with similar velocity-size relation 
to the present-day disk galaxies, albeit larger velocity dispersion. 
The results seem to imply that star-forming galaxies at $z\sim2$ are likely to be 
disk galaxies with thicker disk than the local disk galaxies. 
Unfortunately, the intrinsic shape of these galaxies at $z\sim2$ had not been studied 
yet to confirm the presumption. 

The intrinsic shape of galaxies is related to distributions of their apparent axial ratios ($b/a$) 
and can be constrained by assuming a tri-axial model with axes $A>B>C$ 
\citep[e.g.,][]{ryden04, padilla08, unterborn08}. 
\citet[][hereafter paper I]{yuma11} studied the intrinsic shape of sBzK galaxies in the 
GOODS-North field by using high-resolution imaging data obtained with 
the Advanced Camera for Surveys (ACS) camera on {\it Hubble Space Telescope} ({\it HST}). 
Among 1029 sBzK galaxies selected down 
to $K_{AB} = 24.0$ mag, 54\% of them showed a single component in ACS/F850LP 
images, of which the effective wavelength corresponds to the rest-frame UV ($\sim3000$ \AA). 
Most of these single-component sBzK galaxies show surface brightness distribution with 
S\'ersic indices of $n\sim1$, suggesting that they have disk-like structure. 
Besides, they found that the sBzK galaxies largely distribute in the same region of
the size-mass diagram as $z\sim0-1$ disk galaxies, suggesting a comparable stellar 
surface mass density. However, paper I found that the sBzK galaxies at $z\sim2$ 
apparently prefer a bar-like or oval shape rather than a round disk seen in the 
local disk galaxies. In case of the sBzK galaxies (paper I), the mean face-on axial ratio ($B/A$) 
is 0.61 and the mean disk thickness $C/A$ is 0.28, while they are $0.90-0.92$ and $0.21-0.22$, 
respectively, for the local disks \citep{padilla08, unterborn08}. 
In other words, despite similar S\'ersic index and comparable stellar surface mass 
density, the sBzK galaxies at $z\sim2$ statistically show a bar-like or oval intrinsic 
shape different from disk galaxies in the local universe that show a round and flat 
disk shape. 

Although the conclusion about the intrinsic shape of sBzK galaxies is quite clear, 
some problems still remain for further investigations. The relevant problem is 
difference of effective wavelengths of the images used to derived the apparent $b/a$ 
between galaxies at $z\sim2$ and those at $z\sim0$. 
The apparent $b/a$ of the sBzK galaxies in the GOODS-N field was determined 
in the ACS/F850LP images corresponding to the rest-frame UV wavelength 
where the luminosity is mainly dominated by star-forming activity. 
In contrast, studies of local galaxies were carried out in the optical wavelength, 
which is dominated by more evolved stellar component. 
Thus their structures can possibly be different.
Recent morphological studies of galaxies at $z\sim2$ demonstrated that 
a galaxy may look different in different wavelengths \citep[e.g.,][]{cameron11, conselice11}. 
\cite{conselice11} studied the stellar-mass selected galaxies at $1< z < 3$ and 
found that galaxies appearing peculiar in the rest-frame UV images often show 
compact morphology in the rest-frame optical images. 
Though this result is partially due to the resolution difference between images in two wavelength ranges, 
the morphologies of galaxies seem to be varied when observing in different wavelengths. 
Accordingly, it is important to study the intrinsic shape of galaxies in the 
rest-frame optical wavelengths. 
In this paper, we examine the intrinsic shape of sBzK galaxies at $z\sim2$ in 
GOODS-South and SXDS fields in both rest-frame UV and optical wavelengths. 
In addition to determine the difference in structure between two different wavelength 
ranges, this work increases at least twice the number of galaxies used to derive 
the intrinsic shape in both wavelengths, which improves the statistical significance 
of the determined shapes. 

This paper is organized as follows. In Section \ref{sec:data}, we provide an overview 
of the data sources and photometric catalogs of objects in both GOODS-South 
and SXDS fields. In Section \ref{sec:photoz}, we describe how we determine the 
photometric redshifts and stellar properties of objects. Section \ref{sec:select} 
is given to a description of our sample selection method and counterparts of sBzK galaxies 
in high-resolution images. Section \ref{sec:galfit} presents structural analysis and 
accuracy of structural parameters. In Section \ref{sec:result}, we present results 
of structural parameters, stellar surface mass density, and intrinsic shape of 
the sBzK galaxies derived in both rest-frame UV and optical wavelengths as well 
as the comparison between different wavelengths. 
In Section \ref{sec:discuss}, we compare our results with those of the local disk galaxies 
and discuss possible scenarios for the origin of the intrinsic shape of the sBzK galaxies. Finally, we summarize 
our results in Section \ref{sec:summary}. We use a standard $\Lambda$CDM cosmology of 
$\Omega_m = 0.3$, $\Omega_{\Lambda} = 0.7$, and $H_0 = 70$km s$^{-1}$ Mpc$^{-1}$ 
and AB magnitude system. 

\section{Data Sources and Photometric Catalogs}\label{sec:data}
\subsection{GOODS-S}
Data in the GOODS-South field were collected as a part of the Great Observatories
Origins Deep Survey (GOODS)\footnote{http://www.stsci.edu/science/goods/}. 
The optical $HST$/ACS data were taken from the v2.0 data 
products of the GOODS $HST$/ACS treasury program \citep{giavalisco04}. 
They consist of four band imaging data: F435W, 
F606W, F775W, and F850LP, hereafter referred to as the \wb, \wv, 
\wi, and \wz~band, respectively. The images cover an area of $\sim160$ 
arcmin$^2$. The pixel scale of the images is 0.\ar030 pixel$^{-1}$. 
The $5\sigma$ limiting magnitudes of the \wb, \wv, \wi, and \wz~images at 1.\ar0 
diameter aperture are 27.9, 27.8, 27,3, and 27.0, respectively. 
The $U$ and $R$ data were obtained with $VLT$/VIMOS 
from the ESO/GOODS program \citep{nonino09}. 
The final $U$ and $R$ images in the area common to the GOODS $HST$/ACS 
observations reach $5\sigma$ limiting magnitudes of 28.61 and 28.06 mag, 
respectively (at 1.\ar0 diameter aperture). 

The NIR $VLT$/ISAAC data in $J$, $H$, $K_s$ bands 
were taken from the GOODS/ISAAC data release final version (2.0) \citep{retzlaff10}. 
The pixel scale of the ISAAC images is 0.\ar150, which is chosen so that 
one ISAAC pixel subtends exactly a block of $5\times5$ ACS pixels. 
The images in each band consist of at least 24 sections\footnote{
There are two additional sections for $J$ and $K_s$ data.}
over the GOODS-S field with different seeing size varying from 0.\ar4 to 0.\ar6. 
The $5\sigma$ limiting magnitudes for point sources of $J=25.0$, 
$H=24.5$, and $K_s=24.4$ mag are reached within 75\% of the survey area \citep{retzlaff10}. 
The mid-infrared (MIR) data were obtained from deep observations with the 
Infrared Array Camera (IRAC)  on the $Spitzer$ Space Telescope ({\rm SST}). 
We used 3.6\micron~and 4.5\micron~images from both DR1 and DR2 provided 
by the $SST$ Legacy Science program. Limiting magnitudes of the 3.6\micron~and 
4.5\micron~images are 24.3 and 24.4, respectively ($5\sigma$ at 2.\ar4 diameter aperture). 
The effective area of the GOODS-S field we study is $\sim160$ arcmin$^2$. 

In order to study the structural parameters of sBzK galaxies in the 
rest-frame optical wavelength, the high-resolution NIR image is necessary. 
At $z\sim2$, the rest-frame optical wavelength ($\sim5000$\AA) is 
redshifted to $\sim1.6$\micron, corresponding to the F160W band. 
We used the publicly available data from the Cosmic Assembly Near-Infrared 
Deep Extragalactic Legacy Survey \citep[CANDELS;][]{candels1, candels2}
\footnote{http://candels.ucolick.org/index.html}. 
This survey consists of five fields, one of which is the GOODS-S field. 
The data were taken from data release v0.5 of both deep JH and wide JH subregions, 
covering about 70\% of the ACS/GOODS-S field with 10 observing epochs. 
Limiting magnitude of the WFC3/F160W image (hereafter \wh) is $\sim25.6-26.0$ 
depending on the observing epochs ($5\sigma$ at 1.\ar0 diameter). 
The pixel scale is 0.\ar060 pixel$^{-1}$. 

\subsection {SXDS}
The deep multi-wavelength data in the SXDS/UDS field are also publicly available. 
The optical $Subaru$/Suprime-Cam images ($B$, $V$, $R_C$, \imag, and \zmag) 
were obtained from the Subaru-XMM Deep Survey \citep[SXDS; ][]{furusawa08} with 
the original pixel scale of 0.\ar202. $5\sigma$ limiting magnitudes at 2.\ar0 diameter 
aperture are 27.7, 27.2, 27.1, 27.0, and 26.0 mag for $B$, $V$, $R_C$, \imag, and \zmag, 
respectively. The NIR data are from DR8 version of the UKIDSS Ultra Deep 
Survey \citep[UDS; ][]{lawrence07}. 
The $J$, $H$, and $K_s$ images with the 
original pixel scale of 0.\ar268 were taken with WFCAM on $UKIRT$. 
The limiting magnitudes are 24.9, 24.2, and 24.6 mag for $J$, $H$, and $K_s$, 
respectively ($5\s$ at 2.\ar0 diameter). The $SST$/ IRAC images with the pixel 
scale of 0.\ar600 were obtained from the $Spitzer$ Public Legacy Survey of the 
UKIDSS Ultra Deep Survey (SpUDS; PI: J. Dunlop). The limiting magnitudes 
are 24.8 and 24.5 for 3.6\micron~and 4.5\micron, respectively. We used the 
overlapped area of these data, which effectively covers $\sim0.67$ deg$^2$. 
The high-resolution NIR images (\wh), which is used to study the 
galaxy structure, were taken from the CANDELS data release v1.0 
covering central $\sim210$ arcmin$^2$ of the original SXDS. The $5\s$ limiting magnitude of 
the \wh~image is 27.1 mag for point sources. It is noteworthy that 
the CANDELS images in the SXDS are approximately one magnitude deeper than 
those we used in the GOODS-S field. 

\subsection{Photometric Catalogs}

The photometric catalog was made by first selecting the objects in the ISAAC/$K_s$ 
images and then making photometry in all images (except for WFC3 data) 
ranging from UV to MIR wavelength 
to construct spectral energy distribution (SED) of each object in the GOODS-S field. 
Due to seeing size difference among the ISAAC images, we first smoothed ISAAC/$K_s$ 
images into one homogenous seeing size, i.e., 0.\ar6. 
The ACS/optical images were put into the same pixel scale as the ISAAC/NIR images 
by summing $5\times5$ pixels of the original images and preserving 
the fluxes. The images were then smoothed to have the FWHM common to the $K_s$-band 
images. Because of the dimension differences of images in different bands, objects 
were extracted separately above $1\sigma$ with 5 connected pixels 
by SExtractor \citep{bertin96}. For the ACS images, 
aperture photometry was carried out separately in each band centering at the position 
extracted in $K_s$ band with 0.\ar8 diameter aperture, which provides the best 
signal-to-noise (S/N) ratio for the PSF homogenized images. 
The total magnitude of objects in ACS images was calculated from the aperture 
magnitude scaled up to the SExtractor's {\tt MAG\_AUTO} in $K_s$ band. 
For VIMOS and ISAAC images, the total magnitude was obtained directly from 
{\tt MAG\_AUTO} of each object. 
For IRAC images, the total magnitude was calculated by using 
the aperture photometry at 2.\ar4 diameter and aperture correction. 
The correction factors were determined by generating the artificial objects 
with given total magnitudes and recovering their aperture photometry (paper I). 

The photometric catalog of objects in SXDS is described by \cite{yabe11}. 
The NIR images are aligned and homogenized to the optical images. 
Aperture photometry of objects in each image was made at the position 
extracted in the $K_s$-band image by using dual-image mode in SExtractor 
with an aperture of 2.\ar0 diameter. Total magnitudes are calculated by applying 
the correction factor, which is determined by scaling the aperture magnitude to the 
{\tt MAG\_AUTO} in $K_s$-band image, to the aperture photometry in each band. 
For the IRAC images, aperture photometry was carried out at 2.\ar4 diameter 
aperture and was corrected to the total photometry in the same manner as 
done in the GOODS-S field. 

\section{Photometric Redshifts and SED Fitting}\label{sec:photoz}
The photometric redshift ($z_{phot}$) of each object in both GOODS-S and SXDS fields 
was determined by using $Hyperz$ \citep{bolzonella00}. 
The observed spectral energy distributions (SEDs) of each individual object, 
which consist of photometry in $UBVRizJHK_s$ and IRAC 3.6 and 4.5\micron~band, 
was fitted to the template SEDs covering E, S0, Sa-Sd, Im, and starburst types. 
The redshift range is set to be $0 < z_{phot} < 6.0$ with a step of $\delta z=0.03$. 
The obtained $z_{phot}$ was compared with the available spectroscopic redshift 
catalogs in both GOODS-S \citep{popesso09, balestra10} and SXDS/UDS 
\citep[Simpson et al. 2011, inprep.; Akiyama et al. 2011, in prep.; ][]{smail08}. 
Although there are catastrophic redshift outliers, the $z_{phot}$ is 
mostly consistent with the $z_{spec}$ with uncertainties $\s_z$ 
(standard deviation of $\Delta z/[1+z_{spec}]$) of 0.05 and 0.03 
in GOODS-S and SXDS/UDS, respectively. 

Stellar mass, star formation rate (SFR), age, and color excess ({\it E(B-V)}) 
of objects in the catalogs were obtained from SED fitting with the standard 
$\chi^2$ minimization. The SED fitting was performed by using the {\it SEDfit}
program \citep{sawicki11}, which is the evolved version of the SED fitting 
program used in \citep{sawicki98}. The main differences are that the new version 
uses \cite{calzetti00} law instead of the \cite{calzetti97} and the population synthesis 
model by \cite{bc03} instead of those by \cite{bc93}. 
BC03 population synthesis code \citep{bc03} was 
used to construct the model SEDs with various star formation histories 
(instantaneous burst, exponential declining SF, constant SF). The Salpeter 
initial mass function \citep[IMF; ][]{salpeter55} with the mass range of $0.1-100$ 
\Msun~and metallicity of 1.0\Zsun~were assumed.\footnote{Metallicity of most star-forming 
galaxies at $z\sim2$ is slightly lower than solar abundance; $\sim0.6$\Zsun~or larger 
for $\sim10^{10}\Msun$ galaxies \citep[e.g., ][]{hayashi09, yoshikawa10, wuyts12}. 
Here we adopted the 1.0\Zsun~model rather than the 0.2\Zsun~model. } 
The dust attenuation law by \cite{calzetti00} was adopted with $E(B-V)$ range of 
$0.0-1.0$ mag (0.02 mag step) and the redshift was fixed to the $z_{phot}$ 
determined by $Hyperz$ to reduce the number of free parameters. 
The uncertainty is obtained based on Monte Carlo simulation at 68\% confidence level. 
The median uncertainties for derived stellar masses are 
$\Delta\log M_{*}/\Msun=0.26$ dex and 0.38 dex for samples in GOODS-S and SXDS, respectively.

\begin{figure*}[t]
\centering
\begin{tabular}{cc}
\includegraphics[clip, angle=-90, width=0.3\textwidth]{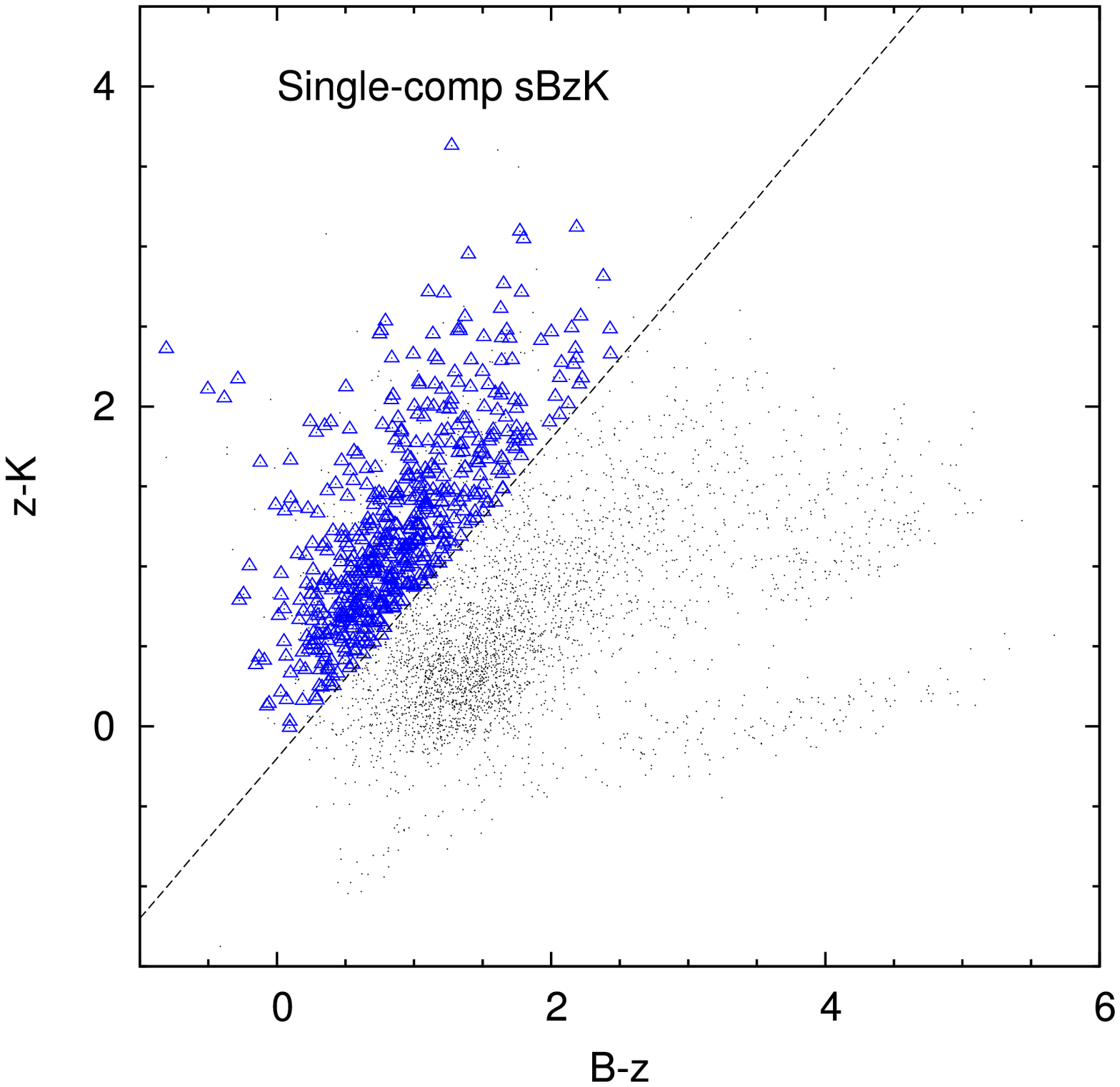} &
\includegraphics[clip, angle=-90, width=0.3\textwidth]{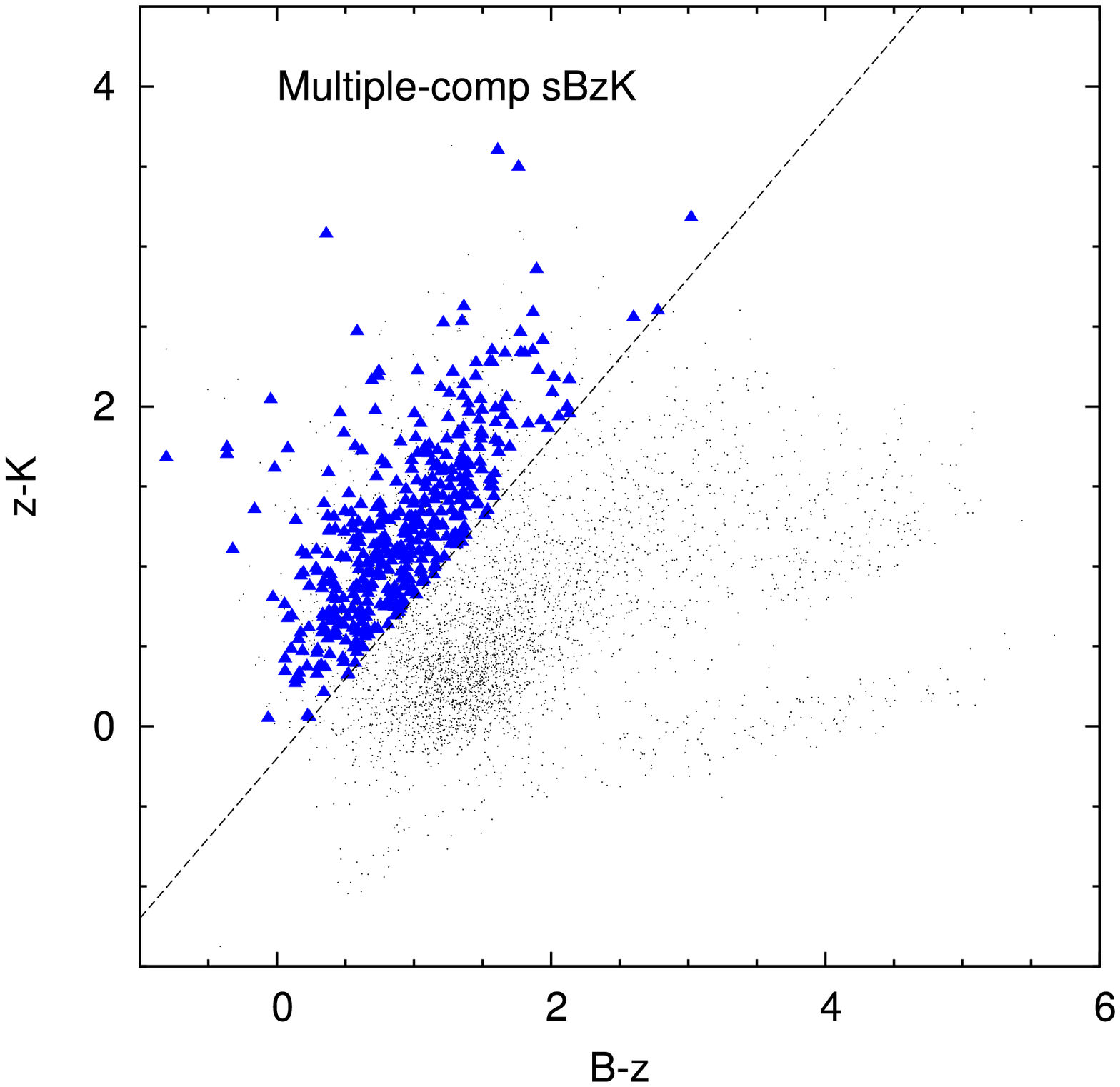} \\
\end{tabular}
\caption{
BzK diagram ($\wb-\wz$ and $\wz-K$) of sBzK galaxies in the GOODS-S field. 
The sBzK criterion is shown in a black dashed line. 
{\it Left}: Blue open triangles are the single-\wz~sBzK galaxies. 
{\it Right}: Blue solid triangles represent those with multiple components in the original \wz~image. 
All objects with $K_s<24.0$ mag are also shown with black dots. 
}
\label{bzk_gds_acsz}
\end{figure*}
\begin{figure*}
\centering
\begin{tabular}{cc}
\includegraphics[clip, width=0.3\textwidth, angle=-90]{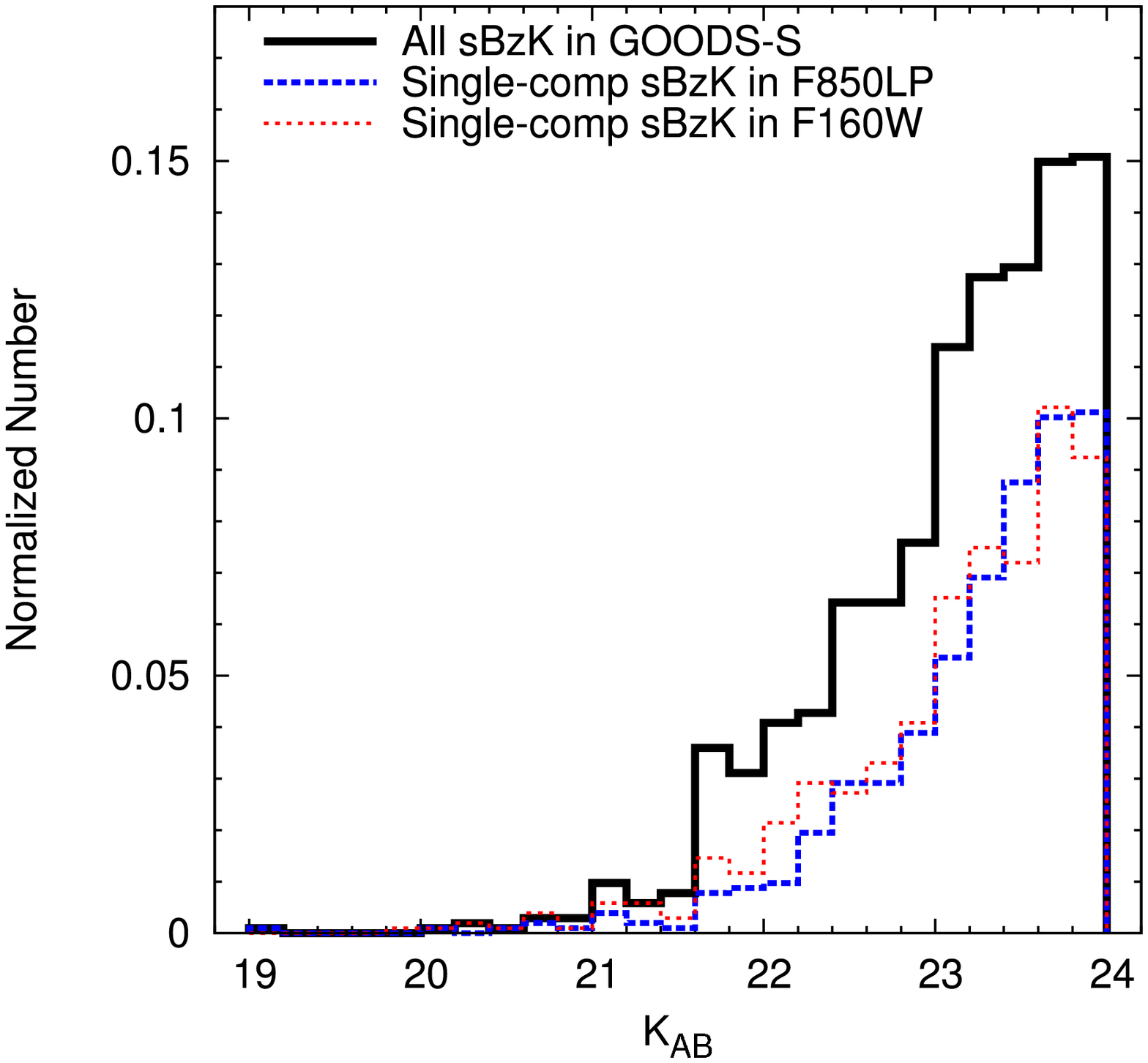} &
\includegraphics[clip, width=0.3\textwidth, angle=-90]{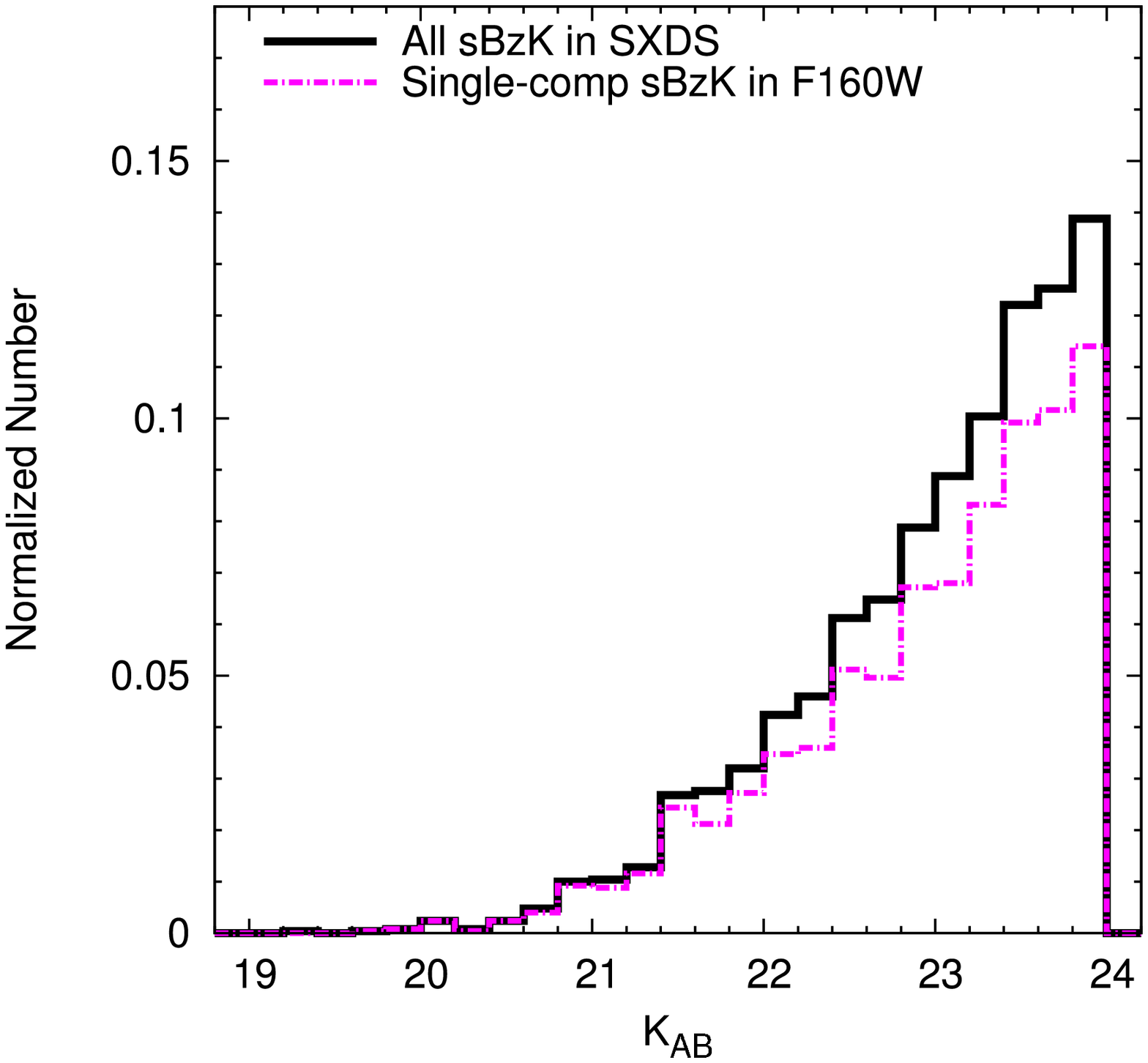} \\
\end{tabular}
\caption{
Normalized histograms of $K_s$-band magnitudes of sBzK galaxies in GOODS-S ({\it left panel}) 
and SXDS ({\it right panel}). The histograms are normalized by the number of all sBzK galaxies in the field. 
(For SXDS, the histograms are normalized by the number of all sBzK galaxies detected in the 
CANDELS field, which is part of the SXDS.) 
{\it Left panel:} Blue dashed and red dotted histograms represent the 
sBzK galaxies in the GOODS-S field appearing as a single component in the \wz~and \wh~images, 
respectively. The black solid histogram is for the entire sBzK sample in the field. 
{\it Right panel:} The single-component sBzK galaxies in SXDS are illustrated with magenta dot-dashed 
histogram, whereas all sBzK galaxies appearing in the CANDELS field of SXDS are shown with black solid line. 
}
\label{histogram_k}
\end{figure*}

\section{Sample Selection}\label{sec:select}
\subsection{Star-Forming BzK Galaxies}
Star-forming galaxies at $z\sim2$ were selected by applying the BzK selection 
method \citep{daddi04} to the photometric catalogs described above down 
to $K_s = 24.0$ mag. Because the response functions used in this paper are 
not identical to those used by \cite{daddi04}, the color corrections were determined 
in both GOODS-S and SXDS field. The stellar spectra by \cite{pickles98} were 
convolved with the response functions including the detector's 
quantum efficiency and atmospheric transmission. Both $B-z$ and $z-K$ colors 
derived from filter system used in the GOODS-S are not significantly different from those 
by \cite{daddi04}. Thus no correction is required for the GOODS-S field. 
In contrast, we adopted the color corrections for the SXDS field as follow:

\begin{eqnarray}
(B-z)_{{\rm Daddi+04}} &=& (B-z)_{SXDS} + 0.3, \\
(z-K)_{{\rm Daddi+04}} &=& (z-K)_{SXDS} + 0.1.
\end{eqnarray}
Then the original sBzK criterion, $BzK \equiv (z-K) - (B-z) \geq -0.2$, 
was applied to construct the sBzK samples in both fields. 
The candidates with $z_{spec} < 1.4$ and $z_{spec} > 2.5$ were excluded 
from the samples. Eventually, we have 1028 and 29835 sBzK galaxies in 
GOODS-S and SXDS, respectively. 

\begin{figure*}[t]
\centering
\begin{tabular}{cc}
\includegraphics[clip, angle=-90, width=0.3\textwidth]{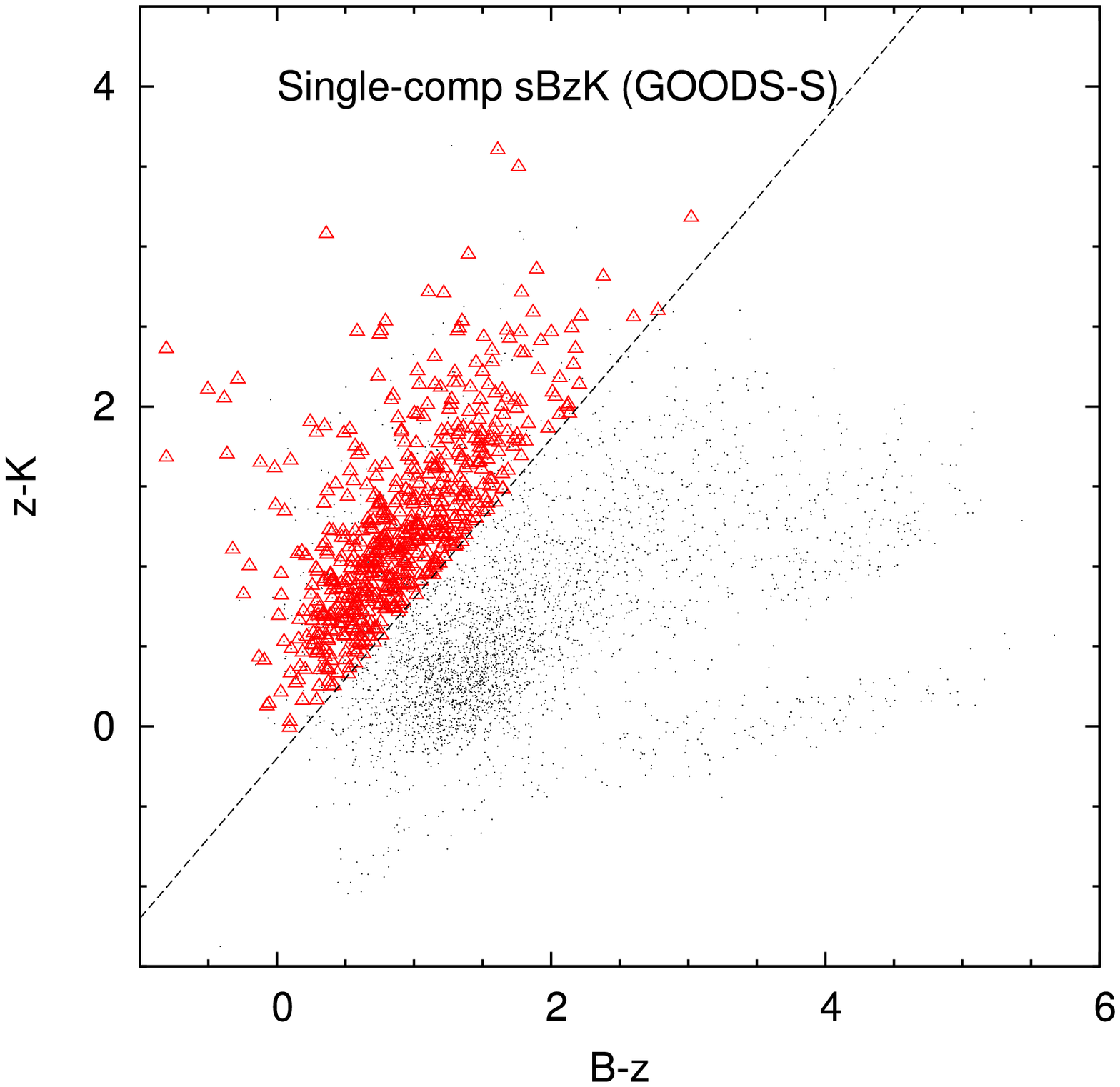} &
\includegraphics[clip, angle=-90, width=0.3\textwidth]{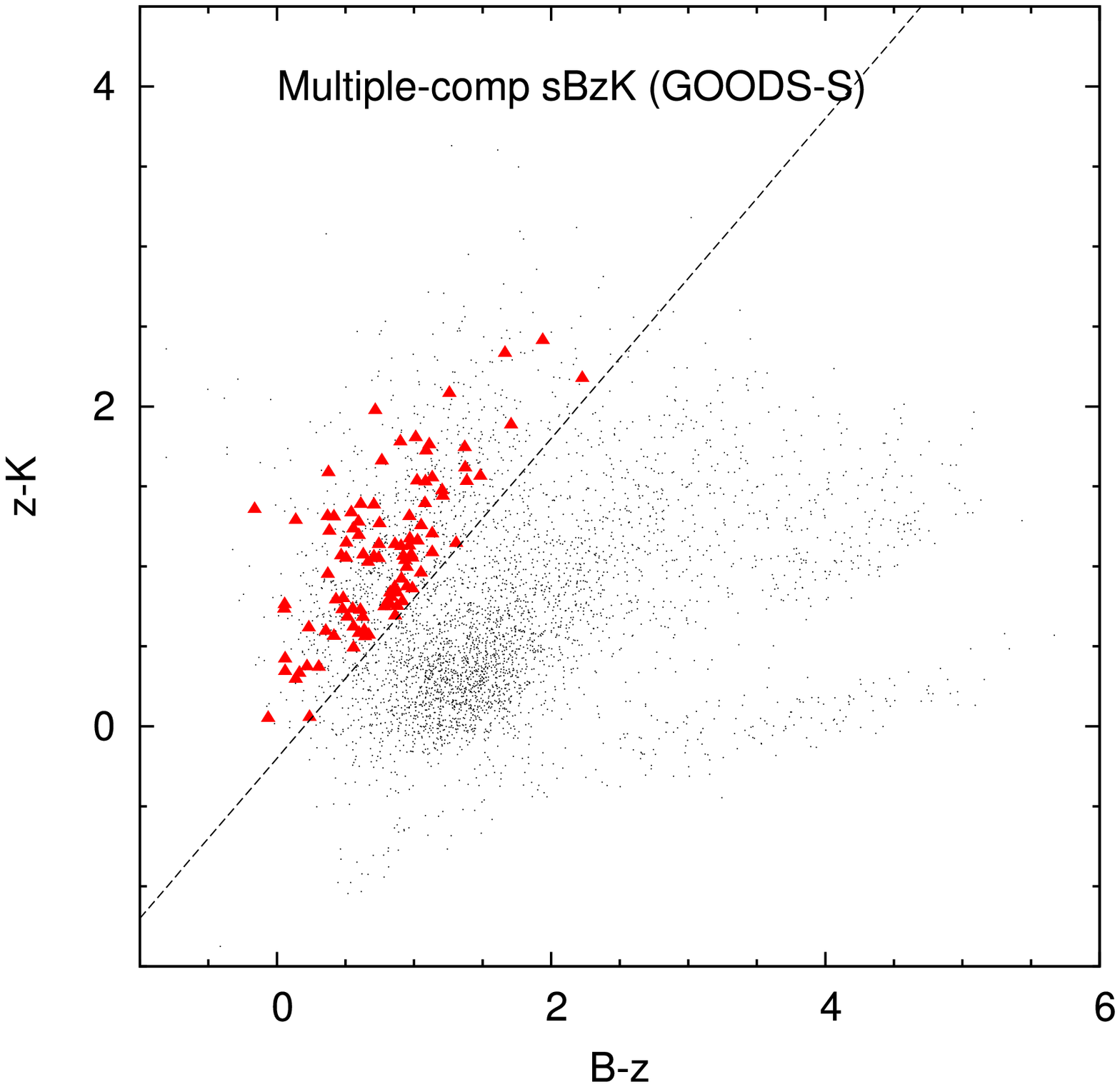} \\
\includegraphics[clip, angle=-90, width=0.3\textwidth]{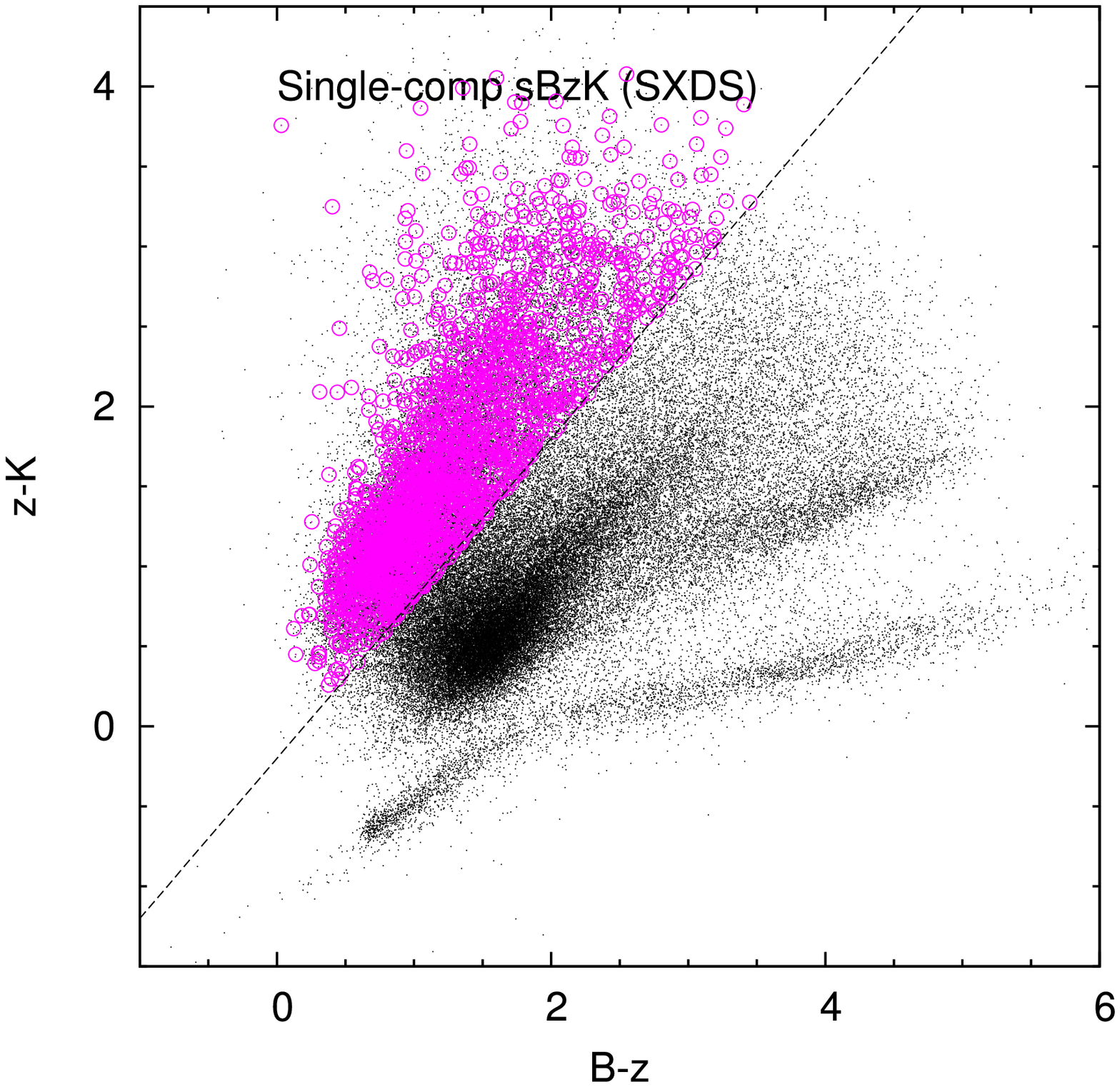} &
\includegraphics[clip, angle=-90, width=0.3\textwidth]{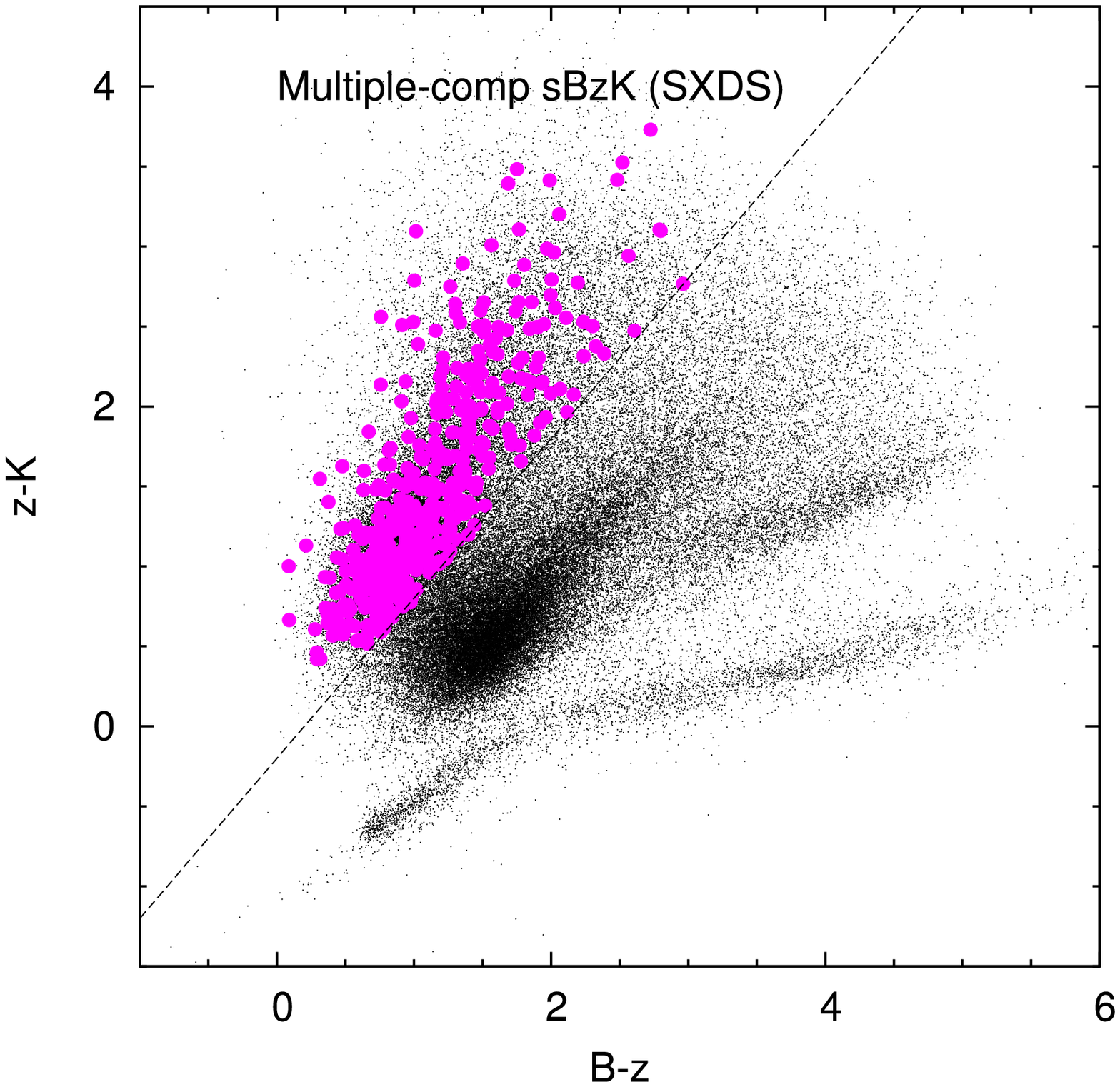} \\
\end{tabular}
\caption{
BzK diagram of the sBzK galaxies detected in the \wh~image in GOODS-S ({\it top panel}) 
and SXDS ({\it bottom panel}). 
The $B-z$ and $z-K$ colors for objects in SXDS were corrected to match the filter system used 
by \cite{daddi04} (see text for more details). The sBzK criterion is shown in a black dashed line. 
{\it top panel:} Red open triangles are the single-\wh~sBzK galaxies in the GOODS-S field, 
while red solid triangles represent the multiple-\wh~galaxies. 
{\it bottom panel:} The single- and multiple-\wh~sBzK galaxies in SXDS are shown with 
magenta open and solid circles, respectively. 
All objects with $K_s<24.0$ mag are also shown with black dots. 
The objects that are above the sBzK criterion but are neither classified as single-component nor 
multiple-component objects are those outside the \wh~images, which cover a part of 
the entire fields of GOODS-S and SXDS.
}
\label{bzk_wfc3h}
\end{figure*}

\subsection{Cross-Identification with $HST$ Images}
\subsubsection{ACS/F850LP}
In order to verify the intrinsic shape of sBzK galaxies in the same rest-frame 
UV wavelength as in the GOODS-N field (paper I), we used the 
\wz~images that are publicly available in the GOODS-S field. 
In section \ref{sec:data}, we constructed the photometric catalog of objects 
selected in $K_s$-band images by using the smoothed ACS images. 
However, we also need to identify galaxies using high-resolution images in 
order to study structure of galaxies. 
The sBzK galaxies are re-identified in the original \wz~image, which corresponds to 
rest-frame $\sim3000$\AA. An object in the \wz~images is identified as a BzK 
counterpart if it is detected above $2\s$ threshold with 5 connected pixels 
in SExtractor and locates within 1.\ar0 radius centered at the sBzK position in $K_s$ band. 
Similar to paper I, the radius was chosen as it is twice of the median 
$1/2$FWHM of the samples in $K_s$ images. The sBzK galaxies are 
divided into 2 groups: the single-component and the multiple-component sBzKs, 
depending on the number of their counterparts within the radius. 
583 (57\%) sources are identified as single-component sBzK galaxies, while 
the multiple-component sample contains 438 galaxies including six objects 
(3 pairs) of (interacting) sBzK galaxies. Note that the ratio of single-component 
sBzK galaxies is in agreement with that (54\%) in GOODS-N (paper I). 
The contamination rate of forground/background galaxies 
within 1.\ar0 radius is estimated to be 14\% for the \wz~image. 
As we select the single- and multiple-component galaxies in both \wz~and \wh~images, 
we explicitly denote the single-component sBzK galaxies in the \wz~images as 
the single-\wz~galaxies and the multiple-component sBzK galaxies 
in the \wz~images as the multiple-\wz~galaxies to avoid any confusion. 
Figure \ref{bzk_gds_acsz} shows BzK diagrams of the single-\wz~and 
multiple-\wz~sBzK galaxies in the \wz~images. 
Distributions of both single- and multiple-\wz~sBzK galaxies are almost the same.
They show similar distribution to the whole sBzK sample in the GOODS-S field as well. 
A histogram of $K_s$ magnitude of the single-\wz~ sBzK galaxies is shown in the 
left panel of Figure \ref{histogram_k} together with the histograms of all sBzK 
galaxies. The figure shows that the single-\wz~galaxies similarly distribute in the 
magnitude range of $21-24$ mag as compared to the total sample in the field. 
The total number of the sBzK galaxies that are able to be identified to either single 
or multiple components in the high-resolution 
\wz~images is 1021 out of all 1028 objects in the GOODS-S field. 
The remaining seven candidates are too faint to be detected in the original \wz~images, 
though they were detected in the $K_s$-band homogenized \wz~images, which were used 
in the sBzK selection process. 

\begin{deluxetable*}{ccccc}
\tabletypesize{\footnotesize}
\tablecaption{Number of sBzK galaxies detected in high-resolution images\label{tab:nobj}}
\tablewidth{0pt}
\tablehead{
\multicolumn{1}{c}{} & 
\multicolumn{3}{c}{GOODS-S} & 
\multicolumn{1}{c}{SXDS} \\
\cline{2-4} \\
\multicolumn{1}{c}{Type of sBzK} & 
\multicolumn{1}{c}{\wz \tablenotemark{a}} & 
\multicolumn{1}{c}{\wh-smoothed \wz} &
\multicolumn{1}{c}{\wh} & 
\multicolumn{1}{c}{\wh} 
}
\startdata
Single-component & 583 ($57\pm2\%$) & 718 ($70\pm2\%$) & 626 ($86\pm1\%$) & 2044 ($82\pm1\%$) \\
Multiple-component & 438 ($43\pm2\%$) & 310 ($30\pm3\%$) & 98 ($14\pm4\%$) & 456 ($18\pm2\%$) \\
\enddata
\tablenotetext{a}{
The number of all sBzK galaxies detected in the original \wz~images is 1021, 
which is not identical to the number of all sBzK in GOODS-S field (i.e., 1028). 
This is because the sBzK galaxies were selected based on the \wz~images homogenized 
to match PSF of the $K_s$-band images instead of the original images (see text). 
}
\end{deluxetable*}

\subsubsection{WFC3/F160W}
The high-resolution WFC3/F160W (\wh) images are used in order to determine the 
structure of the sBzK galaxies in the rest-frame optical wavelength range. 
The \wh~ images have the effective wavelength corresponding to the 
rest-frame $\sim5300$\AA~at $z\sim2$. We applied the definition identical to that used for 
the ACS/F850LP images in order to classify the single or multiple components, i.e., 
the sBzK counterparts in the high-resolution images are identified if being detected 
above $2\sigma$ and 5 connected pixels within 1.\ar0 radius from the sBzK center. 
626 (86\%) single-component and 98 multiple-component sBzK galaxies are 
obtained in the GOODS-S, and we have 2044 (82\%) single-component and 456 
multiple-component galaxies in the SXDS. 
Note that because the WFC3/F160W data are not available for the whole area of the GOODS-S 
and SXDS fields, only some fractions of the sBzK galaxies are used in this paper. 
The contamination rate for the \wh~images is 7\% within 1.\ar0 radius. 
Although most of sBzK galaxies located in the \wh~images ($>80$\%) are classified as 
a single component in both fields, the single-component fraction somehow does not agree 
to each other within $1\sigma$ uncertainty\footnote{$\sigma^2 \equiv (1-f)*f/N$, 
where $f$ denotes the fraction and $N$ is for number of the sample.}, i.e., $86\pm1\%$ 
for GOODS-S and $82\pm1\%$ for SXDS. However, they are still in agreement with 
each other within $2\sigma$ range. 
We refer to the single- and multiple-component sBzK galaxies in the 
\wh~images as the single-\wh~and the multiple-\wh~sBzK galaxies, respectively. 
The BzK diagrams of both single- and multiple-\wh~sBzK galaxies in both fields are 
illustrated together with the whole sBzK sample in Figure \ref{bzk_wfc3h}. 
The single-\wh~ sBzK galaxies in both fields seem to show similar distributions in the BzK diagram 
as compared to the entire sample of sBzK galaxies. 
$K_s$-band magnitude histograms of the single-\wh~ sBzK galaxies in GOODS-S and SXDS 
are shown in the left panel (red dotted) and right panel (magenta dot-dashed) of the 
Figure \ref{histogram_k}, respectively. Histograms of the single-\wh~galaxies in both 
fields are similar to those of the whole sBzK galaxies shown in the black solid lines. 
These imply that the single-\wh~sBzK galaxies can be good 
representatives for the entire populations at least in term of their colors and rest-frame 
optical magnitudes. 

\subsubsection{High Fraction of Single-Component sBzK Galaxies in WFC3/F160W}
As described above, the sBzK galaxies are categorized as single-component or 
multiple-component galaxies according to their appearance in the high-resolution images 
both in the rest-frame UV (\wz) and the rest-frame optical (\wh) wavelengths. 
In the GOODS-S field where high-resolution images are available in both rest-frame UV and 
optical wavelength, 57\% of all sBzK galaxies are identified as single-\wz~galaxies, while 
86\% of the galaxies located in the \wh~images are the single-\wh~samples. 
There are two plausible reasons responsible for the difference in fraction of single-component 
objects in different wavelengths. The first reason is the difference in the resolution of 
images between in \wz~and in \wh~bands. The \wz~images have about two times better 
resolution than the \wh~images; FWHM of the PSF in \wz~and \wh~images are 
$\sim$0.\ar1 and $\sim$0.\ar18, respectively. The second reason is that the difference 
is real. The rest-frame UV component of a galaxy is mainly dominated by star-forming activity, 
whereas the relatively low-mass, long-age stellar component tends to dominate the rest-frame 
optical wavelengths. Consequently, components of one galaxy seen in the rest-frame 
UV (i.e., \wz) are possibly just the star-forming regions in the galaxy, not the entire galaxy. 

In order to examine which reason is responsible for this different fraction, we first 
smoothed the original \wz~image to the \wh~resolution. Then we re-detected and 
re-selected the single-component and multiple-component sBzK galaxies from the 
\wh-smoothed \wz~images. The number and percentage of the single- and 
multiple-component sBzK galaxies are summarized in Table \ref{tab:nobj}. 
As seen in the table, fraction of the single-\wz~galaxies increases significantly 
after homogenizing the \wz~images to match the PSF of the \wh~images; 
however, the fraction is still less than that for the single-\wh~sample. 
The results indicate that the difference in appearance of sBzK galaxies 
between in the rest-frame UV and optical wavelength is partially due to 
the difference in resolution of images obtained with ACS and WFC3. 
On the other hand, more than half of the single-\wh~galaxies seen as multiple 
components in the \wz~images indicate real difference between the morphology of 
galaxies in the rest-frame UV and that in the rest-frame optical wavelength. 

\begin{figure*}
\centering
\includegraphics[clip, width=0.8\textwidth]{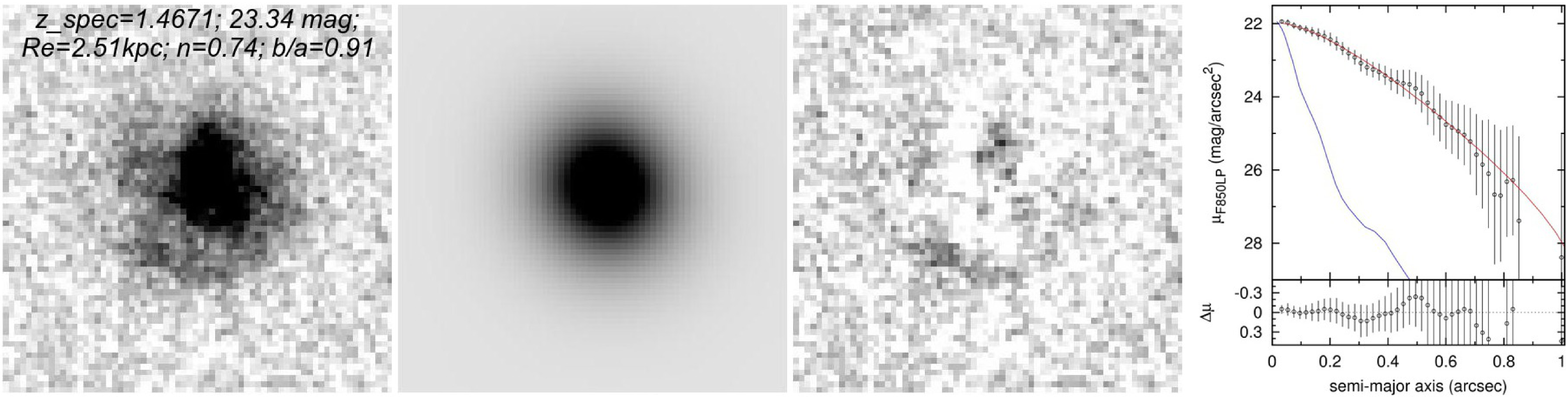}\\
\includegraphics[clip, width=0.8\textwidth]{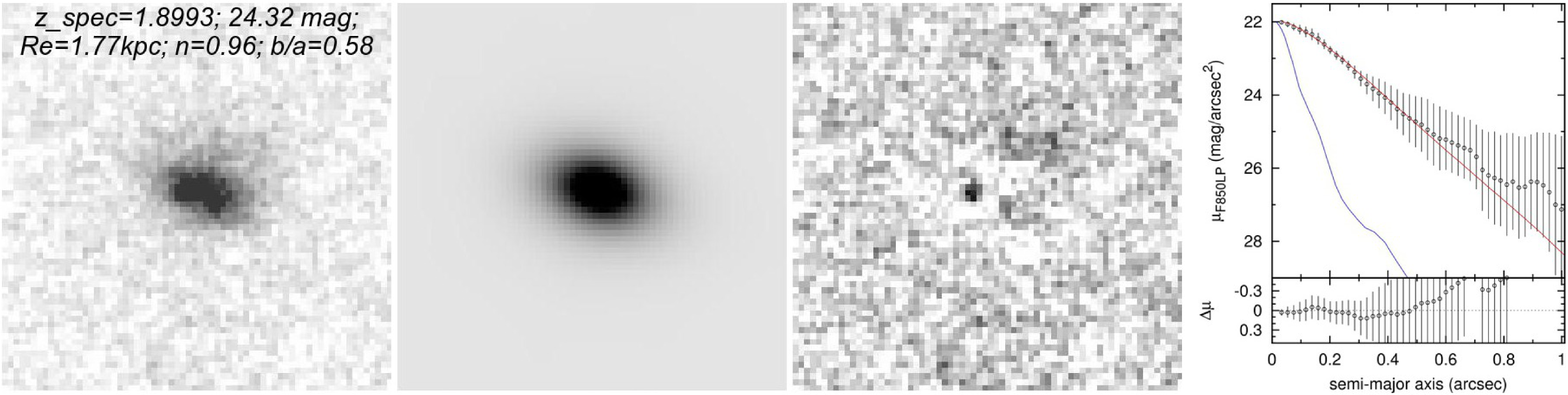}\\
\includegraphics[clip, width=0.8\textwidth]{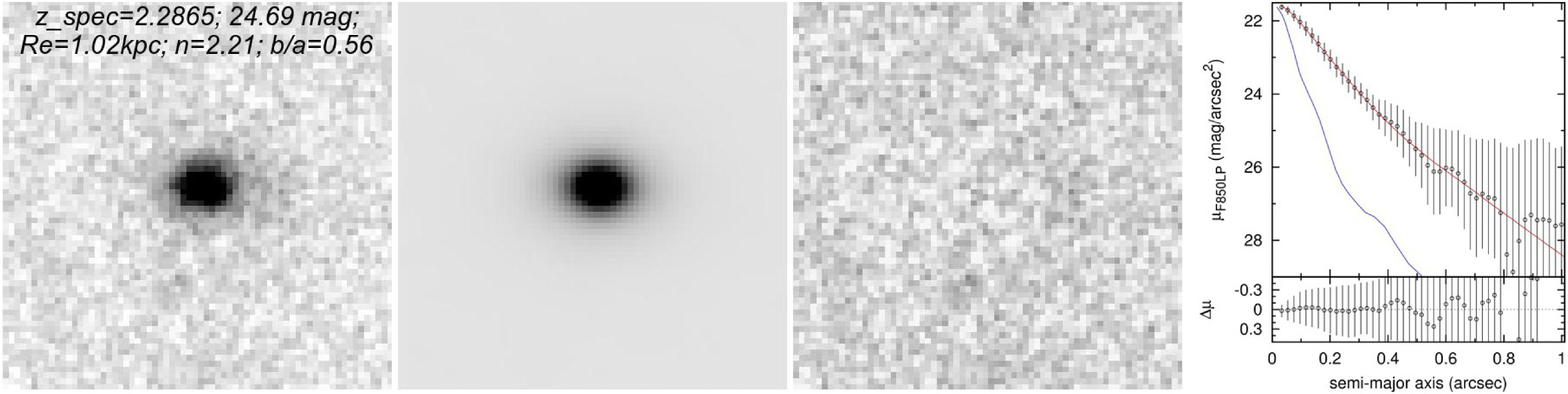}\\
\includegraphics[clip, width=0.8\textwidth]{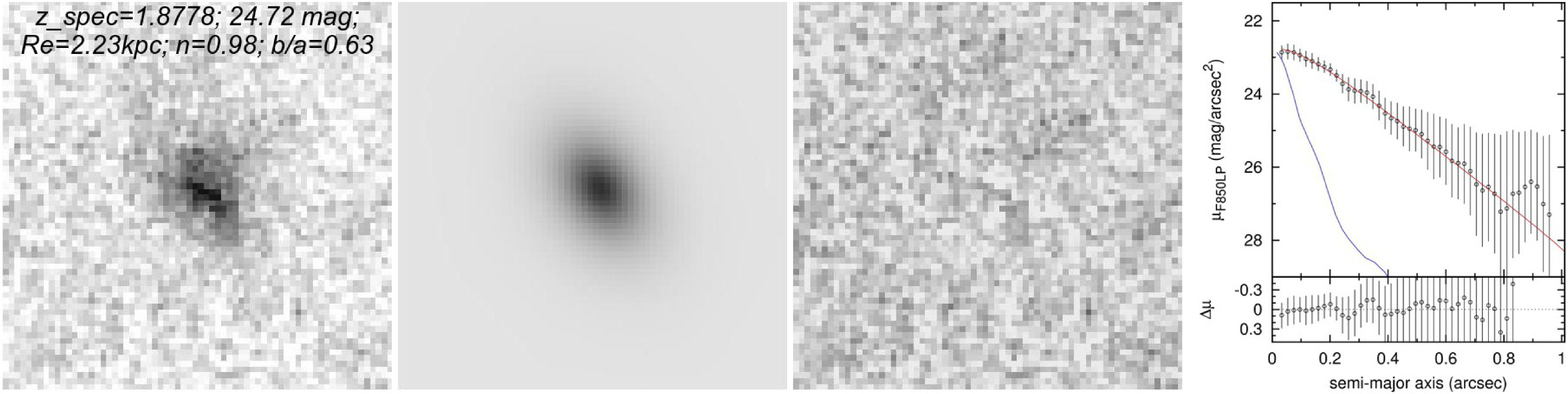}\\
\includegraphics[clip, width=0.8\textwidth]{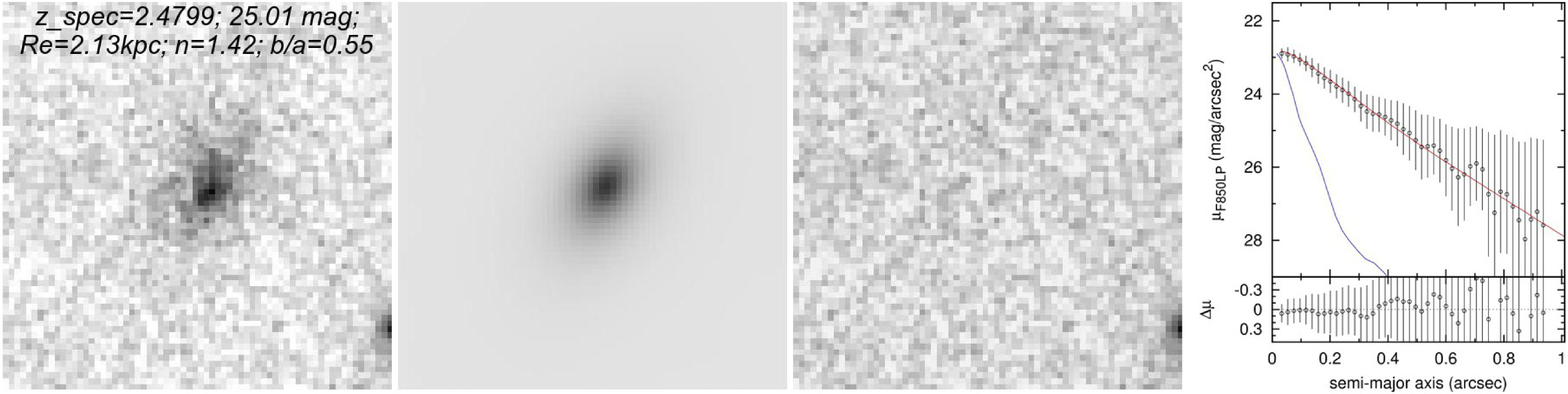}\\
\caption{Examples of 2D surface brightness modeling with GALFIT for the single-\wz~sBzK galaxies 
in the GOODS-S field. The observed \wz~image, the best-fitting model constructed by GALFIT, 
and the residual (the observed image subtracted by the model) are arranged from the leftmost panel. 
The residual images are shown with the narrower scale range to clarify the residual part. 
North is up; east is to the left. The size of each panel is 2.\ar0$\times$2.\ar0 corresponding to 
$\sim17\times17$ kpc at $z\sim2$. The rightmost panel shows the 
azimuthally averaged 1D surface brightness profile (black open circles with errors) with the profile of 
best-fitting model (red solid line) and normalized PSF (blue dashed line). The surface brightness 
of the residual is also shown in the bottom of the rightmost panel. 
}
\label{fig:acsz_gds}
\end{figure*}

\begin{figure*}
\centering
\includegraphics[clip, width=0.8\textwidth]{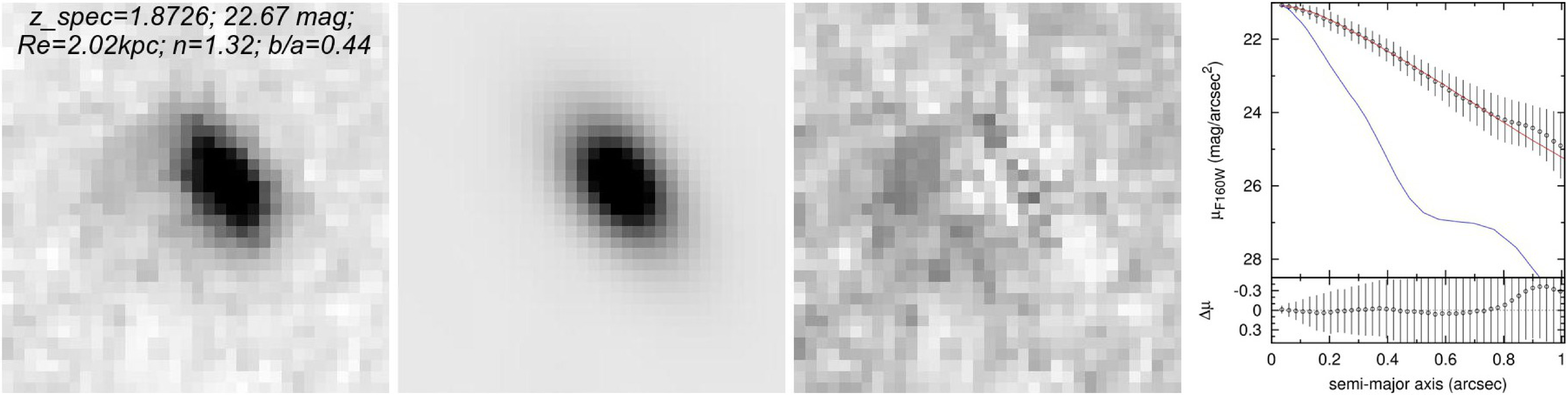}\\
\includegraphics[clip, width=0.8\textwidth]{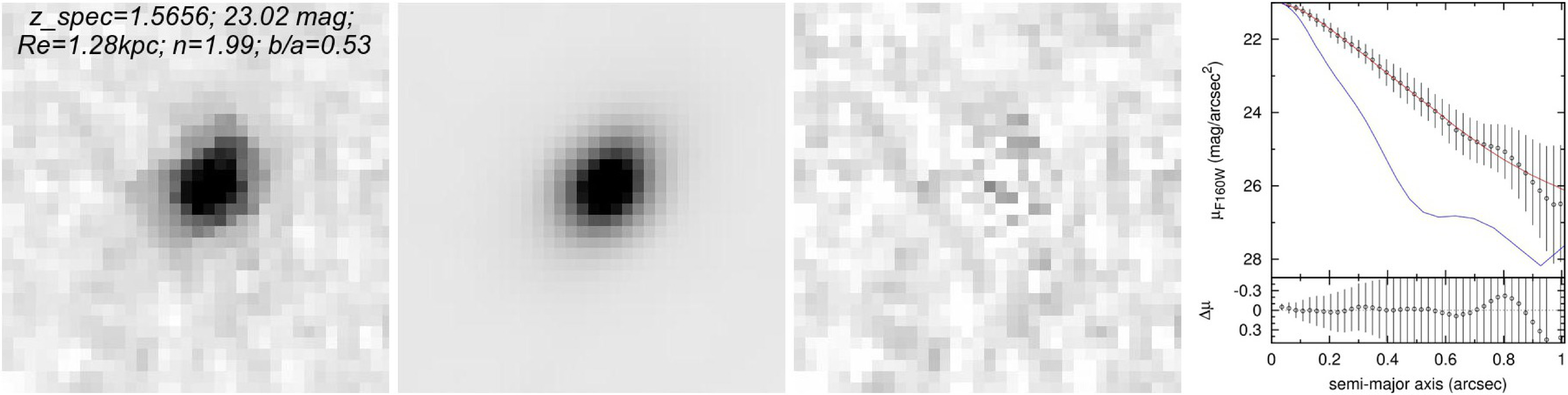}\\
\includegraphics[clip, width=0.8\textwidth]{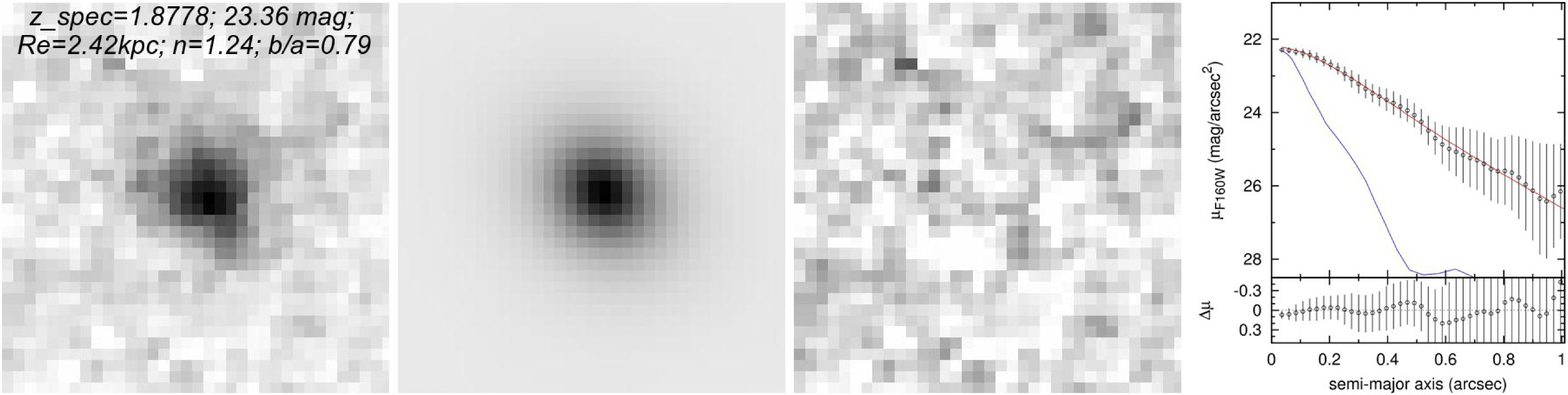}\\
\includegraphics[clip, width=0.8\textwidth]{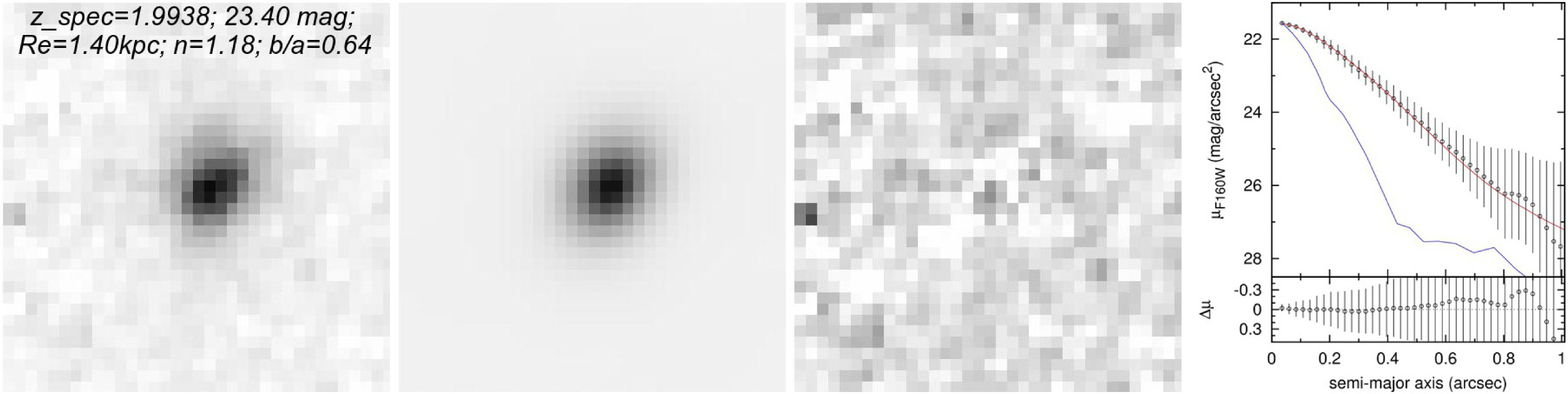}\\
\includegraphics[clip, width=0.8\textwidth]{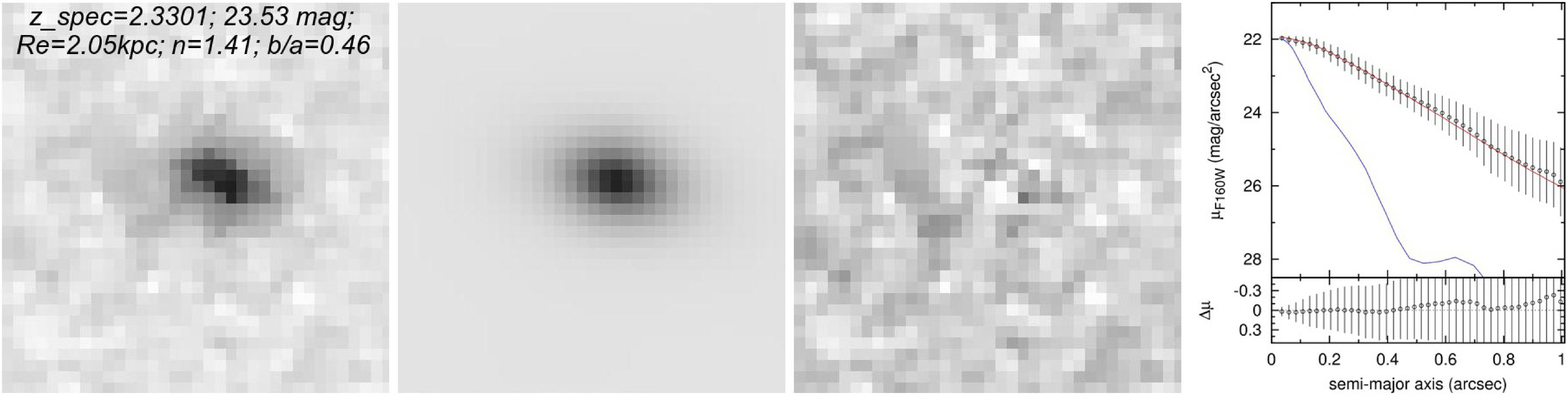}\\
\caption{Same as Figure \ref{fig:wfc3h_gds} but for the \wh~images of the single-\wh~sBzK galaxies in 
the GOODS-S field. Note that the slightly differences among PSFs shown here are due to the differences 
of adopted stars in the image taken at each epoch. 
}
\label{fig:wfc3h_gds}
\end{figure*}

\begin{figure*}
\centering
\includegraphics[clip, width=0.8\textwidth]{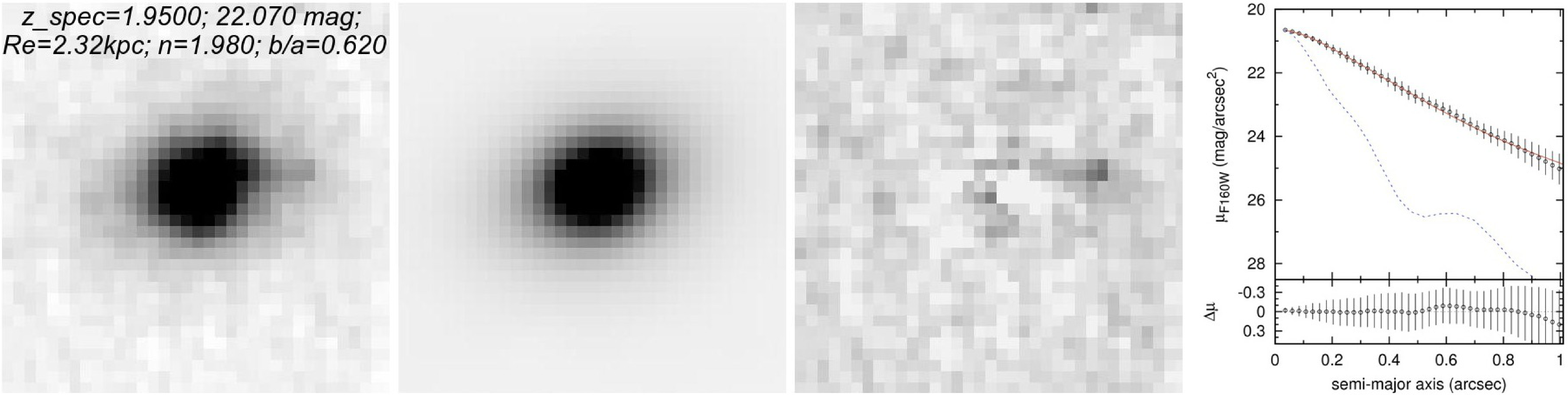}\\
\includegraphics[clip, width=0.8\textwidth]{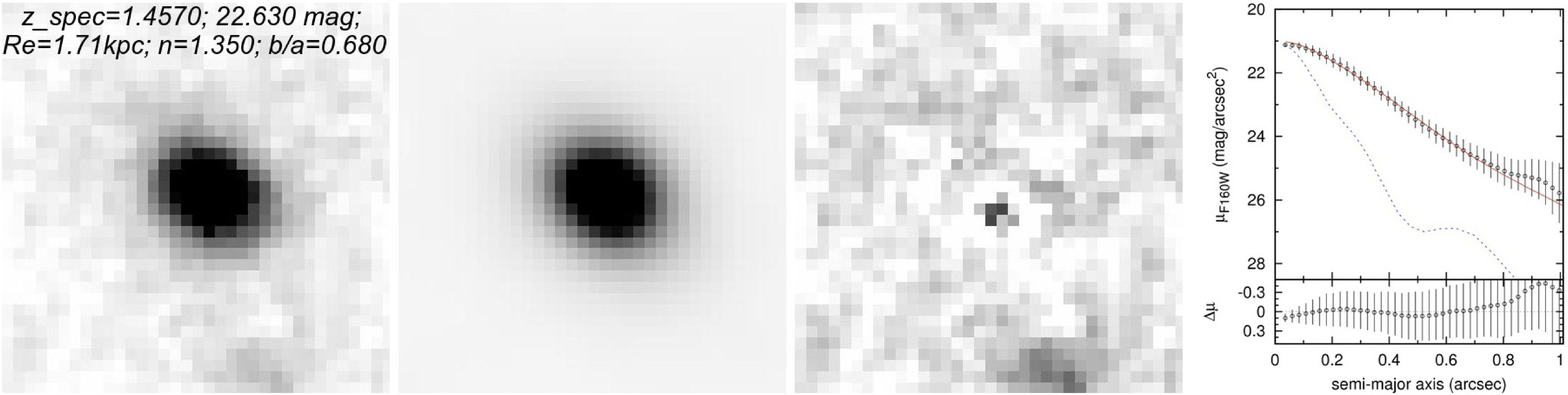}\\
\includegraphics[clip, width=0.8\textwidth]{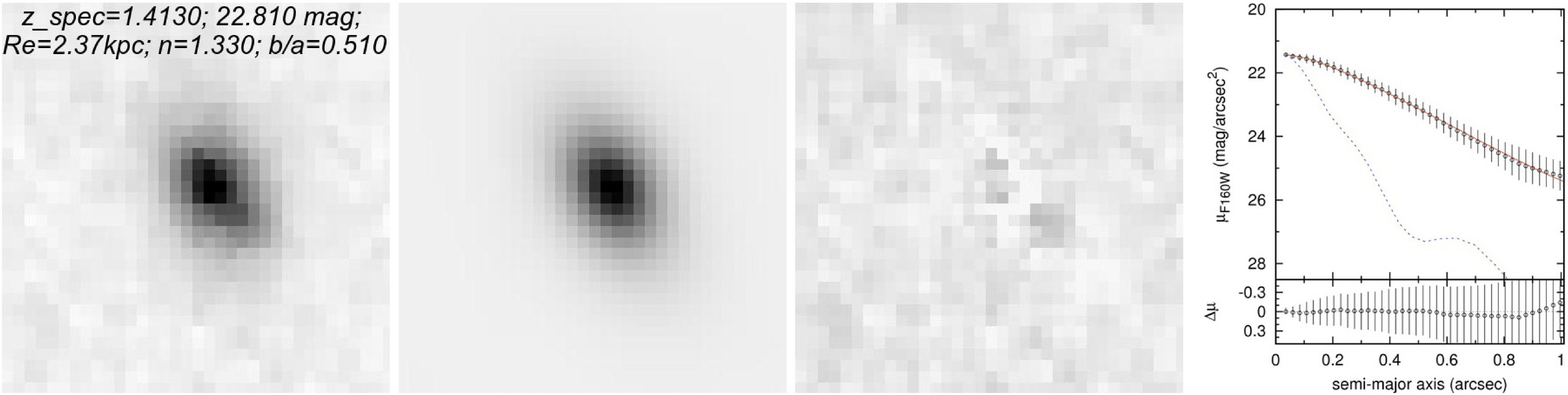}\\
\includegraphics[clip, width=0.8\textwidth]{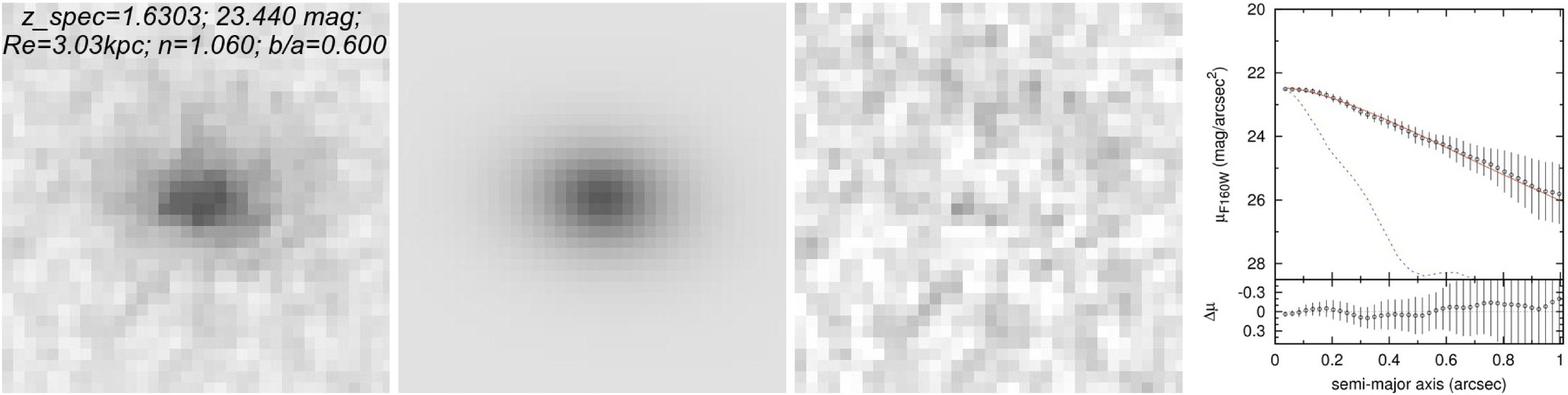}\\
\includegraphics[clip, width=0.8\textwidth]{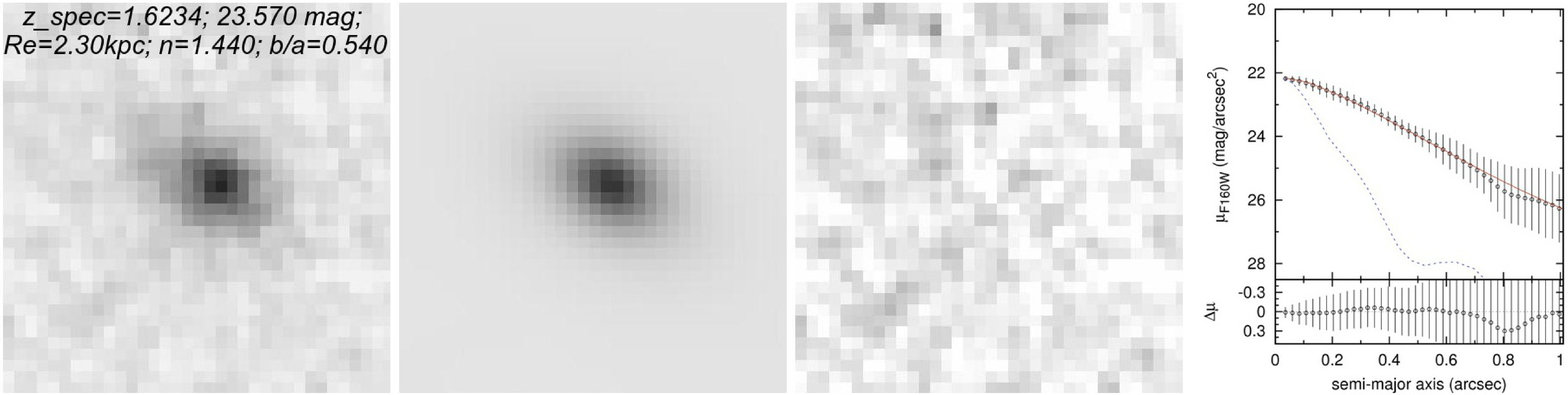}\\
\caption{Same as Figure \ref{fig:wfc3h_gds} but for the \wh~images of the single-\wh~sBzK galaxies in 
the SXDS field.
}
\label{fig:wfc3h_sxds}
\end{figure*}

\section{Structural Analysis}\label{sec:galfit}

We carried out the structural analysis in the same process as done in paper I. 
Briefly, the structural parameters of the sBzK galaxies are obtained 
by fitting their two-dimensional surface brightness distributions in \wz~or \wh~images 
with a single S\'ersic profile \citep{sersic63, sersic68} using GALFIT version 3.0 
\citep{peng10}. The S\'ersic profile is formulated by 
$I(r) = I_e \exp{\{-k_n[(\frac{r}{r_e})^{1/n} -1 ]\}}$, where $I_e$ is surface brightness 
at an effective radius $r_e$ and $k_n$ is a variable parameter coupled with 
S\'ersic index $n$ so that $r_e$ covers half of the total flux. 
The de Vaucouleurs and exponential profiles are special cases of the S\'ersic profile 
when $n=4$ and $n=1$, respectively. Seven free parameters of each object, 
i.e., position (x, y), magnitude, $n$, $r_e$, axial ratio ($b/a$), and position angle (PA), 
are determined by $\chi^2$ minimization. Initial values of all parameters 
are given from SExtractor outputs, except for the S\'ersic index which is initially set 
at $n=1.5$. As explained in paper I, the GALFIT results do not significantly depend 
on the initial S\'ersic index. Sigma images required for calculating the $\chi^2$ and 
fitting errors are constructed from variance maps produced during the reduction 
process. 

\begin{deluxetable}{ccc}
\tabletypesize{\footnotesize}
\tablecaption{Number of the single-component sBzK galaxies before and after GALFIT analysis 
\label{tab:galfit_nobj}
}
\tablewidth{0pt}
\tablehead{
\multicolumn{1}{c}{} &
\multicolumn{2}{c}{Number of sBzK galaxies (After/before GALFIT)\tablenotemark{a}} \\
\cline{2-3}\\
\multicolumn{1}{c}{Observed Fields} & 
\multicolumn{1}{c}{Single-\wz} & 
\multicolumn{1}{c}{Single-\wh} 
}
\startdata
GOODS-S & 467/573 (82\%)\tablenotemark{b} & 550/626 (88\%)  \\
SXDS & $-$\tablenotemark{c} & 1878/2044 (92\%)  \\
\enddata
\tablenotetext{a}{
The objects before GALFIT are those used in GALFIT process. 
The objects after GALFIT are those with $\chi_{\nu}^2 < 1.0$ 
and errors within 30\% of the derived parameter values. 
}
\tablenotetext{b}{
Although there are 583 single-\wz~sBzK galaxies, 10 of them are at the image edge which 
cannot be used to study the structural parameters. 
}
\tablenotetext{c}{
The ACS/F850LP images are not available in the SXDS field. 
}
\end{deluxetable}

\begin{figure*}[t]
\centering
\begin{tabular}{ccc}
\includegraphics[clip, angle=-90, width=0.3\textwidth]{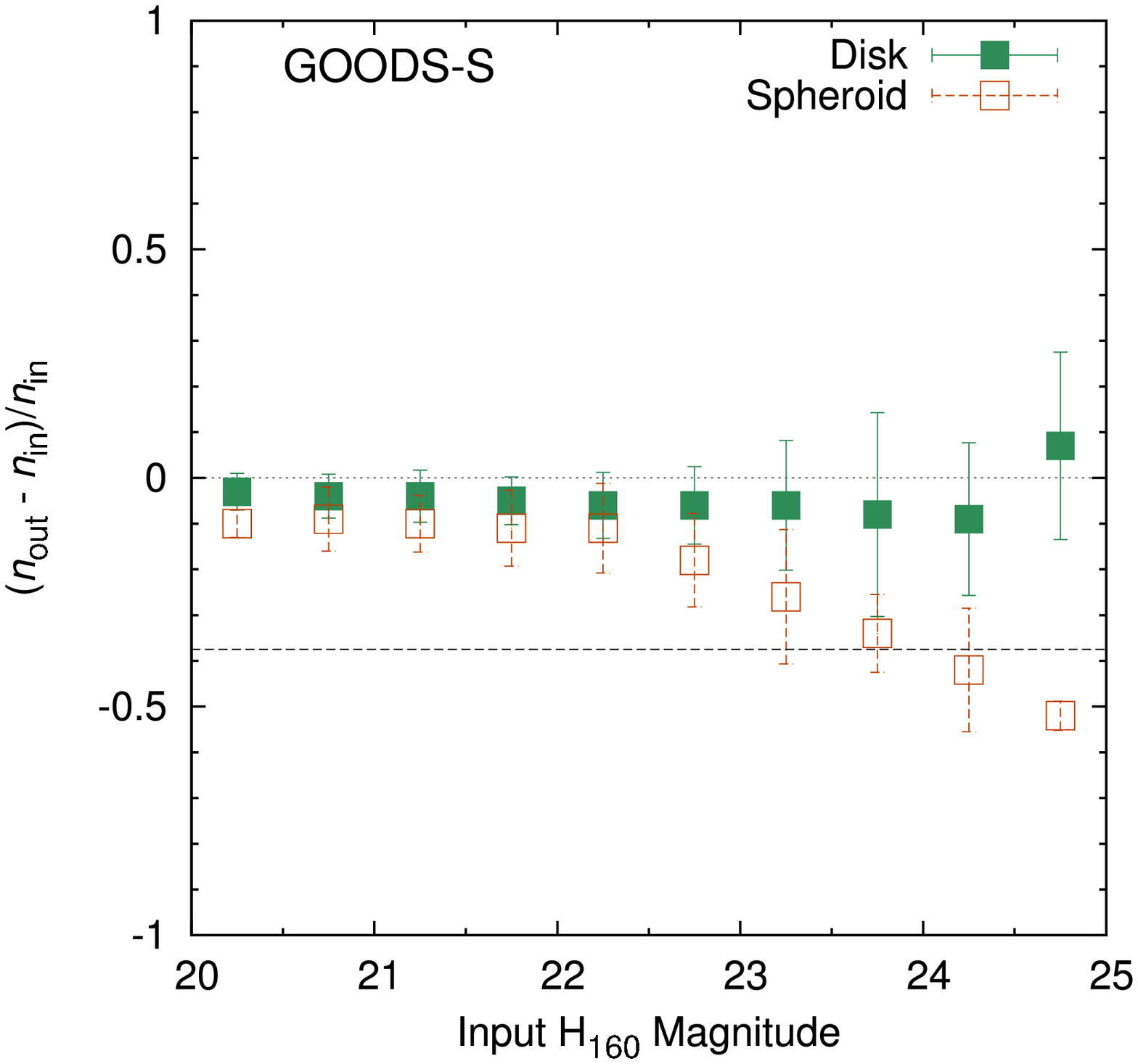} &
\includegraphics[clip, angle=-90, width=0.3\textwidth]{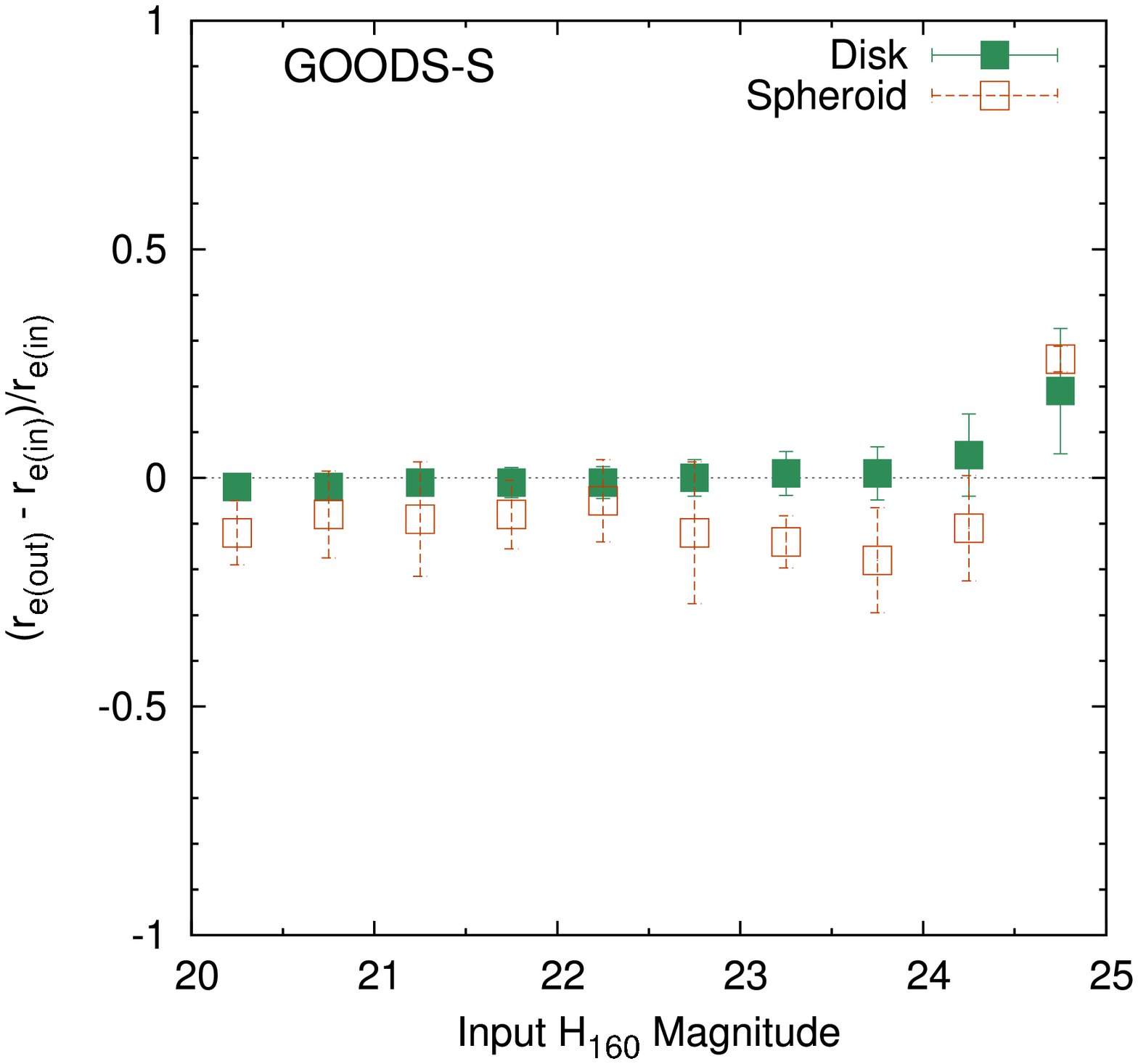} &
\includegraphics[clip, angle=-90, width=0.3\textwidth]{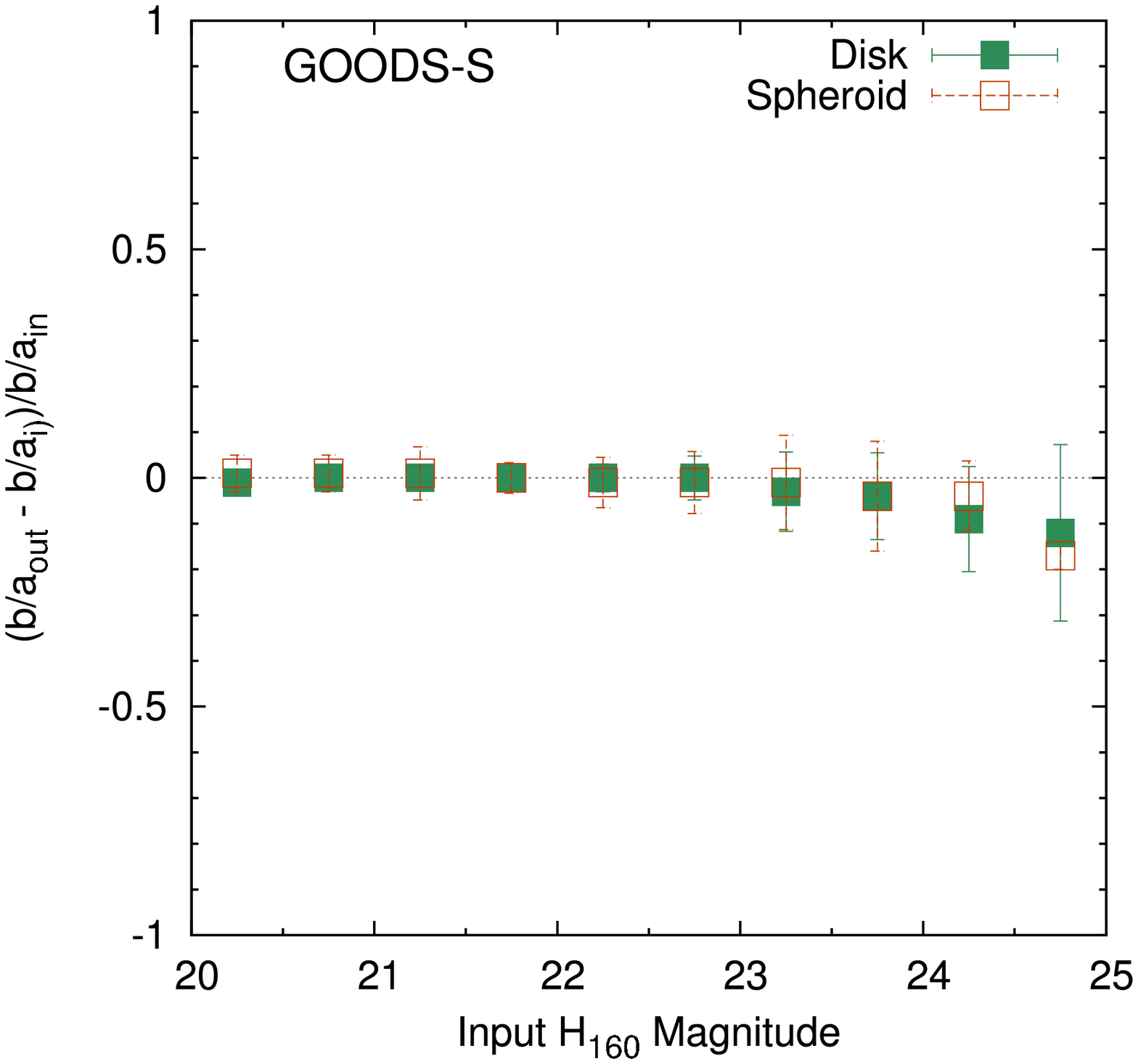} \\
\end{tabular}
\caption{Accuracy in determining structural parameters by GALFIT based on Monte Carlo simulations 
plotted as a function of input magnitudes for the GOODS-S field. From left to right panels, the figure 
shows the accuracy of S\'ersic index ($n$), effective radius ($r_e$), and axial ratio ($b/a$). 
Green solid squares represent the accuracy calculated for the artificial objects with an exponential profile 
($n=1$), while orange open squares show the accuracy for those with a de Vaucouleurs profile ($n=4$). 
Error bars correspond to $1\sigma$ of distributions. 
The horizontal dashed line at $(n_{out} - n_{in})/n_{in} = -0.375$ 
in the left panel indicates where a spheroid can be misclassified by GALFIT as a disk. 
}
\label{fig:sim_gds}
\end{figure*}

The analysis was executed for all single-component sBzK samples, i.e., 
the single-\wz~\& single-\wh~sBzK galaxies in the GOODS-S and the single-\wh~sBzK 
galaxies in the SXDS. We excluded objects with $\chi^2_{\nu}$ larger than one 
and containing inconvincibly large fitting uncertainties, i.e., objects with parameter 
errors more than 30\% of the parameter values. The excluded galaxies are mostly 
too faint to obtain reliable results. The resultant numbers of sBzK galaxies 
left after GALFIT analysis are listed in Table \ref{tab:galfit_nobj}. Over 80\% of the single-component 
sBzK galaxies in both fields show good fitting results of GALFIT. Examples of the GALFIT 
results are demonstrated in Figures \ref{fig:acsz_gds}, \ref{fig:wfc3h_gds}, and \ref{fig:wfc3h_sxds}, 
for the single-\wz, single-\wh sBzK galaxies in the GOODS-S and the single-\wh~sBzK in the 
SXDS, respectively. The figures show the observed images, best-fitting models, and residuals, 
along with one-dimensional light profile visualizing the agreement between the observed and 
modeled profiles. 

\subsection{Accuracy in Determining Structural Parameters}\label{sec:accuracy}

In paper I, we carried out the Monte Carlo simulations to examine the accuracy of 
the derived structural parameters in the \wz~images in GOODS-N field and found that the 
estimation of S\'ersic index becomes less reliable at $\wz\geq24.5$ mag. The effective 
radii derived for disk profiles are robustly recovered down to $\wz=26.0$ mag and 
more accurate than those in the spheroid case. 
Since the \wz~images in the GOODS-S are 0.5 mag deeper than those in the GOODS-N, 
we conservatively assume the same accuracy as estimated in paper I. 

\begin{figure*}
\centering
\begin{tabular}{ccc}
\includegraphics[clip, angle=-90, width=0.3\textwidth]{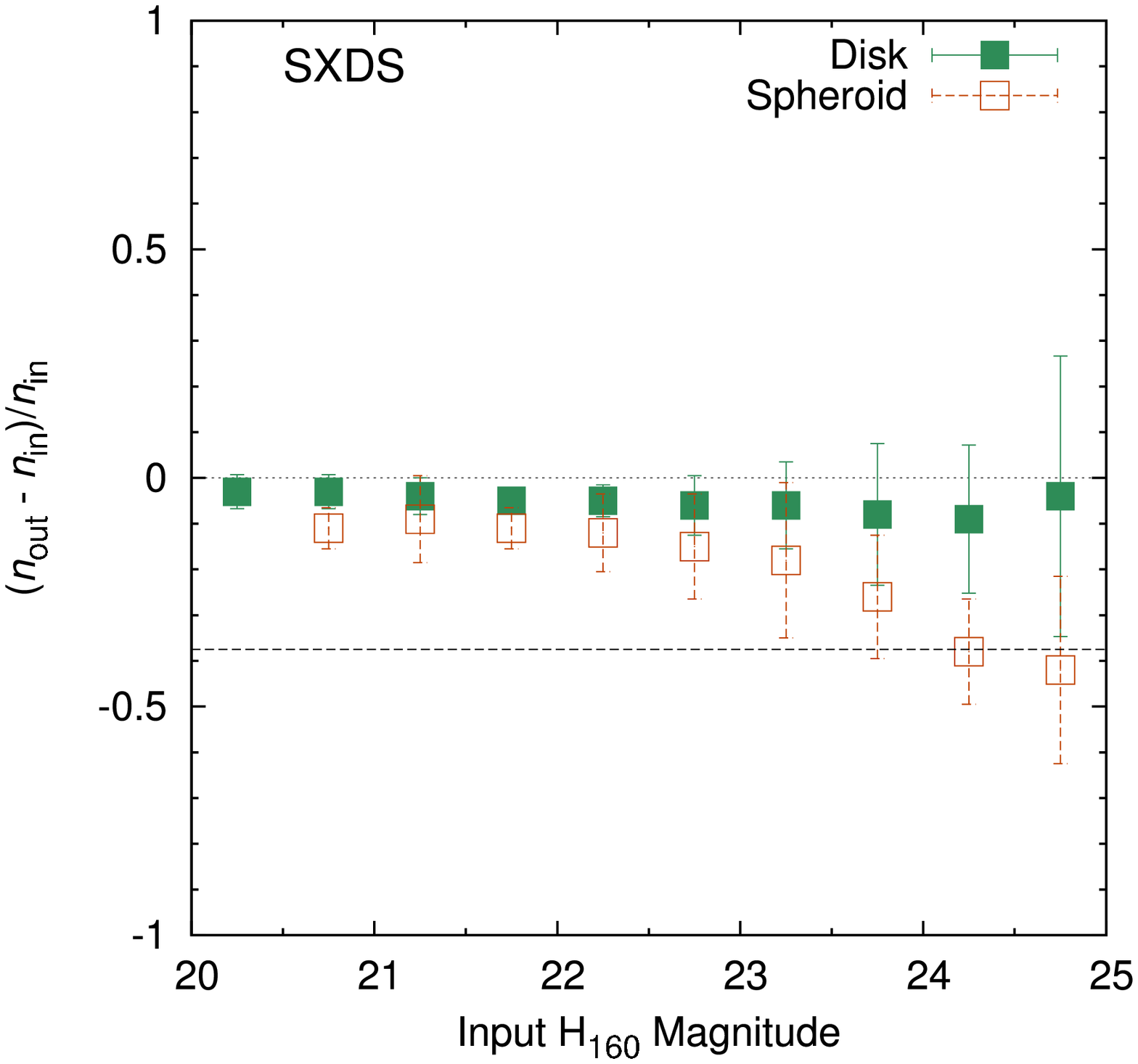} &
\includegraphics[clip, angle=-90, width=0.3\textwidth]{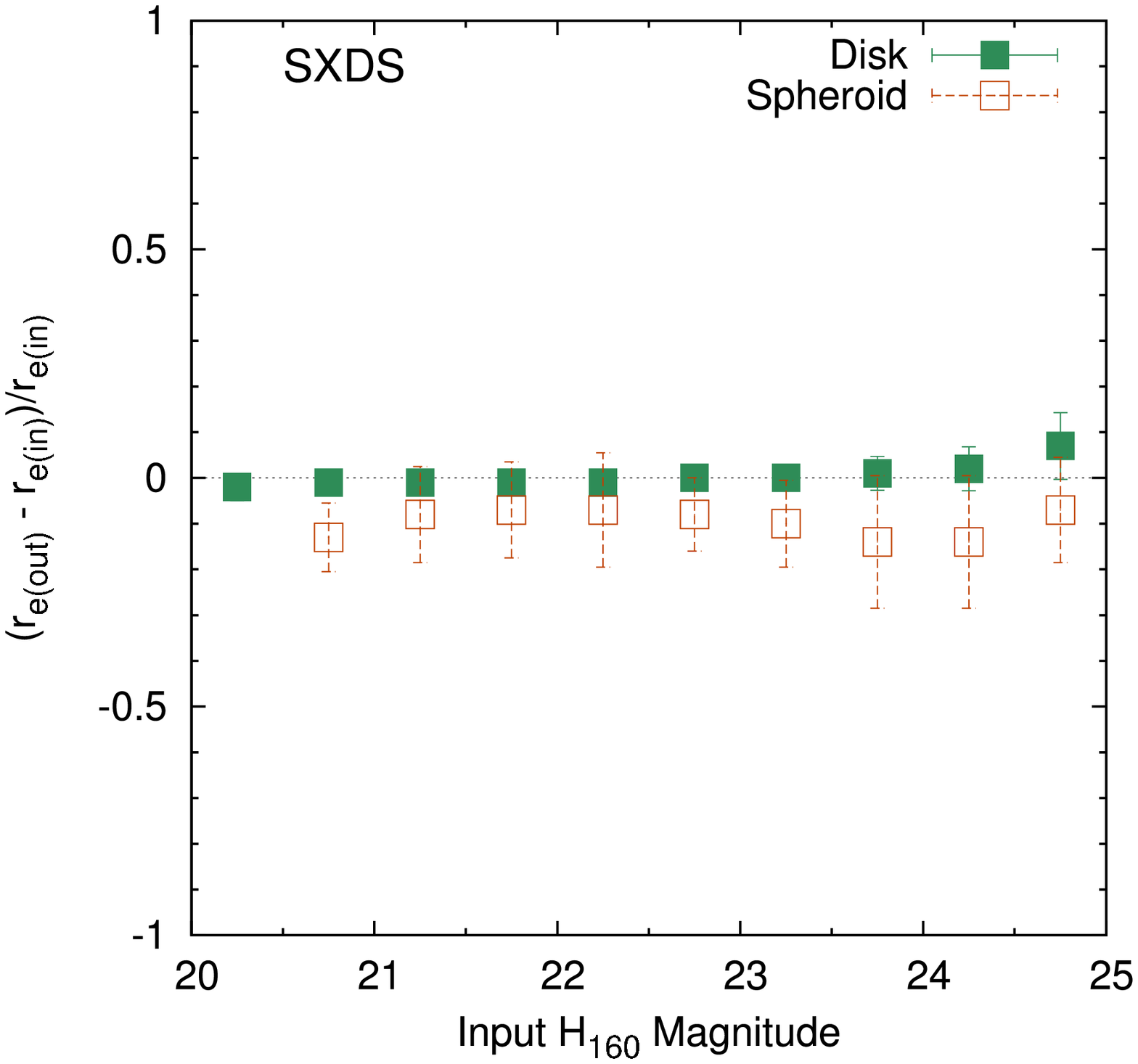} &
\includegraphics[clip, angle=-90, width=0.3\textwidth]{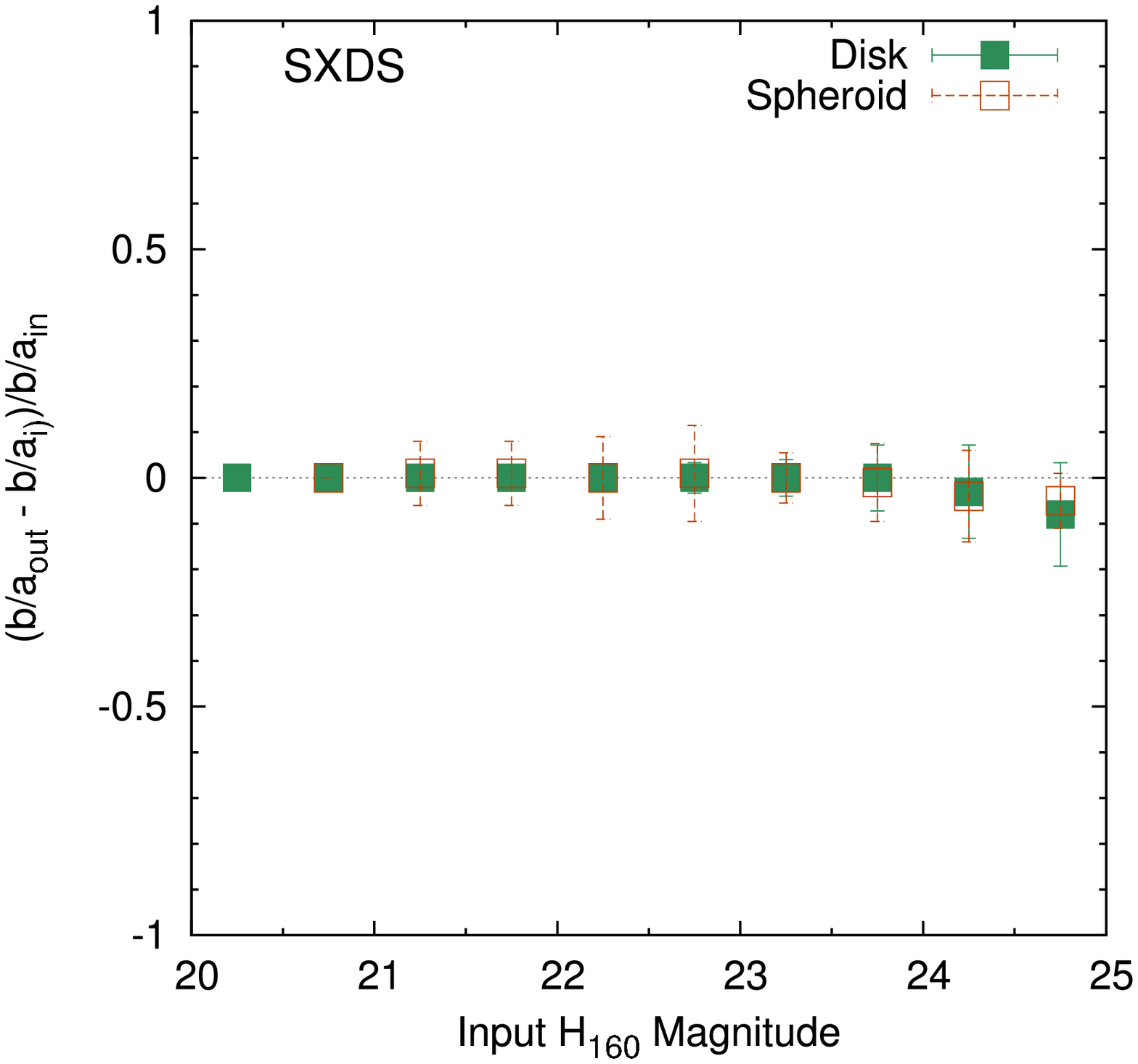} \\
\end{tabular}
\caption{Same as Figure \ref{fig:sim_gds} but for the SXDS field. }
\label{fig:sim_sxds}
\end{figure*}

In case of the \wh~images, we separately generated $\sim8000$ artificial objects in both 
GOODS-S and SXDS with random magnitude, effective radius, ellipticities, and position 
angles of $20-25$ mag, 0.\ar1$-$1.\ar0, $0.1-1.0$, and $0-180^{\circ}$, respectively. 
The ranges of the input parameters were chosen to be the same as those of the sBzK galaxies. 
The artificial objects, which are equally constructed for disks ($n=1$) and spheroids ($n=4$), 
were convolved with PSF of \wh~images and inserted into the original \wh~images. Then 
we performed the GALFIT and selected the good fits in the identical way to that used 
for the sBzK galaxies; i.e., $\chi_{\nu}^2<1.0$ and errors $<30$\%. 
The percentages of successful fittings for the \wh~artificial objects in both GOODS-S (86\%) 
and SXDS (92\%) fields are very similar to those for the single-\wh~sBzK galaxies (Table \ref{tab:galfit_nobj}). 

Figure \ref{fig:sim_gds} shows the accuracy in determining the structural parameters from 
Monte Carlo simulations in the GOODS-S field by GALFIT program along with $1\sigma$ 
distributions of the recovered structural parameters. The left panel shows the 
S\'ersic index accuracy of both disk-like ($n=1$; green solid square) and spheroid-like 
($n=4$; orange open square) artificial objects. It is shown that the recovered S\'ersic indices 
are systematically underestimated in most cases. It is not significant in case of 
simulated disks; the median difference between the input and recovered S\'ersic indices is 
no more than 10\% at all magnitude ranges. In contrast, the S\'ersic index estimation for 
the simulated spheroids becomes less accurate at fainter magnitudes; their recovered S\'ersic 
indices go below $n<2.5$ at $\wh>24.0$ mag. In other words, we are able to distinguish 
the disk-like sample from the spheroid based on the S\'ersic index criterion $n=2.5$ down 
to $\wh<24$ mag in the GOODS-S. Similar to the simulations made in the \wz~images (paper I), 
the effective radius of a disk profile is better recovered than that of the spheroid one. 
The middle panel shows that the effective radii of the simulated disks are well recovered 
down to the faint magnitudes; however, they are systematically underestimated in case of spheroid 
about $10-20\%$ except for the faintest magnitude bin. However, this underestimation is not severe 
as we show in the next section that most of our samples are disk-like, 
of which the effective radii can be accurately recovered. 
The other important parameter that we obtain from GALFIT is the apparent axial ratio ($b/a$), 
which will be used later in determining the intrinsic shape of the galaxies. 
The resulting accuracy in determining the $b/a$ is shown in the right panel of Figure \ref{fig:sim_gds}. 
The $b/a$ ratios are robustly recovered in both types of profiles down to $\wh=24$ mag, at which 
the recovered $b/a$ begins to be underestimated. 

Similarly, the accuracy in recovering the structural parameters of the artificial objects inserted in 
the SXDS is summarized in Figure \ref{fig:sim_sxds}. The simulations performed in 
both GOODS-S and SXDS fields show the similar accuracy in all parameters over most of the 
magnitude range. However, as the images in the SXDS are about one magnitude deeper than 
those in the GOODS-S, the accuracy calculated at the faintest magnitude bins (i.e., $\wh>24.0$ mag) 
in the SXDS is roughly better than that in the GOODS-S. 
The S\'ersic indices (left panel) are generally underestimated in both disk and spheroid profiles, 
but negligible ($<10.0$\%) in the disk case. The figure shows that the disk-like and spheroid profiles are 
accurately distinguishable down to $\wh<24.0$ mag. Similar to the GOODS-S, the middle 
panel indicates that the effective radii are robustly derived in the disk profile, while they are 
averagely $10-20\%$ underestimated in the case of spheroids. The axial ratios $b/a$ are 
accurately determined in both profiles over the magnitude range as seen in the right panel 
of Figure \ref{fig:sim_sxds}. 

Surface brightness of galaxies can be different due to their inclination; edge-on galaxies tend 
to show higher surface brightness than the face-on galaxies at the same total magnitude. 
The sample selection can possibly be biased toward edge-on galaxies, which consequently 
results in lower a peak of $b/a$ distribution. This selection bias was already tested in paper I 
by generating $\sim10,000$ artificial objects with the random inclination angles (axial ratios) 
and uniformly distributed magnitudes of $21-24$ mag in the $K_s$-band image. 
Their effective radii were assumed to be in the same range as the observed galaxies selected 
from the image at each magnitude bin. They found that the detection rate remains constant 
above 95\% regardless of inclination angle or axial ratio down to $K_s=23.0$ mag (Figure 10 in paper I). 
For the faintest magnitude bin ($23-24$ mag), the detection rate reaches 100\% at $b/a=0.1-0.2$ 
(edge-on case) and then becomes constant at $\sim90\%$ from $b/a=0.3$ to $b/a=1.0$, 
which is a face-on case. Therefore, there seems to be no significant bias toward selecting mostly edge-on galaxies. 
As we use the same $K_s<24$ mag limit as that in paper I, we assume that brighter surface brightness 
of edge-on galaxies does not cause any bias on our sample selection. 

\begin{figure*}
\centering
\begin{tabular}{cc}
\includegraphics[clip, angle=-90, width=0.3\textwidth]{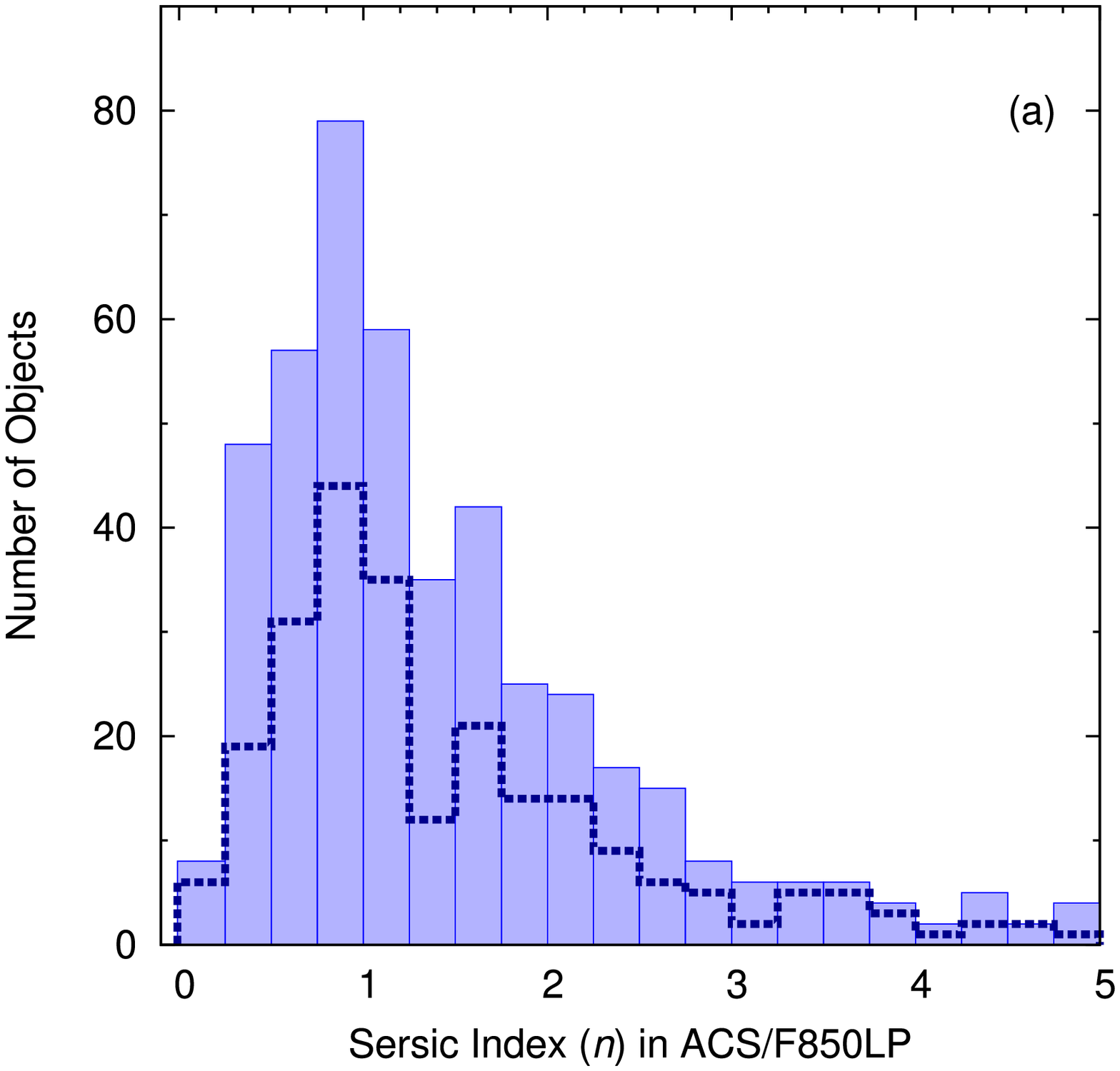} &
\includegraphics[clip, angle=-90, width=0.3\textwidth]{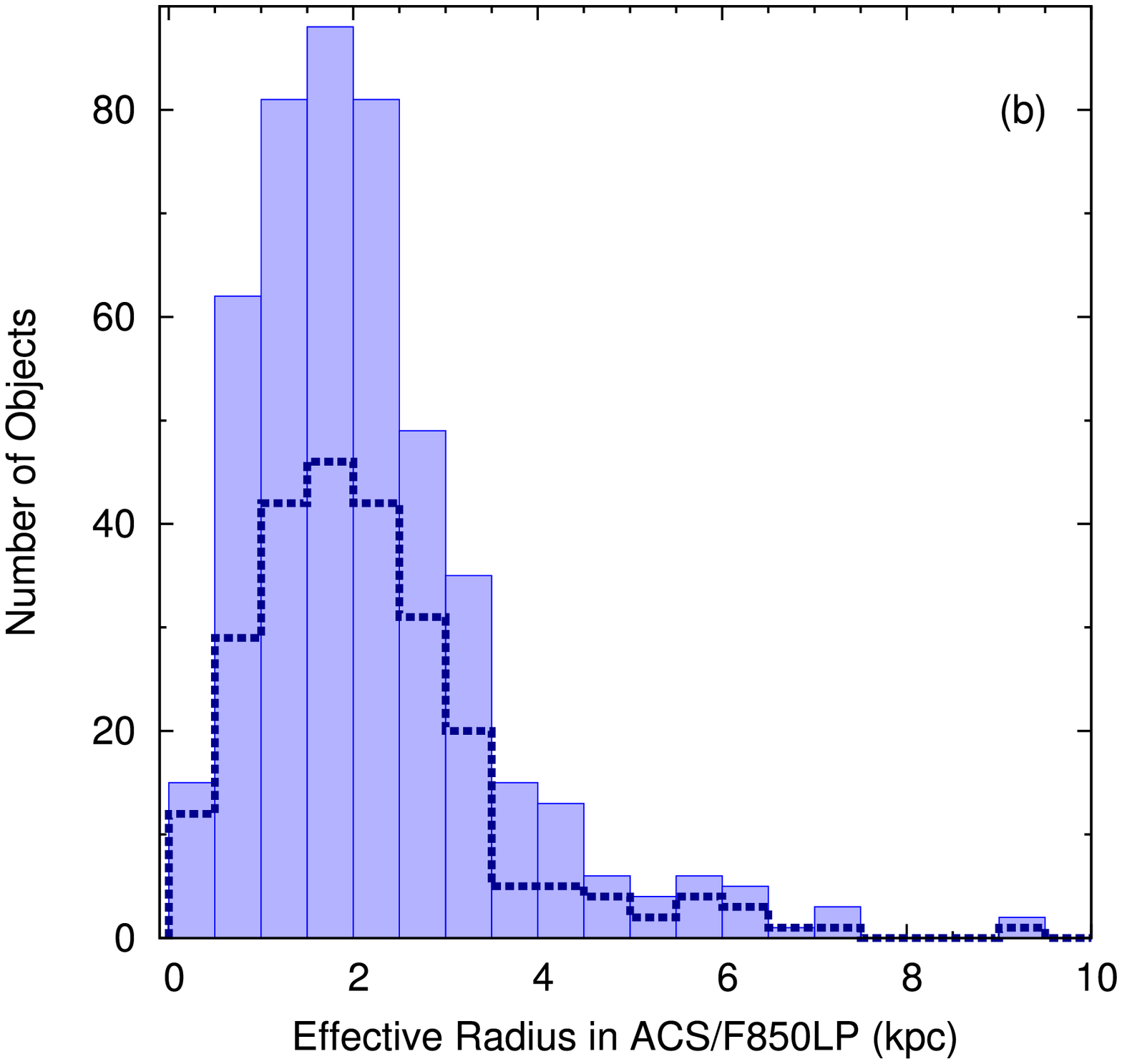} \\
\includegraphics[clip, angle=-90, width=0.3\textwidth]{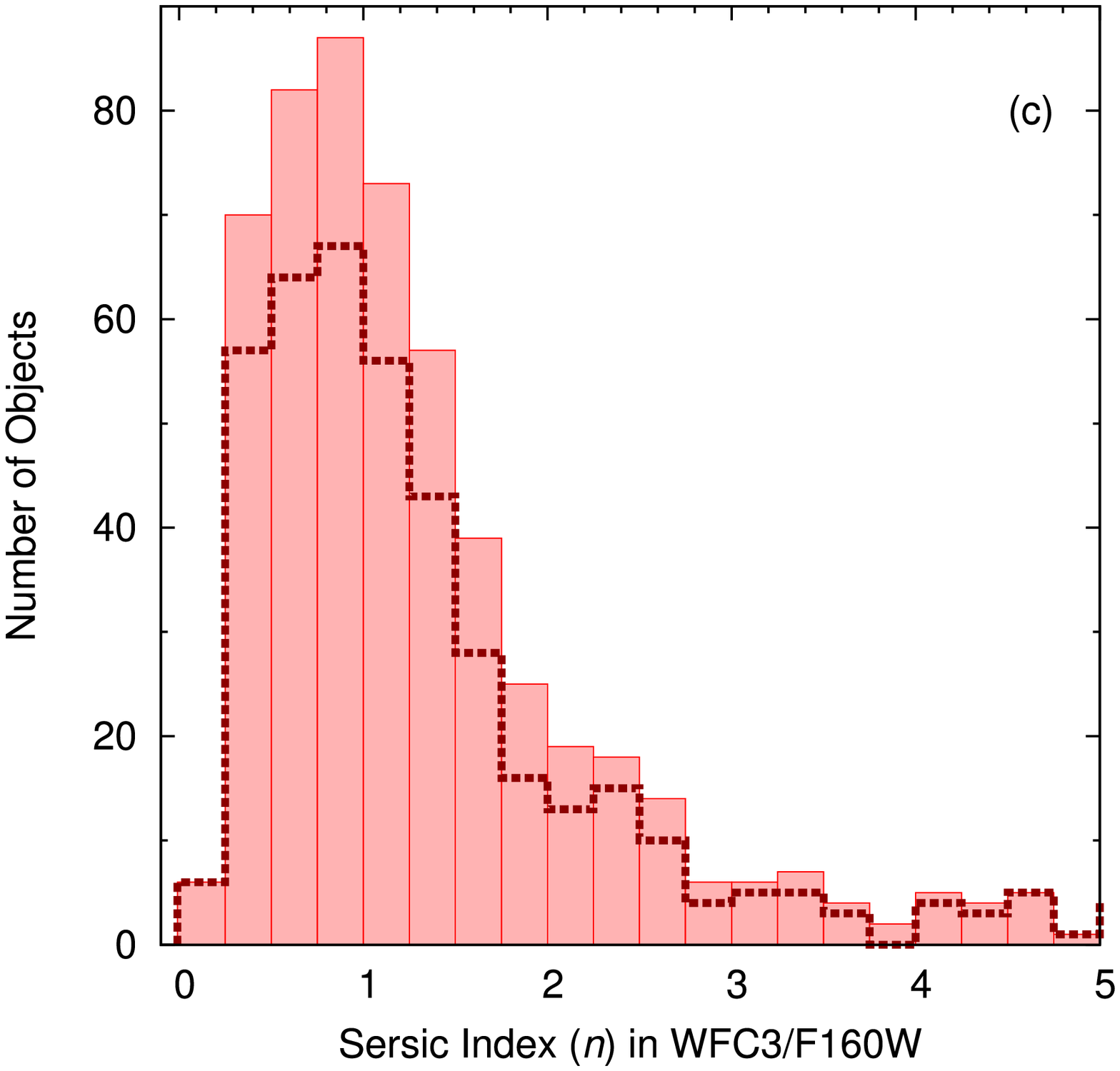} &
\includegraphics[clip, angle=-90, width=0.3\textwidth]{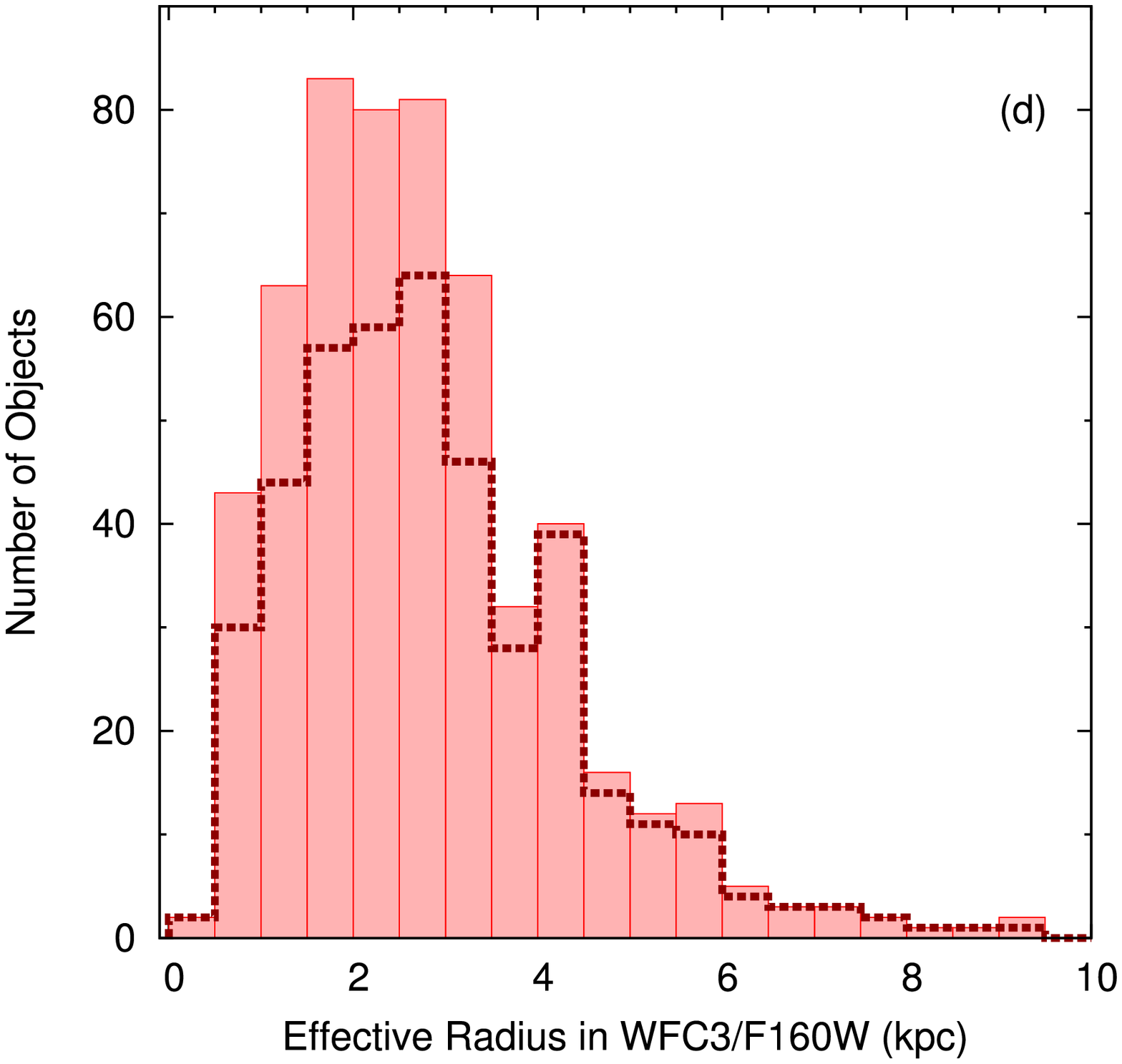} \\
\end{tabular}
\caption{Histograms of the derived S\'ersic indices ($n$) and effective radii ($r_e$) of the single-\wz~
(top panels) and the single-\wh~(bottom panels) sBzK galaxies in the GOODS-S field. The solid histograms 
represent all single-\wz~(blue) and single-\wh~(red) sBzK galaxies, whereas the open 
histograms are for the galaxies with $\wz<24.5$ mag and $\wh<24.0$ mag for the single-\wz~and the 
single-\wh~sBzK galaxies, respectively. The effective radii are described in unit of kpc calculated by 
assuming the photometric redshift or spectroscopic redshift (if available) for each object. 
}
\label{fig:galresult_gds}
\end{figure*}

\section{Results}\label{sec:result}
\subsection{Structural Parameters}
\subsubsection{GOODS-S}

In this section, we present results of GALFIT of the single-\wz~and the single-\wh~sBzK galaxies 
in the GOODS-S field. 
Distributions of the derived S\'ersic indices and effective radii are shown in Figure \ref{fig:galresult_gds}. 
The effective radii are indicated in the unit of kpc calculated by assuming the photometric redshift 
or spectroscopic redshift (if available) of each object. 
In the figure, we also show the histograms for the brighter sample; galaxies with $\wz<24.5$ mag (top panels) 
and $\wh<24.0$ mag (bottom panels). 
It is seen in Figure \ref{fig:galresult_gds} (a) that most of the single-\wz~sBzK galaxies 
have $n\sim1$ regardless of the magnitude limit. 
The median $n$ for all and the brighter single-\wz~sBzK galaxies are the same at $n=1.17$. 
Figure \ref{fig:galresult_gds} (b) shows that the effective radii of most single-\wz~galaxies are in 
the range of $r_e =1.0-3.0$ kpc, regardless of magnitude limit. 
The median $r_e$ are 1.89 and 1.91 kpc for all single-\wz~galaxies and the brighter samples, respectively. 

Figures \ref{fig:galresult_gds} (c) and (d) show the $n$ and $r_e$ histograms for the single-\wh~
sBzK galaxies. Similar to those in the \wz~images, the S\'ersic indices of the single-\wh~sBzK galaxies 
mostly distribute around $n\sim1$ with the median values of $n=1.09$ and $n=1.08$ for all samples 
and those with $\wh<24.0$ mag, respectively (Figure \ref{fig:galresult_gds} (c)). 
Figure \ref{fig:galresult_gds} (d) shows that the effective radii of the single-\wh~galaxies are typically around 
$r_e=1.5-4.0$ kpc. 
The median $r_e$ for all single-\wh~sBzK galaxies is 
$2.52$ kpc, similar to the median of $r_e=2.64$ kpc for the brighter sample. 

We performed Kolmogorov-Smirnov (K-S) tests to the $n$ and $r_e$ distributions of 
all and the brighter sample and found that we could not reject the null hypothesis that the $n$ and 
$r_e$ histograms of both samples are drawn from the same populations at more than 
$98\%$ confidence levels. We need to use as large number as possible of the sBzK galaxies 
in order to determine their intrinsic shape with sufficient statistical accuracy. 
Since the distributions of both $n$ and $r_e$ are consistent regardless of the magnitude limit, 
we keep using the whole samples of both single-\wz~and single-\wh~sBzK galaxies to determine 
the intrinsic shape in section \ref{sec:discuss}. We divide the sBzK galaxies into disk-like and 
spheroid-like samples according to their S\'ersic index. The disk-like galaxies are those with 
S\'ersic indices of $0.5 \leq n < 2.5$, whereas the spheroid ones are those with $n\geq 2.5$. 
In both cases of the single-\wz~and single-\wh~sBzK galaxies, majority of them (72\%) are similarly 
classified as disk-like galaxies. 

\begin{figure*}
\centering
\begin{tabular}{cc}
\includegraphics[clip, angle=-90, width=0.3\textwidth]{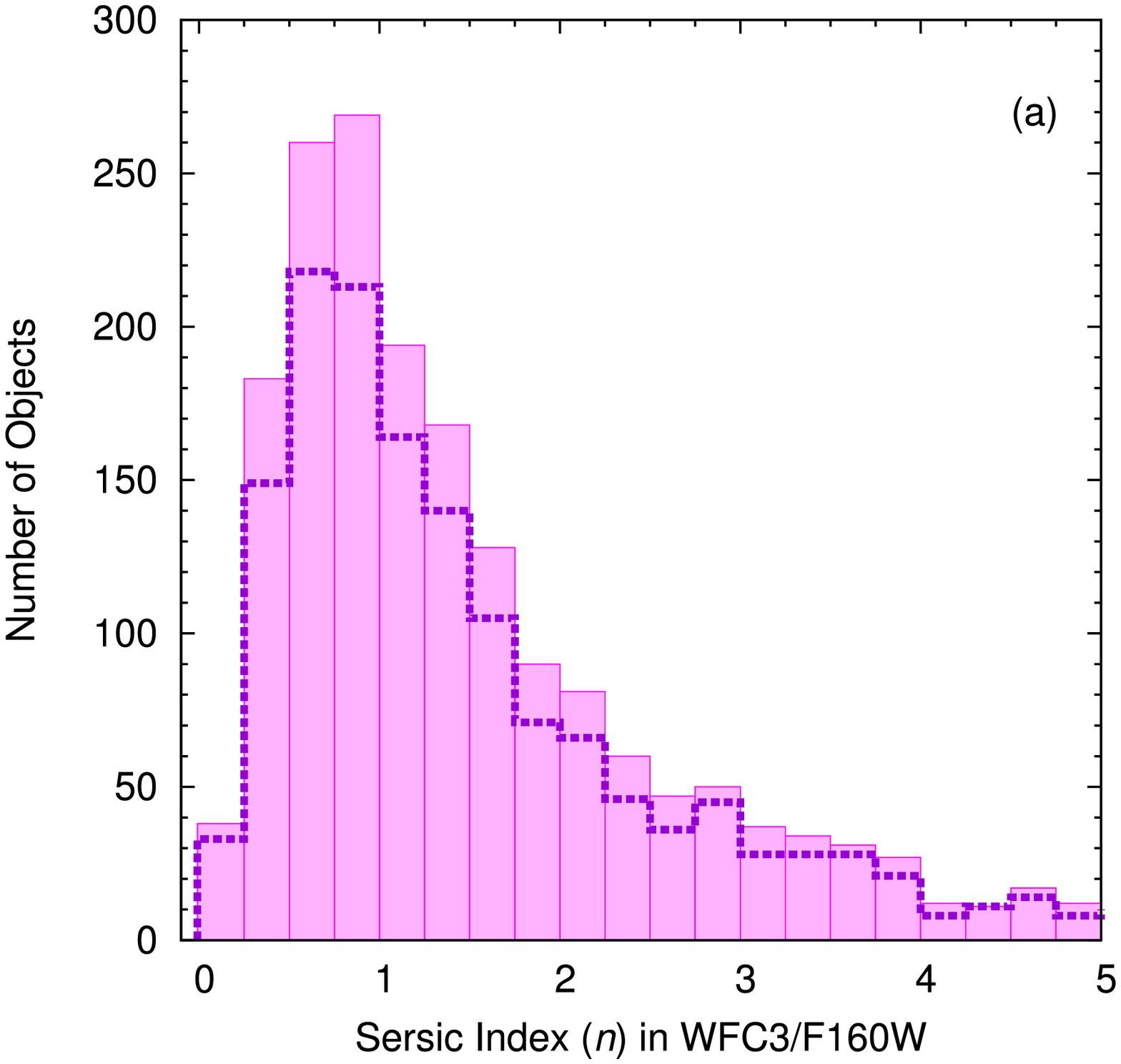} &
\includegraphics[clip, angle=-90, width=0.3\textwidth]{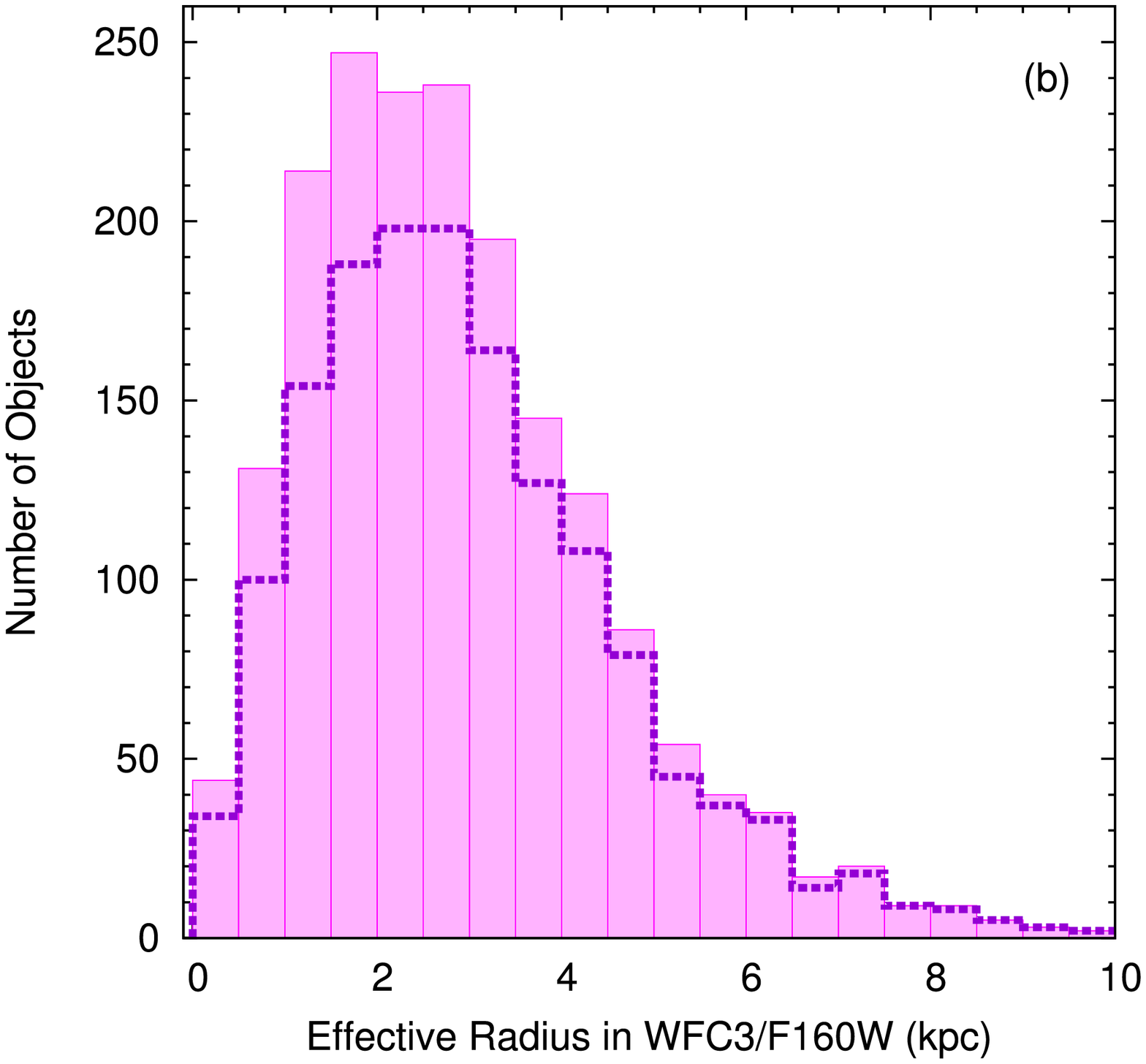} \\
\end{tabular}
\caption{Histograms of the derived S\'ersic index ($n$) and effective radius ($r_e$) of the single-\wh~sBzK 
galaxies in the SXDS field. Solid magenta and open violet histograms show samples of all and $\wh<24.0$ 
mag single-\wh~sBzK galaxies in the field, respectively. 
}
\label{fig:galresult_sxds}
\end{figure*}

\subsubsection{SXDS}

We illustrate the histograms of the S\'ersic indices ($n$) and the effective radii ($r_e$) of the single-\wh~
sBzK galaxies in Figures \ref{fig:galresult_sxds} (a) and (b), respectively. Figure \ref{fig:galresult_sxds} (a) 
shows that the single-\wh~sBzK galaxies in the SXDS mainly have the S\'ersic indices of $n\sim1$ 
in both all and the brighter sample. The median values of both samples ($n=1.23$) are very close 
to those obtained in GOODS-S. 
Using the same criteria as those in the GOODS-S, we categorized 67\% of our all 
single-\wh~sBzK galaxies in the SXDS as the disk-like sample. Figure \ref{fig:galresult_sxds} (b) 
also shows that the effective radii range between $r_e=1.5$ and $r_e=4.0$ kpc in both cases. 
The median $r_e$ of the whole sample is $2.62$ kpc, which is similar to that of $\wh<24.0$ sBzK galaxies 
(median $r_e=2.76$ kpc) and almost equal to those in the GOODS-S (median $r_e=2.52$ kpc). 

\begin{figure*}
\centering
\begin{tabular}{cc}
\includegraphics[clip, angle=-90, width=0.3\textwidth]{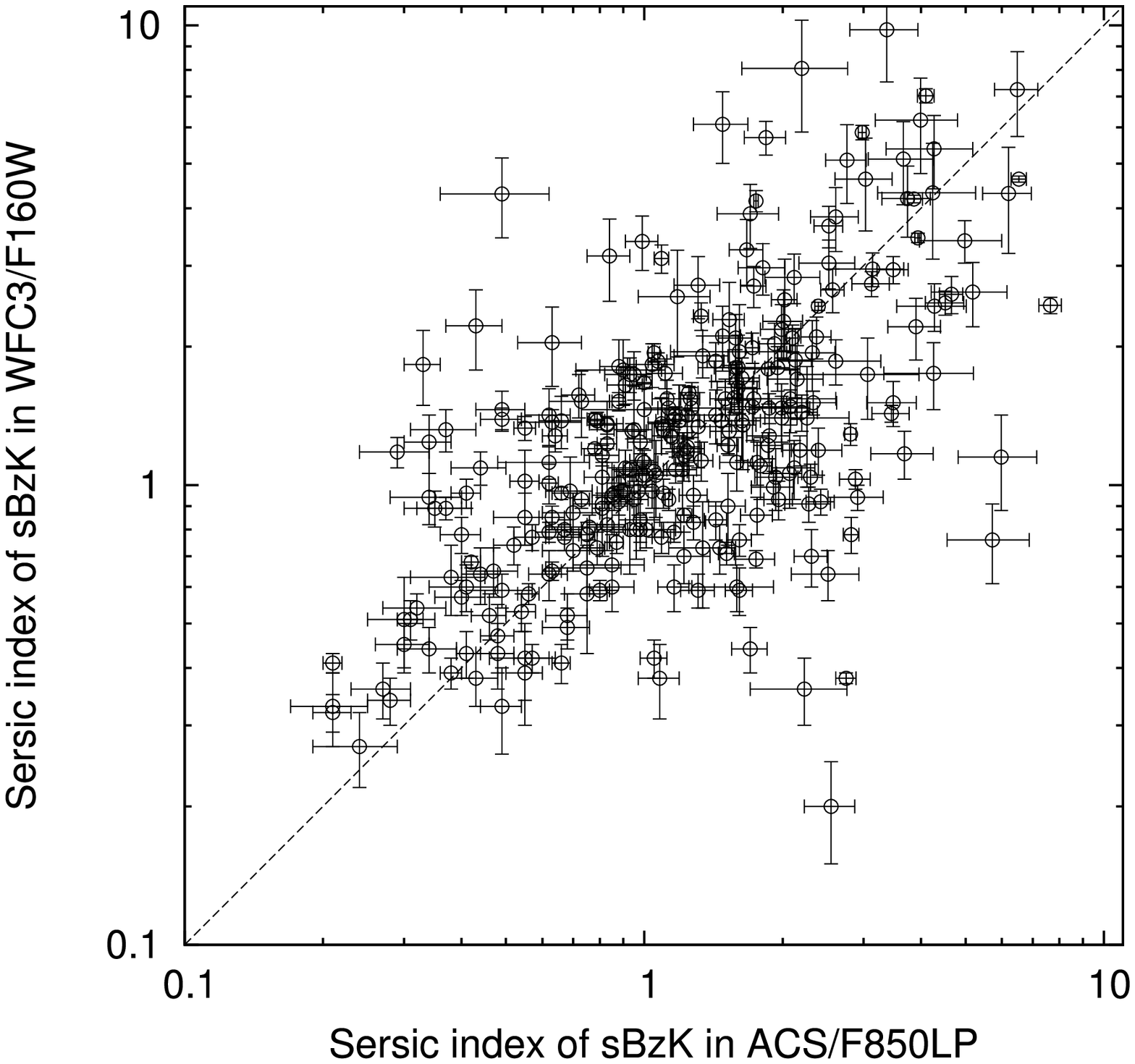} &
\includegraphics[clip, angle=-90, width=0.3\textwidth]{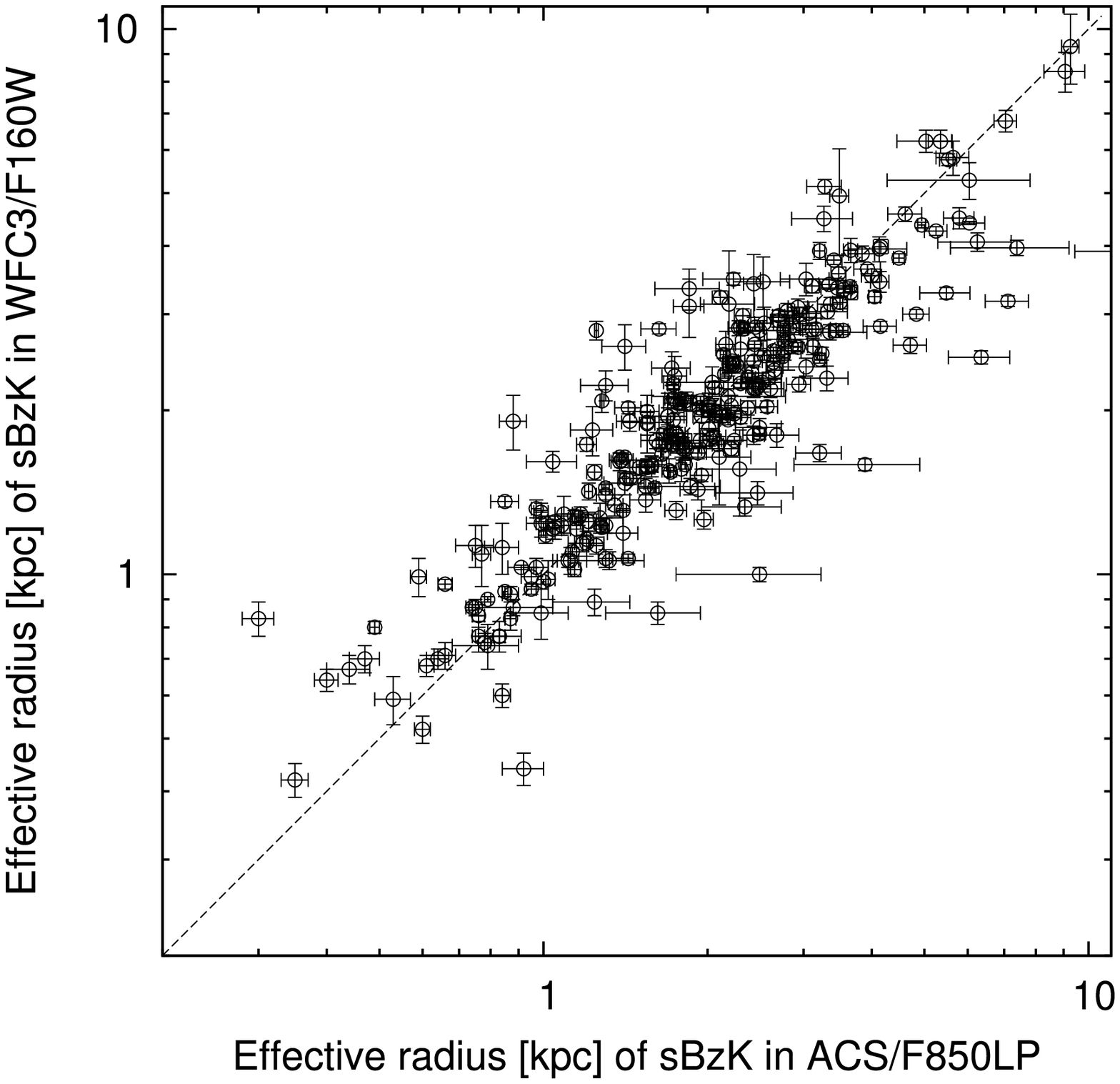} \\
\includegraphics[clip, angle=-90, width=0.3\textwidth]{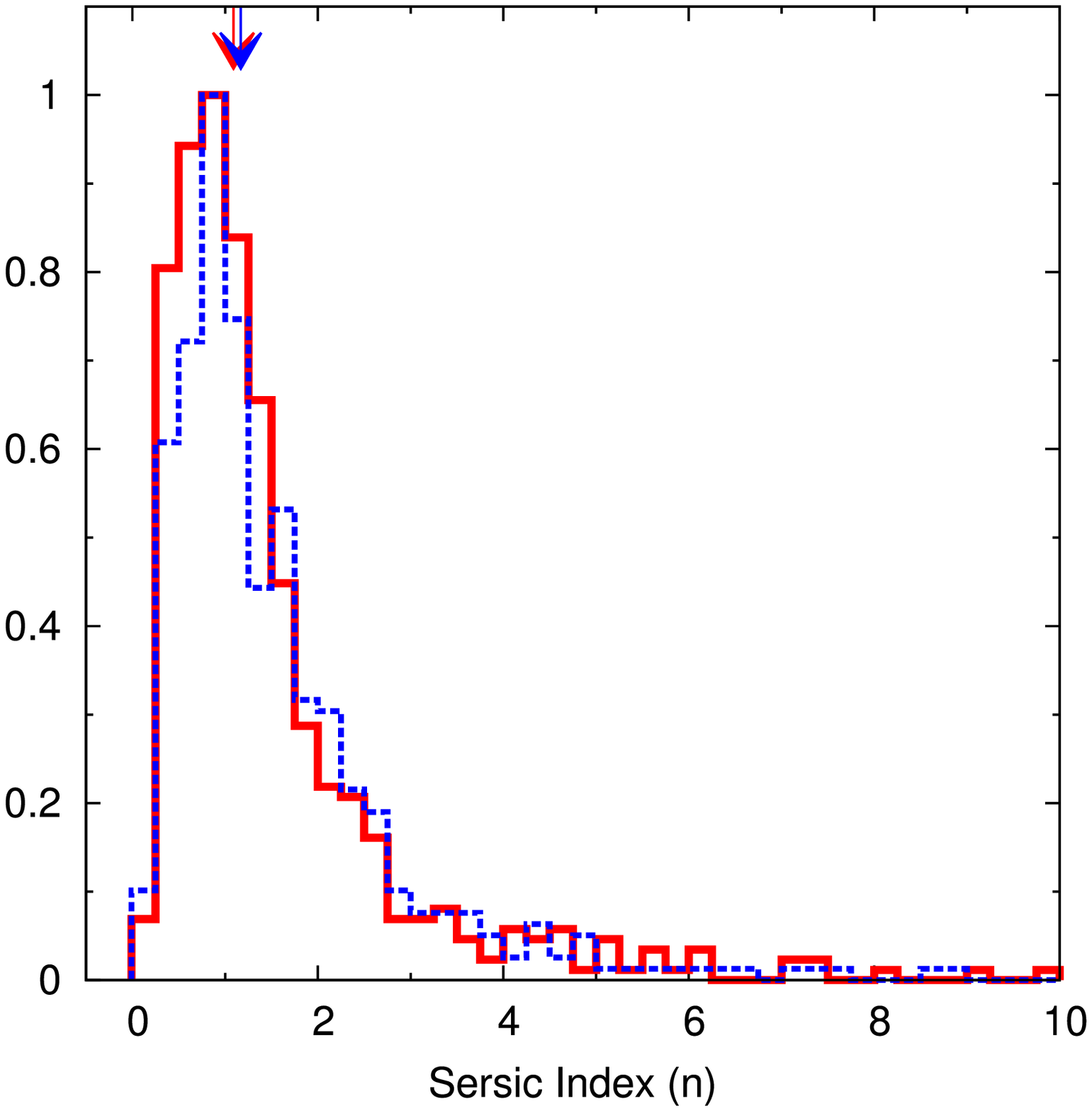} &
\includegraphics[clip, angle=-90, width=0.3\textwidth]{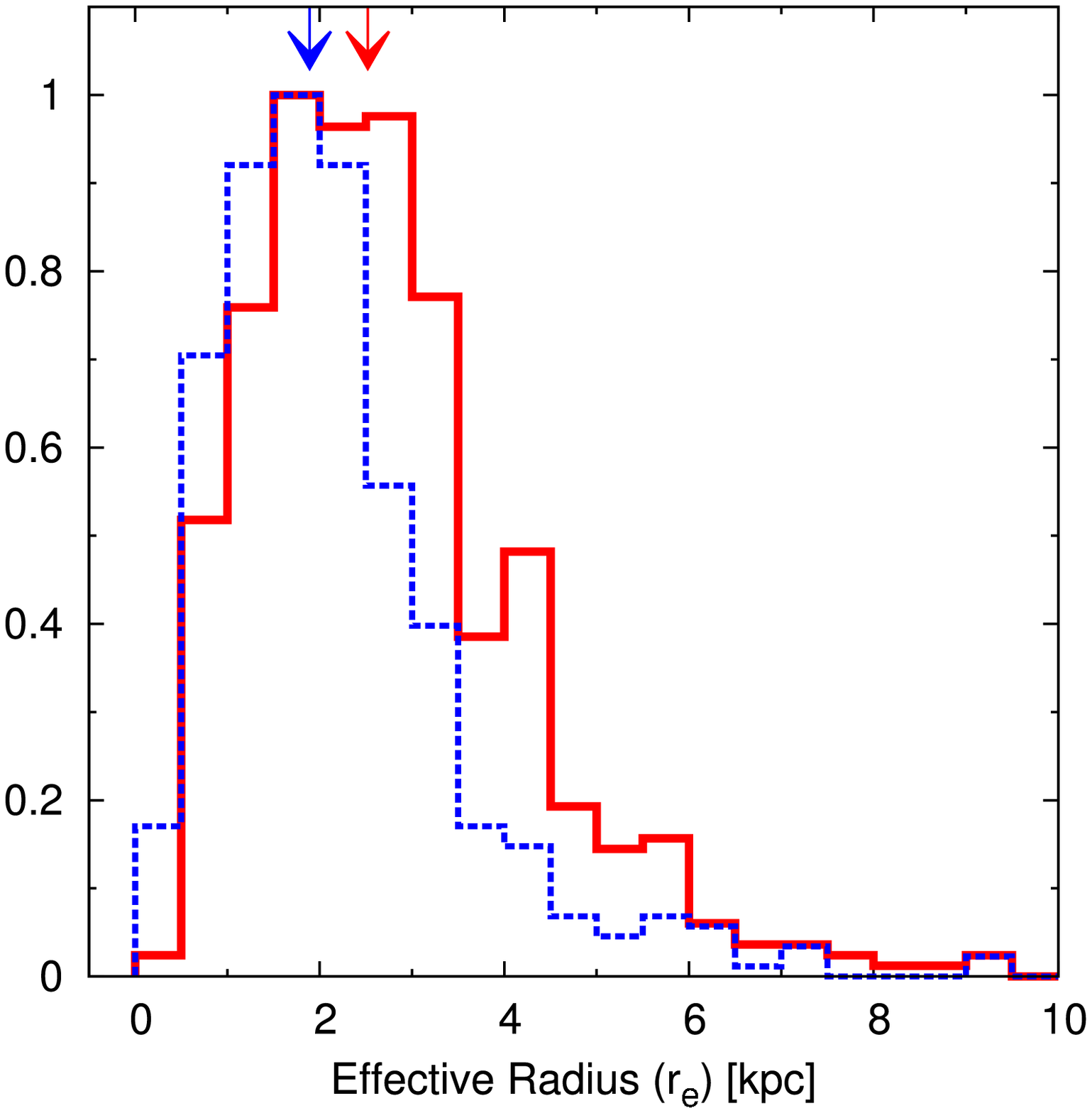} \\
\end{tabular}
\caption{
{\it Top panels}: Comparison of S\'ersic indices ({\it left}) and effective radii ({\it right}) of the 
single-component sBzK galaxies in GOODS-S detected in both \wz~and \wh~images. 
{\it Bottom panels}: Histograms of S\'ersic indices ({\it left}) and effective radii ({\it right}) of all 
single-component sBzK galaxies in GOODS-S. Blue and red histograms indicate the S\'ersic 
indices or the effective radii of the galaxies derived from \wz~and \wh~images, respectively. 
Distributions are normalized to peak at unity. The median values are also shown by arrows with 
corresponding colors. 
}
\label{compareresults}
\end{figure*}

\subsubsection{Comparison between the Rest-frame UV and Optical Wavelengths}

As mentioned in section \ref{intro}, a star-forming galaxy may have different structure when 
being observed in different wavelengths. Star-formation activity probably dominates the 
rest-frame UV structure; therefore, studying the rest-frame optical structure is crucial 
to understand the galaxy morphology in the wavelength where more evolved stellar component 
dominates. In this section, we compare the structural parameters obtain from GALFIT between 
the rest-frame UV and optical wavelengths. Comparison of S\'ersic indices and effective radii 
of the single-component sBzK galaxies between in rest-frame UV and optical wavelengths in 
the GOODS-S field is shown in Figure \ref{compareresults}. 
Top left panel shows the S\'ersic index comparison for the galaxies detected in both \wz~and 
\wh~images. It is important to note that there are 296 galaxies that are detected as a single 
component in both images. S\'ersic indices derived from both wavelength ranges of most 
galaxies are in agreement with each other. $1\sigma$ scatter of S\'ersic index ratio ($n_{opt}/n_{UV}$) 
is $\pm0.2$. Similar comparison but for the effective radius is shown 
in the top right panel. The effective radii derived in both wavelength range seem 
to be almost equal with little scatter. 

Bottom panels show histograms of all single-component sBzK galaxies detected in either 
\wz~or \wh~images. The bottom panel shows that the S\'ersic index distributions of 
the sBzK galaxies in both wavelengths are averagely very similar and their median 
values are comparable with slightly larger S\'ersic index in the rest-frame UV wavelength. 
The median S\'ersic indices are 1.17 and 1.09 in the rest-frame UV and rest-frame optical 
wavelengths, respectively. 
Histogram of the effective radii is shown in the bottom right panel. The effective radii of all 
single-\wz~galaxies distribute in narrower range and are apparently smaller 
than those of all single-\wh~samples. Their median $r_e$ values are 1.89 and 2.52 
in the rest-frame UV and optical wavelengths, respectively. The ratio of median $r_e$ in 
rest-frame optical to that in rest-frame UV is 1.33, which agrees well with the ratio of 1.37 
found in the star-forming galaxies at the similar redshifts (BM/BX and LBGs) by \cite{swinbank10}. 
The effective radii in the rest-frame optical wavelength are also roughly in agreement with 
those derived for the sBzK and BM/BX galaxies at the same redshift by \cite{overzier10} and \cite{mosleh11}, respectively. 
It suggests the similar structures between the sBzK galaxies and the BM/BX or 
LBGs at $z\sim2$. 

The ratio of median $r_e$ seems to contradict with the $r_e$ comparison for 
individual galaxy that appears as a single component in both wavelength range 
(57\% of all single-\wh~sBzK; top right panel) with the median $r_{e,opt}/r_{e,UV}$ of 1.0. 
This is mainly because the galaxies with larger effective radii in the 
\wh~image appear as multiple components in the \wz~image; $\sim38\%$ of all single-\wh~sBzK 
galaxies are those classified as multiple-\wz~galaxies and almost all of them have $r_{e,opt}$ larger 
than the median $r_{e,UV}$. It is worth to recall that higher fraction of single-\wh~sBzK against the 
single-\wz~sBzK galaxies are partly due to the poorer resolution in the WFC3/F160W image as 
compared to the ACS/F850LP. 

\begin{figure}
\centering
\includegraphics[width=0.4\textwidth, clip]{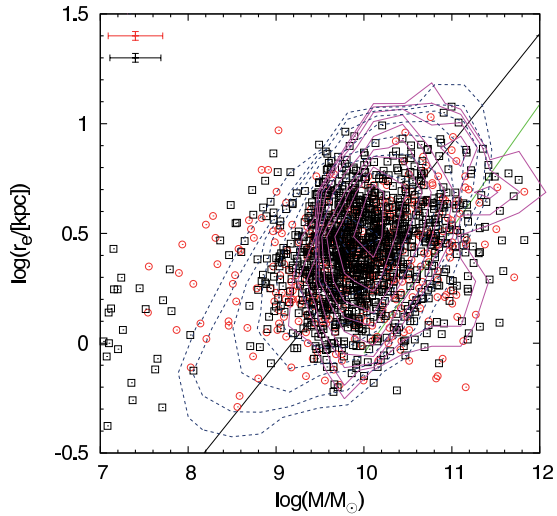}
\caption{Size-mass relation of the single-\wh~sBzK galaxies with $0.5\leq n < 2.5$ both in 
GOODS-S (red open circles) and in SXDS (black open squares). 
The average uncertainties of the stellar mass and effective 
radius are shown in the top left corner of the figure with corresponding colors. 
Dashed and solid contours 
illustrate size-mass distributions of disk galaxies at $z=0$ and $z=1$, respectively \citep{barden05}. 
The black solid line is an average surface mass density for disk galaxies at $z=0-1$. 
\citep[$\log\Sigma_M = 8.50$ with $q=0.5$; ][]{barden05}. The green line shows the typical 
size-mass relation of the elliptical galaxies in the local universe \citep{shen03}. 
Note that the stellar masses shown in the figure are all corrected to Salpeter IMF.}
\label{massden}
\end{figure}

\subsection{Stellar Surface Mass Density}

Figure \ref{massden} shows the size-mass relation of the single-\wh~sBzK galaxies with 
$0.5\leq n < 2.5$ in both GOODS-S and SXDS fields and local galaxies. Because we derived the 
effective radii of the sBzK galaxies in the \wh~images (i.e., the rest-frame optical images), 
the size-mass relation of the sBzK galaxies is able to be directly compared with the relation 
of galaxies in the local universe, where the effective radii were measured in optical wavelengths. 
The stellar masses of local galaxies were corrected into Salpeter IMF with 
$\log(M_{Salpeter})-\log(M_{Kroupa})\sim 0.15-0.2$ dex depending on $U-V$ colors \citep{akiyama08, rudnick03}. 
As seen in the figure, most of the sBzK galaxies with $0.5\leq n < 2.5$ 
locate in the same region as the local disk and $z\sim1$ disk galaxies locate. The result is similar 
to what we found for the single-component sBzK galaxies in GOODS-N (paper I), confirming that 
the sBzK galaxies at $z\sim2$ have comparable stellar surface mass density to the disk galaxies 
at $z\sim0-1$. However, there are some outskirts that show larger surface mass density, which 
are closer to the typical values of the local elliptical galaxies \citep[green line; ][]{shen03}. 
Note that the galaxies with stellar mass of $10^7-10^8\Msun$ are those with the estimated 
$photoz$ lower than 1.4. 

\begin{figure*}
\centering
\begin{tabular}{cc}
\includegraphics[clip, angle=-90, width=0.3\textwidth]{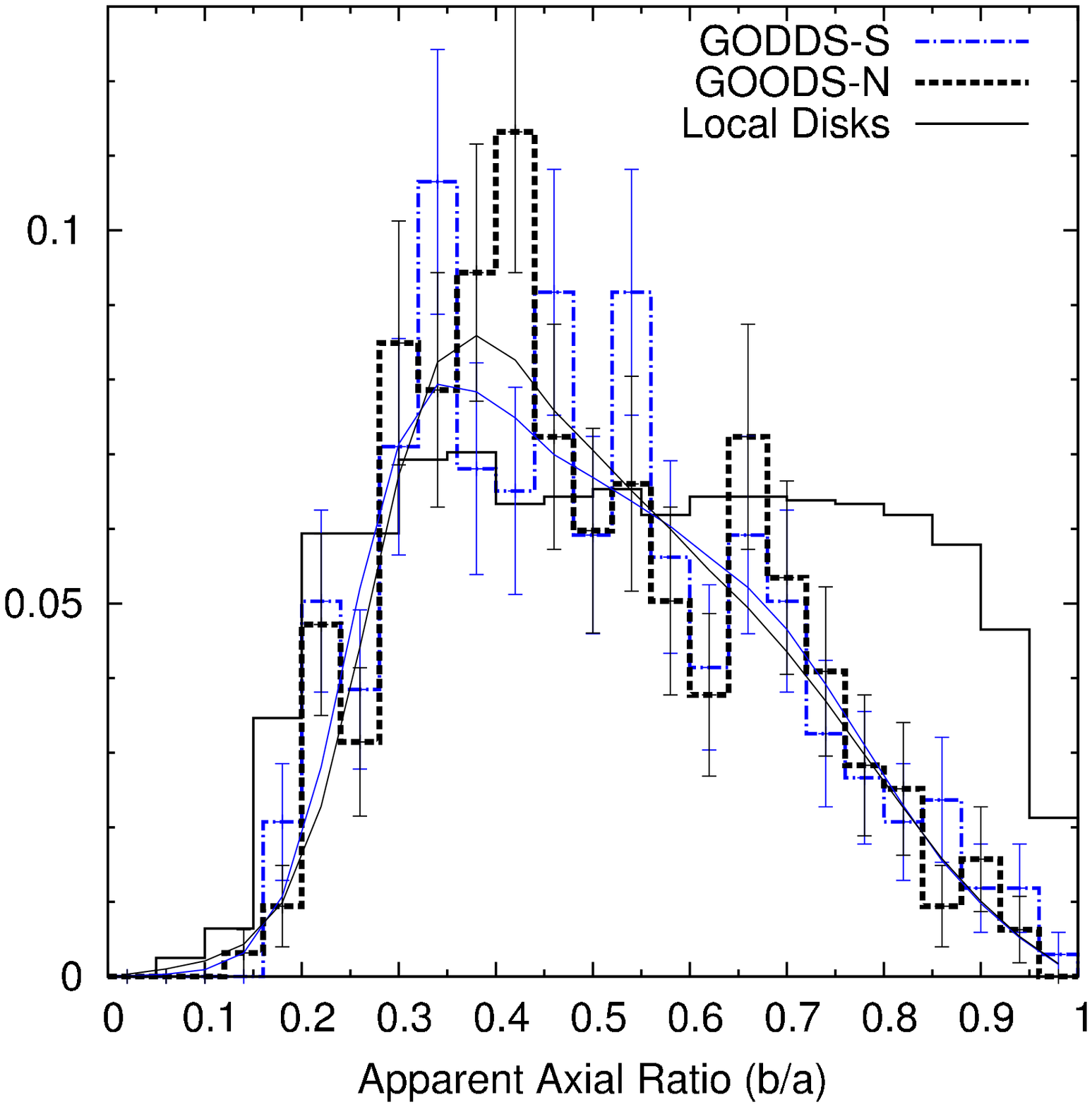} &
\includegraphics[clip, angle=-90, width=0.35\textwidth]{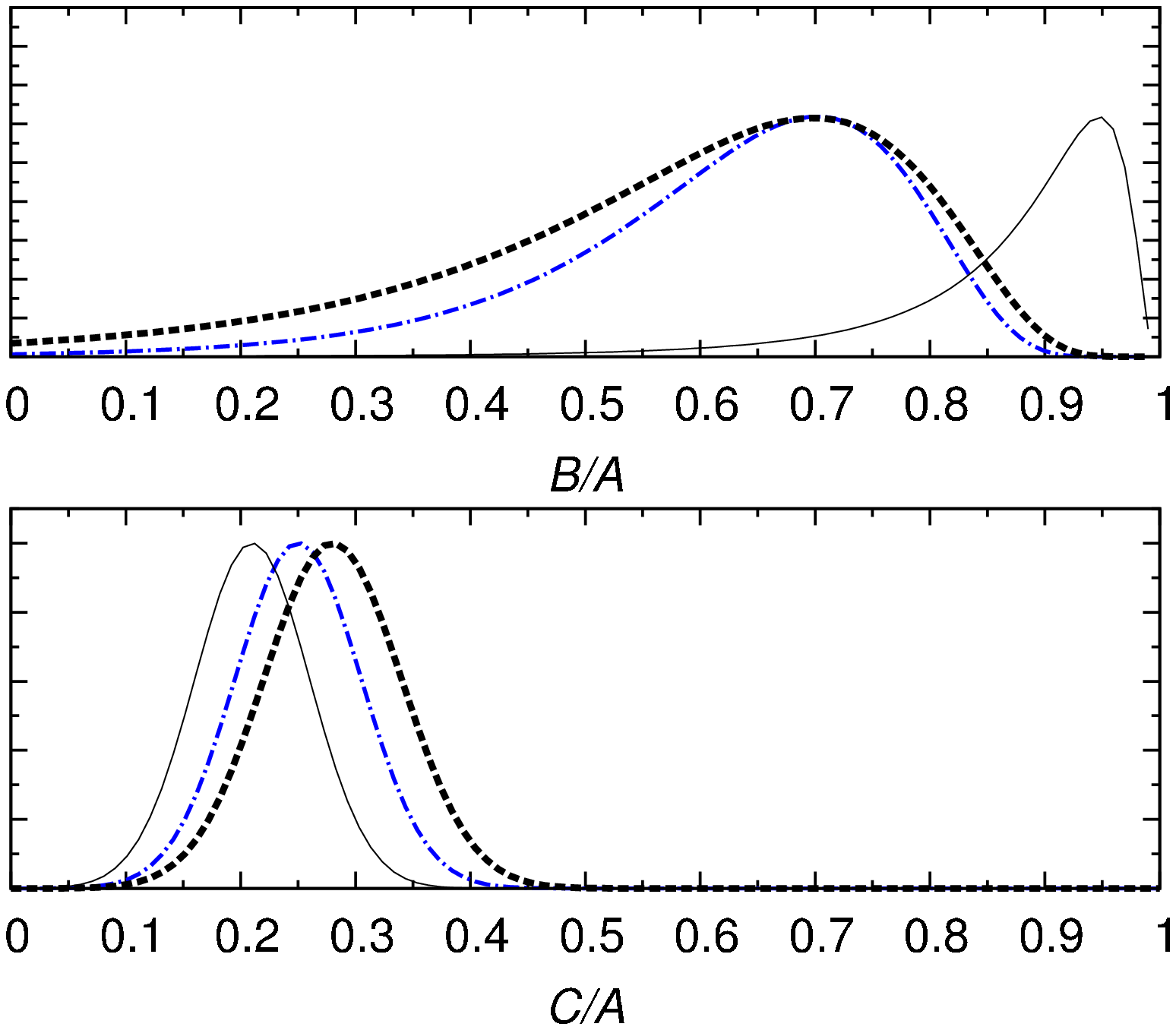} \\
\end{tabular}
\caption{{\it Left panel:} Normalized distributions of the apparent axial ratio ($b/a$) of the single-\wz~
sBzK galaxies in the GOODS-N (black dashed) and GOODS-S (blue dot-dashed) fields. 
The best-fitting models are shown with solid lines with corresponding colors. 
Distribution of local disk galaxies by \cite{padilla08} is also shown with thin solid histogram. 
{\it Right panel:} Distributions of the best-fitting intrinsic axial ratio: $B/A$ distribution 
on the top right panel and $C/A$ distribution on the bottom right panel with colors 
identical to the left panel. Y-axis is in an arbitrary unit. }
\label{qhistcompareuv}
\end{figure*}

\subsection{Axial Ratio Distribution and Intrinsic Shape of sBzK Galaxies}

Although we found that the sBzK galaxies in both wavelengths show the S\'ersic indices around one 
and have stellar surface mass density comparable to disk galaxies at $z\sim0-1$, it is still important 
to see whether or not their structure is a round disk as seen in the local disk galaxies. 
Since the intrinsic shape of galaxies can be statistically examined from their observed 
distributions of apparent axial ratios ($b/a$), we show the $b/a$ distribution and the resulting 
intrinsic shape of the sBzK galaxies in this section. 

\subsubsection{Rest-Frame UV in GOODS-S}\label{subsec:restuvshape} 

Histograms of the apparent $b/a$ of the single-\wz~galaxies with $0.5\leq n < 2.5$ 
in GOODS-S and GOODS-N fields (paper I) are shown in the left panel of figure 
\ref{qhistcompareuv}. Distribution of local disk galaxies by \cite[][$\sim300,000$ in SDSS]{padilla08} 
is also shown. The distribution of samples in GOODS-S peaks at $b/a\sim0.35$ and declines significantly 
toward $b/a=1.0$. This trend agrees with that found in GOODS-N. We performed the 
K-S test and found that $b/a$ distribution of the single-\wz~sBzK galaxies are drawn from 
the same distribution with $92\%$ confidence level. However, they are obviously different 
from the distribution of the local disk, which is relatively flat in the range of $b/a\sim0.2-0.8$. 
It is implied that structure of the sBzK galaxies in the rest-frame UV differs from that of the local disks. 

\begin{deluxetable}{cccc}
\tabletypesize{\footnotesize}
\tablecaption{Grids of Intrinsic Model Parameters
\label{tab:param}
}
\tablewidth{0pt}
\tablehead{
\multicolumn{1}{c}{Parameter} &
\multicolumn{1}{c}{Minimum Value} &
\multicolumn{1}{c}{Maximum Value} & 
\multicolumn{1}{c}{Step}
}
\startdata
$\mu$ ($B/A$)\tablenotemark{a} & $-4.0$ (0.98) & $-0.1$ (0.1)& 0.15\\
$\sigma$ & 0.20 & 2.00 & 0.15 \\
$\mu_\gamma$ & 0.10 & 0.98 & 0.02 \\
$\sigma_\gamma$ & 0.01 & 0.35 & 0.02\\
\enddata
\tablenotetext{a}{
The corresponding $B/A$ is written in the parenthesis. 
}
\end{deluxetable} 

In order to study the intrinsic shape of galaxies, a tri-axial model is adopted with axes 
$A>B>C$. The face-on ellipticity ($\epsilon \equiv 1-B/A$) is assumed to be described by 
lognormal distribution with mean $\mu$ and dispersion $\s$, while the edge-on 
thickness ($C/A$) is described by Gaussian distribution with mean $\mg$ and 
standard deviation $\sg$. One set of these 4 parameters provides one apparent-$b/a$ 
distribution after repeatedly computing the apparent $b/a$ at various random viewing angles. 
The parameters $\mu$, $\s$, $\mg$, and $\sg$ were constrained by comparing the model 
distributions of $b/a$ with those of the observed samples. The ranges and steps of the parameters 
are summarized in Table \ref{tab:param}. The best-fitting parameter set was obtained by $\chi^2$ 
minimization method. The observed $b/a$ distribution was first compared with the above parameter 
sets to obtain the primary result. Then we refitted the observed distribution with parameter set 
with finer grid (at least half of the current step) centered at the primary best-fitting result. 
The fitting uncertainties are calculated 
at 68\% confidence level based on Monte Carlo realization. We derived the best-fitting set 
of parameters and repeated the process 500 times by varying the observed distribution 
of $b/a$ ratios within their Poisson noises. 

\begin{deluxetable*}{c c cccc}
\tabletypesize{\footnotesize}
\tablewidth{0pt}
\tablecaption{Best-Fitting Results of Intrinsic Shape for sBzK Galaxies
\label{tab:shaperesult}}
\tablehead{
\colhead{Rest-Frame Wavelength} & \colhead{Field} & \colhead{$\mu$} & \colhead{$\s$} & \colhead{$\mg$} & \colhead{$\sg$}\\
}
\startdata
\multirow{3}{*}{UV (\wz~image)} & GOODS-S & $-1.05^{+0.15}_{-0.25}$ & $0.40^{+0.45}_{-0.40}$ & $0.25^{+0.03}_{-0.03}$ &  $0.055^{+0.020}_{-0.010}$ \\
						  & GOODS-N\tablenotemark{a} & $-0.95^{+0.20}_{-0.15}$ & $0.50^{+0.45}_{-0.15}$ & $0.28^{+0.03}_{-0.04}$ &  $0.060^{+0.020}_{-0.015}$ \\
						  \cmidrule{2-6}
						  & GOODS-N$+$S & $-1.05^{+0.10}_{-0.10}$ & $0.45^{+0.10}_{-0.10}$ & $0.26^{+0.02}_{-0.02}$ & $0.055^{+0.015}_{-0.010}$\\
\hline
\multirow{3}{*}{Optical (\wh~image)} & GOODS-S & $-1.30^{+0.20}_{-0.10}$ & $0.40^{+0.30}_{-0.20}$ & $0.26^{+0.06}_{-0.02}$ & $0.080^{+0.040}_{-0.030}$\\
							& SXDS & $-1.30^{+0.10}_{-0.10}$ & $0.50^{+0.20}_{-0.10}$ & $0.27^{+0.03}_{-0.02}$ & $0.080^{+0.000}_{-0.010}$\\
							 \cmidrule{2-6}
							& GOODS-S $+$ SXDS & $-1.30^{+0.10}_{-0.00}$ & $0.50^{+0.10}_{-0.20}$ & $0.27^{+0.01}_{-0.02}$ & $0.080^{+0.000}_{-0.010}$\\
							\hline
Local Disks ($r$-band image)\tablenotemark{b} & SDSS DR6 & $-2.33\pm0.13$ & $0.79\pm0.16$ & $0.21\pm0.02$ & $0.05\pm0.015$ \\
\enddata
\tablenotetext{a}{
\citet[][paper I]{yuma11}
}
\tablenotetext{b}{
\cite{padilla08}. 
}
\end{deluxetable*}

The final best-fitting results of the single-\wz~sBzK galaxies in GOODS-S are summarized in 
Table \ref{tab:shaperesult}. For comparison purpose, we also show the results of the 
single-\wz~galaxies in GOODS-N by paper I in the table. Distributions of the best-fitting 
parameters are also shown in right panel of Figure \ref{qhistcompareuv}. 
The peak (i.e., mode) of the $B/A$ distribution is determined from $\mu$ and $\sigma$ 
parameters, while $\mg$ and $\sg$ parameters corresponds to 
a peak and width of the $C/A$ distribution, respectively. The single-\wz~galaxies in 
GOODS-S show similar $B/A$ distribution to those in the GOODS-N with a slightly larger 
width of GOODS-N sample (top right panel of Figure \ref{qhistcompareuv}). 
The best-fitting $\mu$ of the single-\wz~galaxies in GOODS-S is $-1.05^{+0.15}_{-0.25}$, 
consistent with that in GOODS-N within $1\sigma$ ($\mu = -0.95^{+0.20}_{-0.15}$). 
Although a slightly larger best-fitting $\sigma$ in the 
GOODS-N samples, the $\sigma$ parameters of the samples in both fields are well 
in agreement with each other within $68\%$ confidence level (Table \ref{tab:shaperesult}). 
The bottom right panel of Figure \ref{qhistcompareuv} shows the slightly different distributions 
of the intrinsic edge-on $C/A$ ratio of the single-\wz~galaxies between in GOODS-S and in GOODS-N. 
However, their best-fitting $\mg$ and $\sg$, which correspond to $C/A$ distribution, agree 
well to each other within $1\sigma$ uncertainties. This indicates that the sBzK galaxies in 
both GOODS-S and GOODS-N are not round ($B/A=1.0$) and seem to be flat 
(mean $C/A=0.25-0.28$) when being observed in rest-frame UV wavelength. 

Since the intrinsic shapes of the single-\wz~sBzK galaxies in both GOODS-S and GOODS-N 
fields are consistent with each other, we combined the samples together and re-derived the 
intrinsic shape in order to improve the statistics. The fitting results are shown in 
Table \ref{tab:shaperesult}. The best-fitting parameters are similar to those determined 
in the separated fields but with smaller range of uncertainties. The results confirm that 
the sBzK galaxies that appear as a single component in the rest-frame UV wavelength 
show a flat (peak $C/A = 0.26$) and bar-like/oval shape (peak $B/A = 0.71$). 

\begin{figure*}
\centering
\begin{tabular}{cc}
\includegraphics[clip, angle=-90, width=0.3\textwidth]{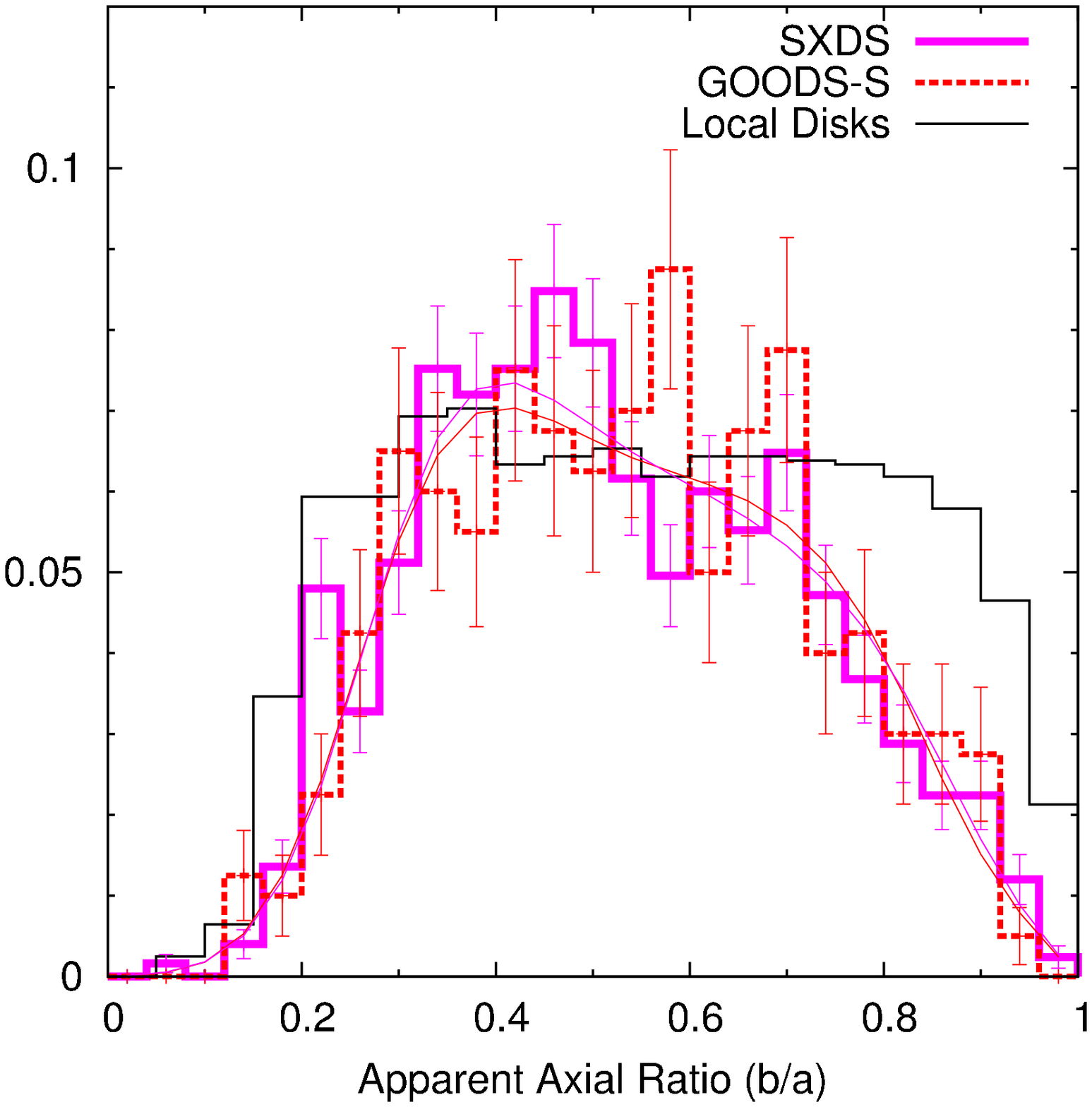} &
\includegraphics[clip, angle=-90, width=0.35\textwidth]{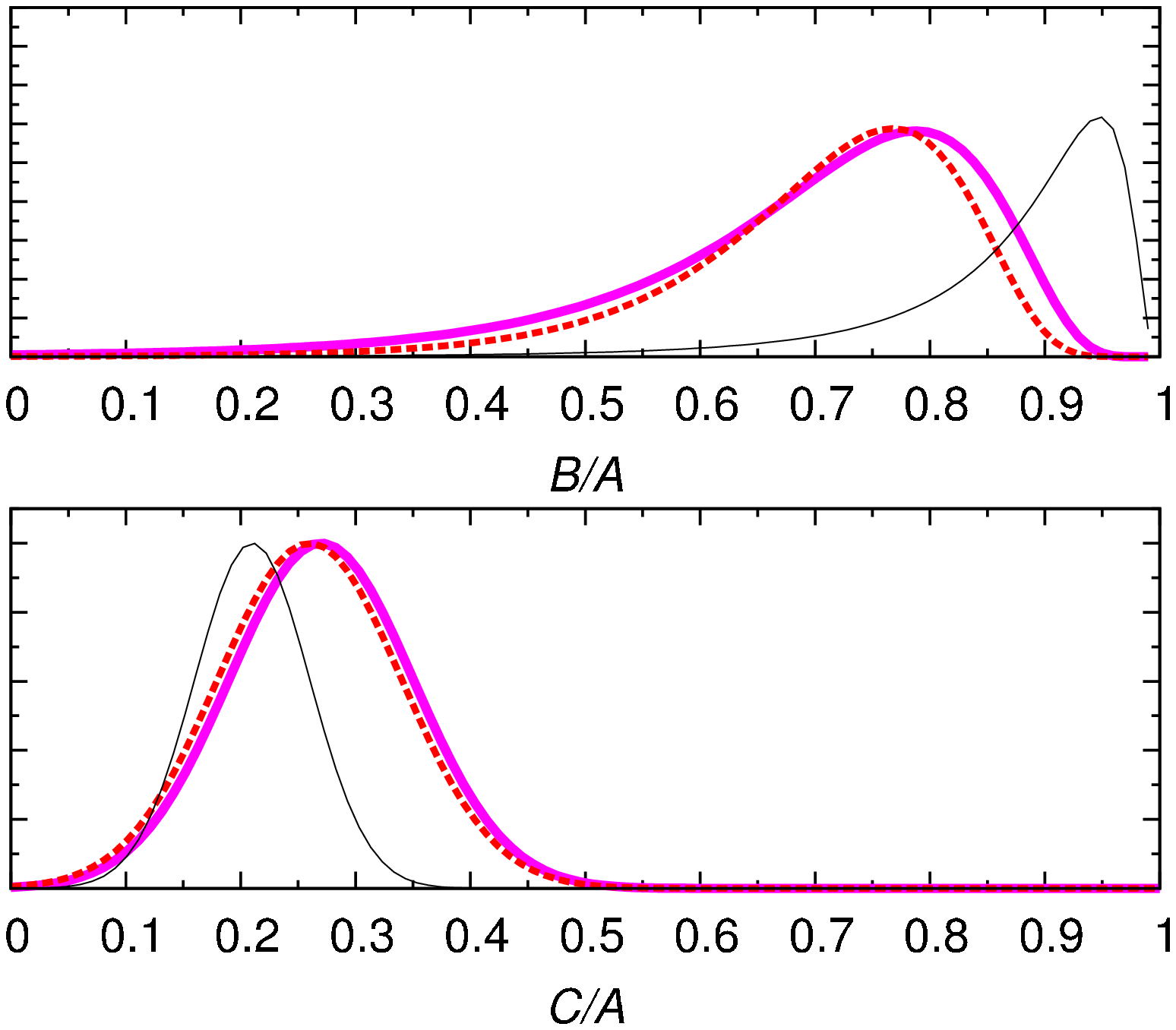} \\
\end{tabular}
\caption{{\it Left panel:} Normalized distributions of the apparent axial ratio ($b/a$) of the single-\wh~
sBzK galaxies at $z\sim2$ in the GOODS-S (red dashed histogram) and in SXDS (magenta solid histogram) 
fields. The best-fitting models are shown with solid lines with corresponding colors. 
Distribution of the local disk galaxies by \cite{padilla08} is also shown with a thin solid histogram. 
{\it Right panel:} Distributions of the best-fitting intrinsic axial ratios: $B/A$ distribution 
on the top right panel and $C/A$ distribution on the bottom right panel with colors identical to the 
left panel. Y-axis is in arbitrary unit.}
\label{qhistcompareopt}
\end{figure*}

\subsubsection{Rest-Frame Optical in GOODS-S and SXDS}\label{subsec:restoptshape}

Thanks to the publicly available data from CANDELS survey, we are able to study 
the structure of sBzK galaxies at $z\sim2$ in the rest-frame optical wavelength. 
The apparent $b/a$ histograms of the single-\wh~sBzK galaxies in GOODS-S and SXDS 
are shown in the left panel of Figure \ref{qhistcompareopt}. The distributions of samples in 
both fields look similar to each other; however, the K-S test indicates that these two histograms 
are drawn from different distribution. In any case, the distributions roughly seem to 
be flat from $b/a\sim0.3$ to $b/a\sim0.7$ and decline continuously toward $b/a=1.0$. 
The distributions of the single-\wh~sBzK galaxies are still different from those of the local 
disk galaxies, which is also shown with the black solid line in the figure. 
\cite{law12} also studied the rest-frame optical morphologies of star-forming galaxies 
at $z=1.5-3.6$ and showed the $b/a$ histogram that peaks at $b/a\sim0.55$. 
Their histogram seems to be similar to that of the sBzK sample in GOODS-S. 

We determined the intrinsic shape of the single-\wh~sBzK galaxies in the identical manner 
as done for the single-\wz~samples. The fitting results are summarized in Table \ref{tab:shaperesult}. 
The best-fitting parameters are also shown in the right panels of Figure \ref{qhistcompareopt} 
as $B/A$ and $C/A$ distributions. Despite statistically significant difference of the $b/a$ distribution, 
the best-fitting $\mu$ of the single-\wh~sBzK galaxies in both fields are the same at $\mu=-1.30$. 
Meanwhile, slightly different $\sigma$ parameters result in 
different $B/A$ distribution as seen in the top right panel of Figure \ref{qhistcompareopt}. 
The best-fitting $\sigma$ value of samples in GOODS-S is still consistent with that in SXDS within 68\% uncertainty. 
The $C/A$ distributions of the sBzK galaxies in both fields show almost the same peak 
with the same width, since they have comparable $\mg$ and $\sg$ values. 

Because the intrinsic shapes derived separately in GOODS-S and SXDS are well in agreement 
with each other within $1\sigma$ uncertainties, we combined the samples and re-derived the 
intrinsic shape to increase the statistical significance. Fitting results of the combined samples 
are listed in Table \ref{tab:shaperesult}. The peak $B/A$ and $C/A$ ratios of the combined 
samples are 0.79 and 0.27, respectively, indicating that the sBzK galaxies are indeed not 
round even when observing in the rest-frame optical wavelength. 
The best-fitting parameters of the combined samples 
are identical to those of the single-\wh~galaxies in SXDS alone. This is presumably due to the 
fact that number of the single-\wh~sBzK galaxies in SXDS is approximately three times 
larger than that in GOODS-S. 

\begin{figure*}
\centering
\begin{tabular}{cc}
\includegraphics[clip, angle=-90, width=0.32\textwidth]{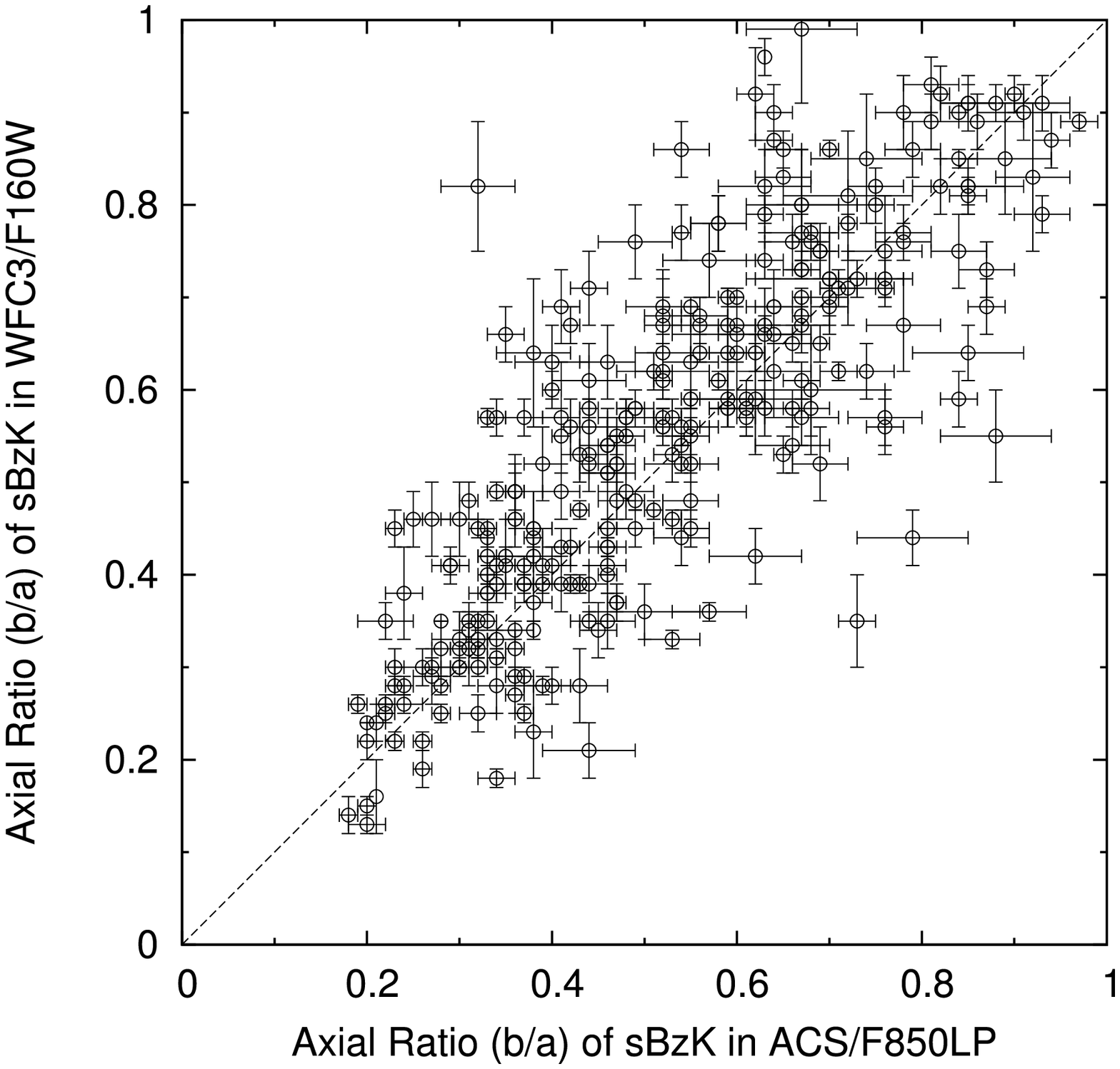} &
\includegraphics[clip, angle=-90, width=0.3\textwidth]{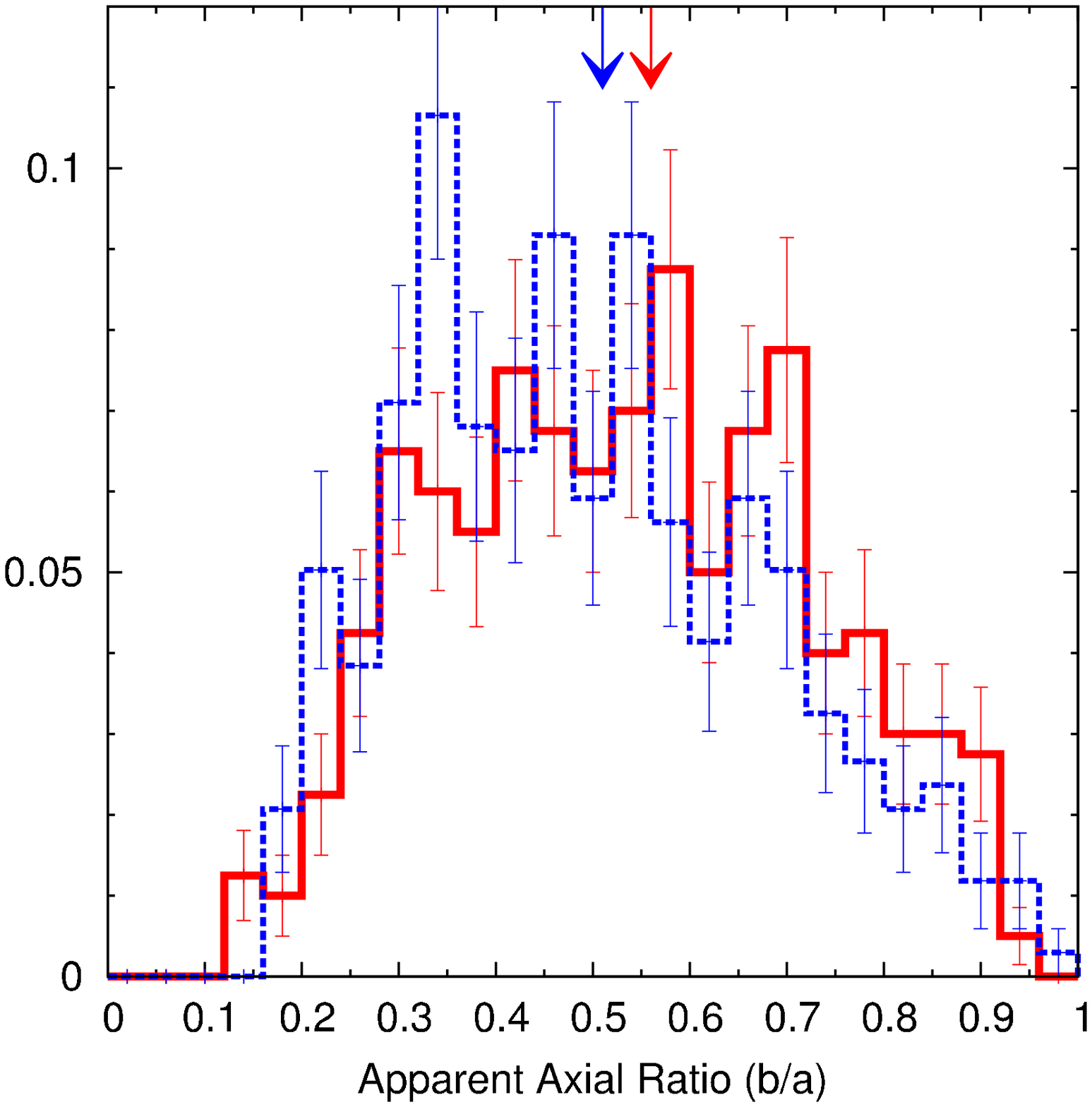} \\
\end{tabular}
\caption{{\it Left panel:} Comparison of apparent axial ratios ($b/a$) of the sBzK galaxies in the 
GOODS-S that appear as a single component in both \wz~and \wh~images. 
{\it Right panel:} Normalized $b/a$ histograms of the single-\wz~(blue dashed histogram) and 
the single-\wh~sBzK galaxies (red solid histogram) in the GOODS-S. 
}
\label{qcompareuvopt}
\end{figure*}

\begin{figure*}
\centering
\begin{tabular}{cc}
\includegraphics[clip, angle=-90, width=0.3\textwidth]{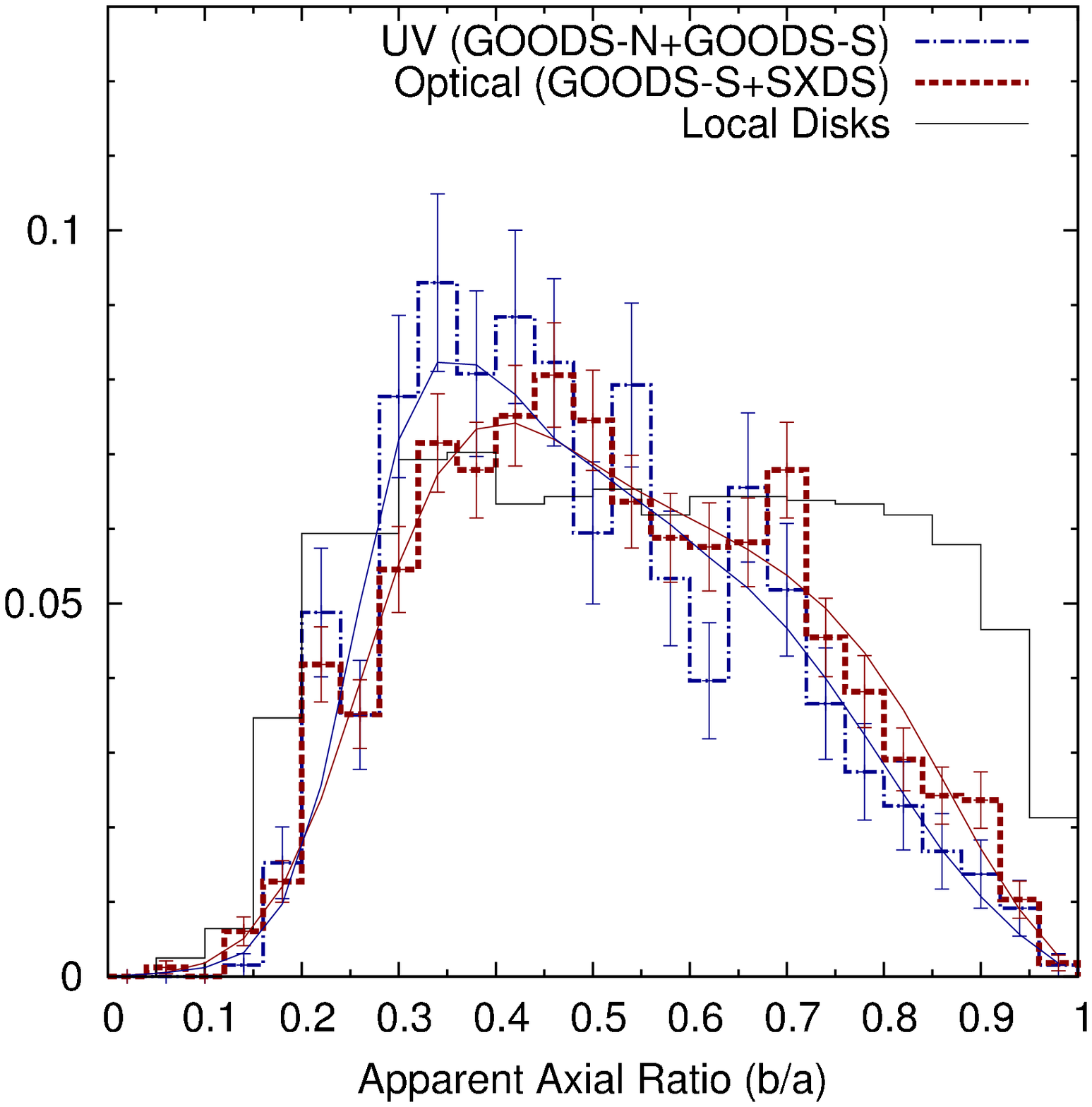} &
\includegraphics[clip, angle=-90, width=0.35\textwidth]{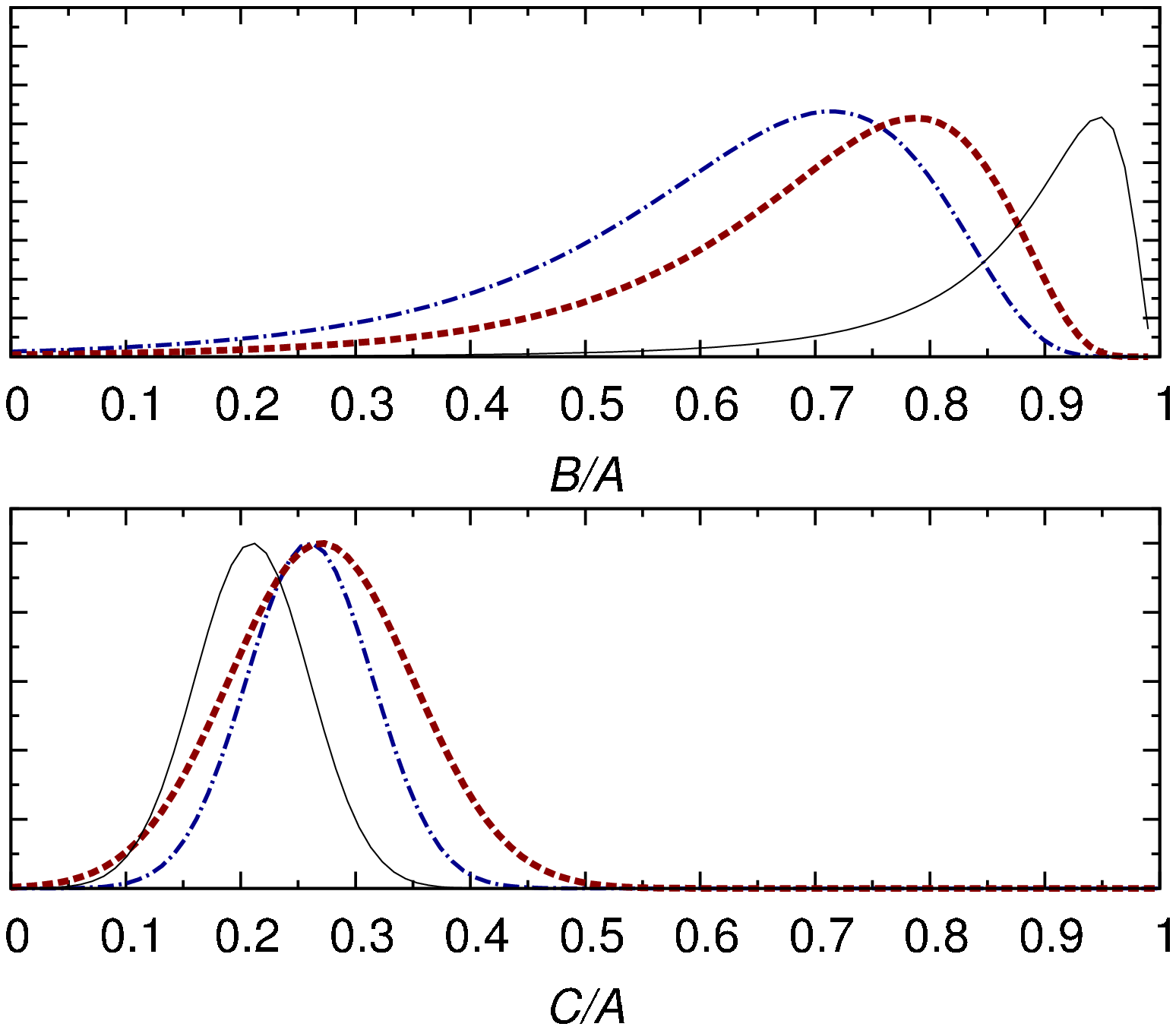} \\
\end{tabular}
\caption{{\it Left panel:} Normalized distribution of the observed axial ratio ($b/a$) of the combined sample 
of the single-\wz~(GOODS-N+GOODS-S) and single-\wh~(GOODS-S+SXDS) sBzK galaxies. 
The best-fitting models are shown with solid lines with corresponding colors. 
Distribution of the local disk galaxies by \cite{padilla08} is also shown with a thin solid histogram. 
{\it Right panel:} Distributions of the best-fitting intrinsic axial ratio: $B/A$ distribution 
on the top right panel and $C/A$ distribution on the bottom right panel with colors 
identical to the left panel. Y-axis is in arbitrary unit. 
}
\label{qhistcombine}
\end{figure*}

\subsubsection{Comparison between Rest-Frame UV and Optical Wavelengths}

In this section, we discuss the structural difference between in the rest-frame UV and 
in the rest-frame optical wavelength in terms of the apparent axial ratios ($b/a$). 
The $b/a$ comparison of sBzK galaxies that appear as a single component in both 
wavelengths in the GOODS-S is shown in Figure \ref{qcompareuvopt}. The left panel shows that 
most objects show larger $b/a$ ratios in the rest-frame optical wavelength (\wh~images) 
than those in the rest-frame UV (\wz~images), though some show the same or even 
lower $b/a$ ratios in the rest-optical images. The comparison of $b/a$ histograms is 
illustrated in the right panel of the figure. It is seen that the $b/a$ histogram is slightly 
different at $b/a>0.5$ in the way that larger fraction of the sBzK galaxies in the rest-frame 
optical wavelength shows $b/a$ ratios more than 0.5. The median $b/a$ value 
in the rest-frame optical images ($b/a=0.56$) is also larger than that in the rest-frame UV 
($b/a=0.51$). This difference may result in the difference in the derived intrinsic shape of 
the sBzK galaxies. 

We investigate the difference of the intrinsic shape of sBzK galaxies between 
in the rest-frame UV and in the rest-frame optical wavelengths by using the combined 
samples in both wavelength ranges, i.e., the single-\wz~galaxies in GOODS-S and GOODS-N 
for rest-frame UV and the single-\wh~galaxies in GOODS-S and SXDS for rest-frame optical, 
to improve statistical significance. Their $b/a$ histograms are shown in the left panel of 
Figure \ref{qhistcombine} along with those of the local disk galaxies by \cite{padilla08}. 
Similar to what we found in the GOODS-S, the $b/a$ distribution of the sBzK galaxies 
observed in the rest-frame optical wavelength shows a slightly larger fraction at $b/a>0.5$ than 
those in the rest-frame UV wavelength. Comparing the histograms with those of the 
local disks, we even see more difference of the histograms at $b/a>0.5$. In paper I, they 
discussed that difference in $b/a$ histogram results in difference in the intrinsic shape (Figure 12 in paper I). 
The $b/a$ histogram at $b/a > 0.5$ affects the intrinsic face-on shape, mainly the $\mu$ parameters. 
Higher fraction at $b/a > 0.5$ histogram indicates rounder face-on shape. 
The effect can be seen in the right panel of Figure \ref{qhistcombine}; the top right panel 
shows the intrinsic $B/A$ distributions derived for the sBzK galaxies in the rest-frame 
UV and optical wavelengths. The $B/A$ ratio of the single-\wz~sBzK galaxies peaks at 
$B/A=0.65$, which is different from that of the single-\wh~sBzK galaxies more than $1\sigma$ uncertainty. 
The same trend of dramatically increase at $b/a=0.1-0.3$ of the $b/a$ histograms results in 
the similar $\mg$ for the samples in both wavelength ranges, whereas 
some excess fraction at $b/a=0.3-0.4$ in the rest-frame UV sample causes 
difference in the derived $\sg$ parameters. 
Their best-fitting $\mg$ are similar at $\mg=0.26-0.27$ or peak $C/A$ of $0.26-0.27$, whereas wider $C/A$ 
distribution can be seen in the single-\wh~galaxies (bottom right panel) due to slightly 
larger $\sg$. However, they are still in agreement within $1\sigma$ confidence level. 
In conclusion, as seen in the right panel of Figure \ref{qhistcombine} and Table \ref{tab:shaperesult}, 
the sBzK galaxies appearing as a single component in rest-frame UV show just slightly 
different $B/A$ or intrinsic face-on shape as compared to those in rest-frame optical wavelength 
with similar $C/A$ or intrinsic edge-on distribution. 

\begin{figure*}
\centering
\includegraphics[clip, angle=-90, width=0.3\textwidth]{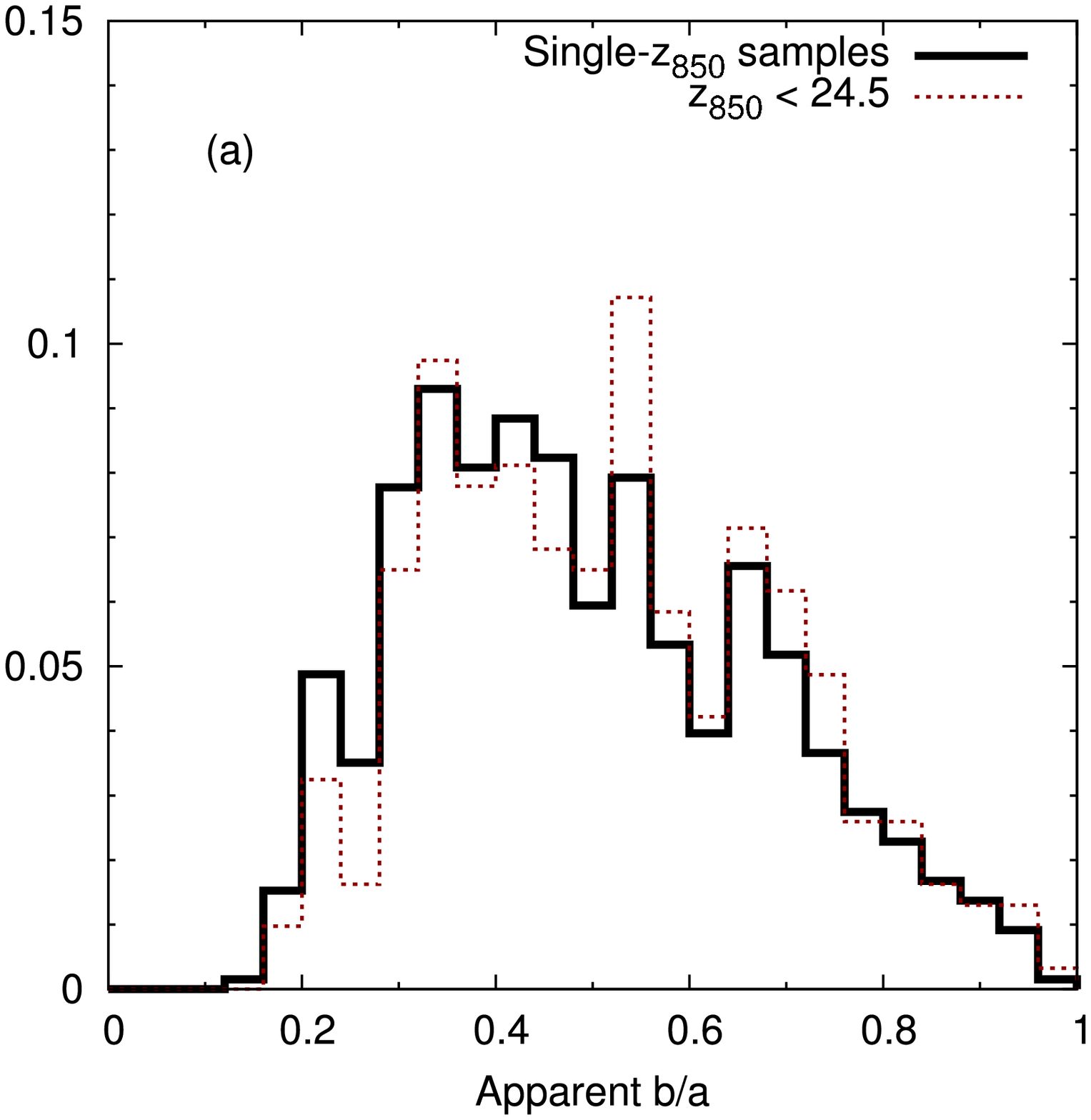} 
\includegraphics[clip, angle=-90, width=0.3\textwidth]{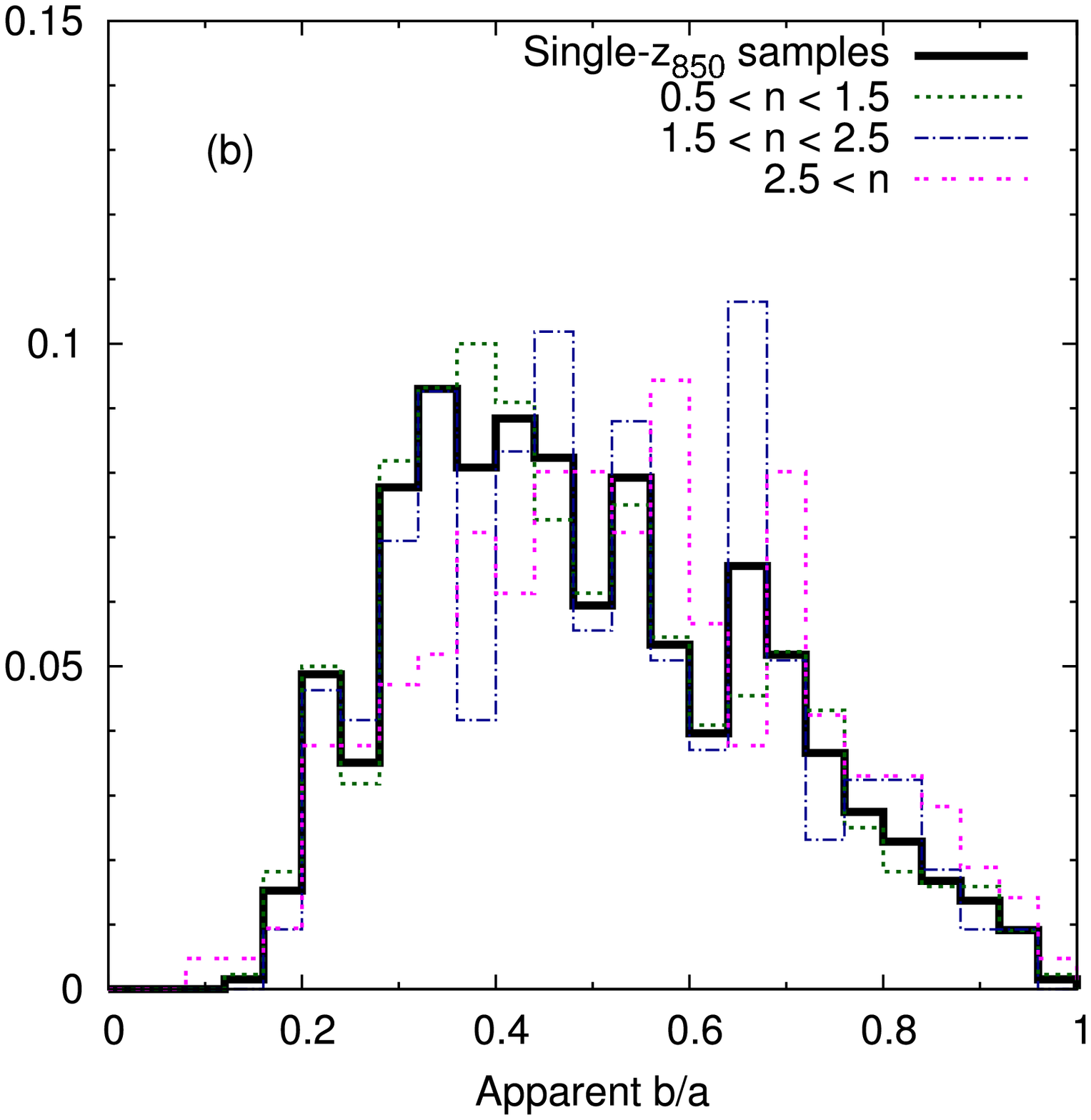} 
\includegraphics[clip, angle=-90, width=0.3\textwidth]{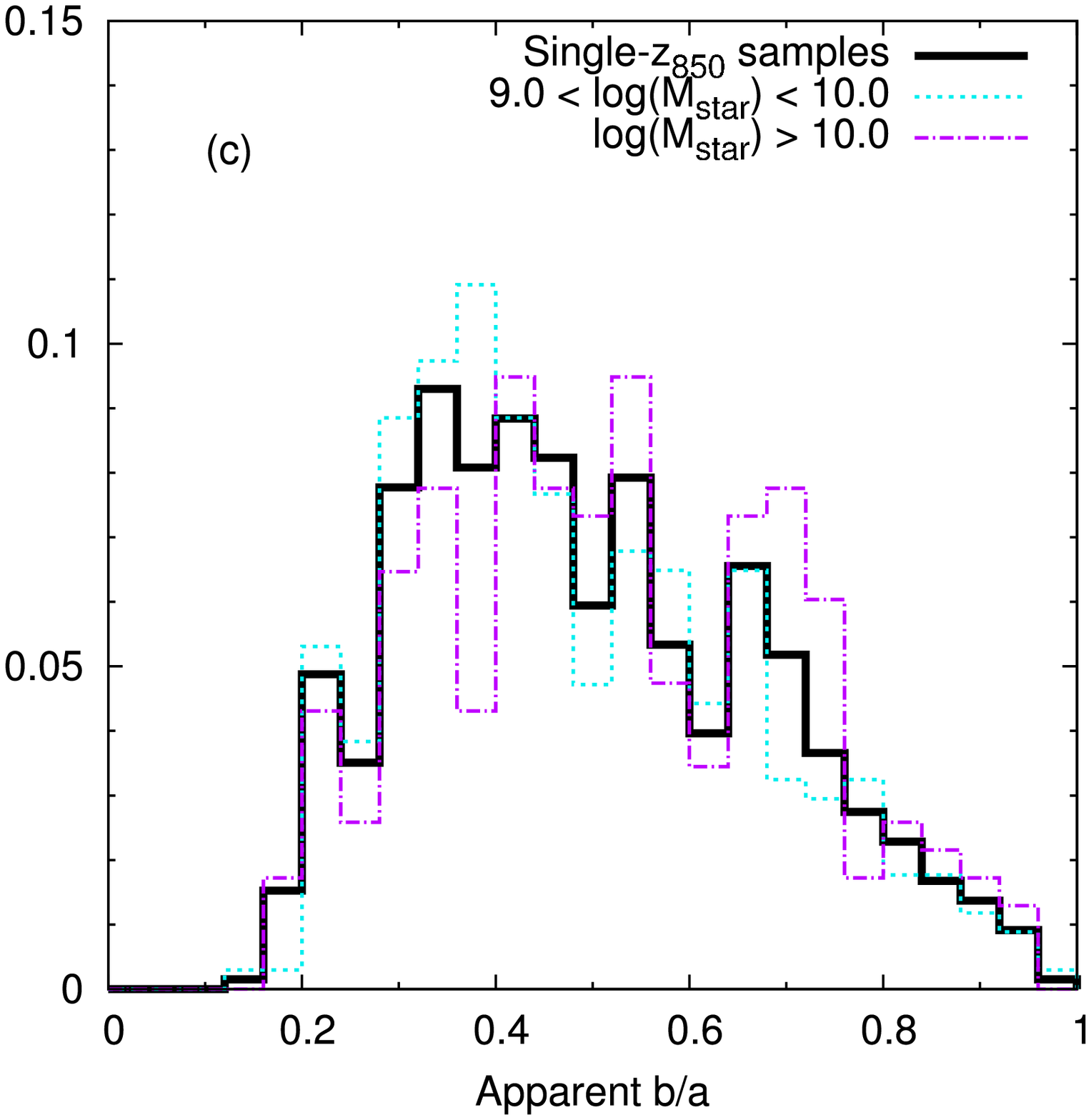} \\
\includegraphics[clip, angle=-90, width=0.3\textwidth]{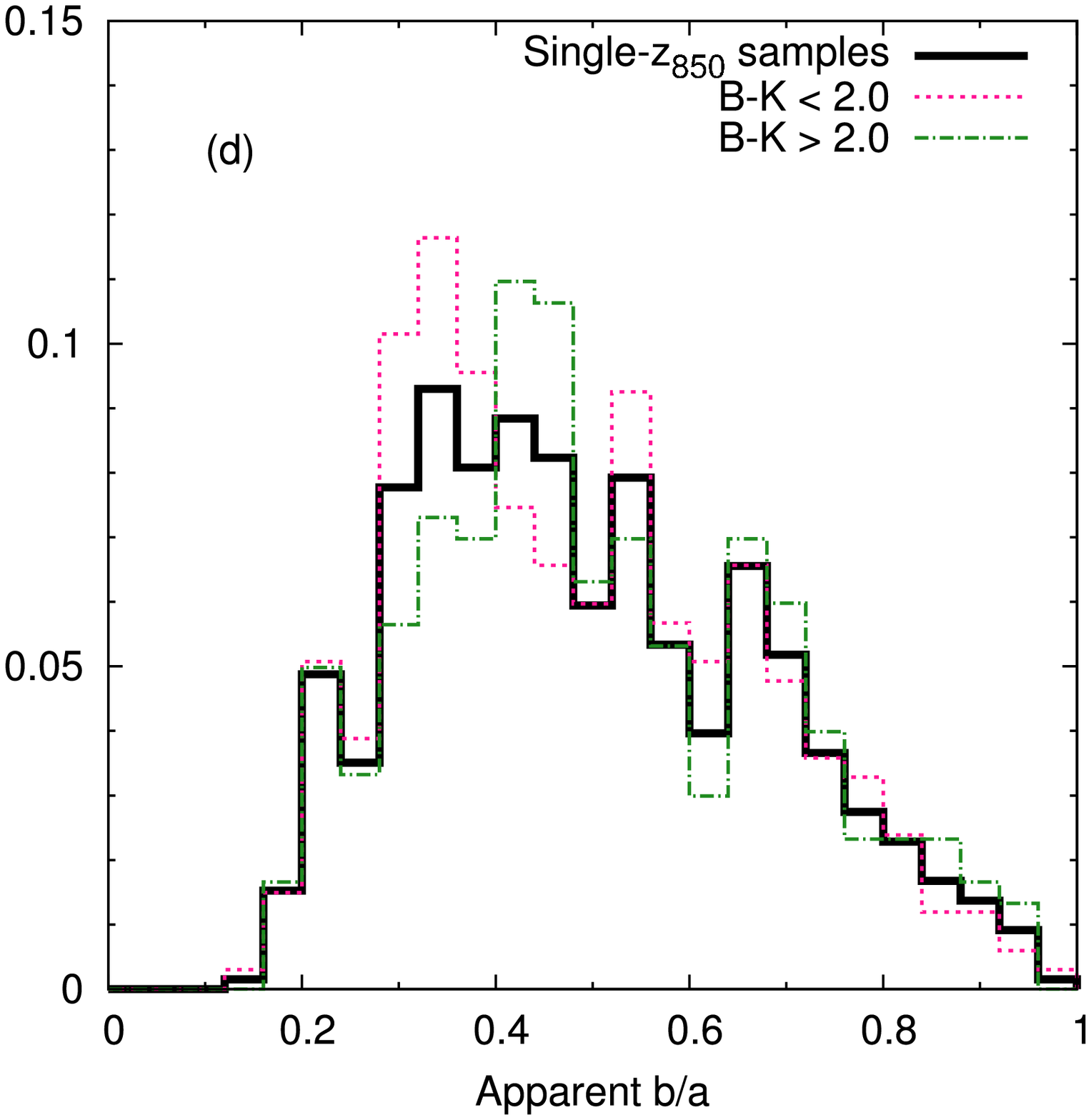} 
\includegraphics[clip, angle=-90, width=0.3\textwidth]{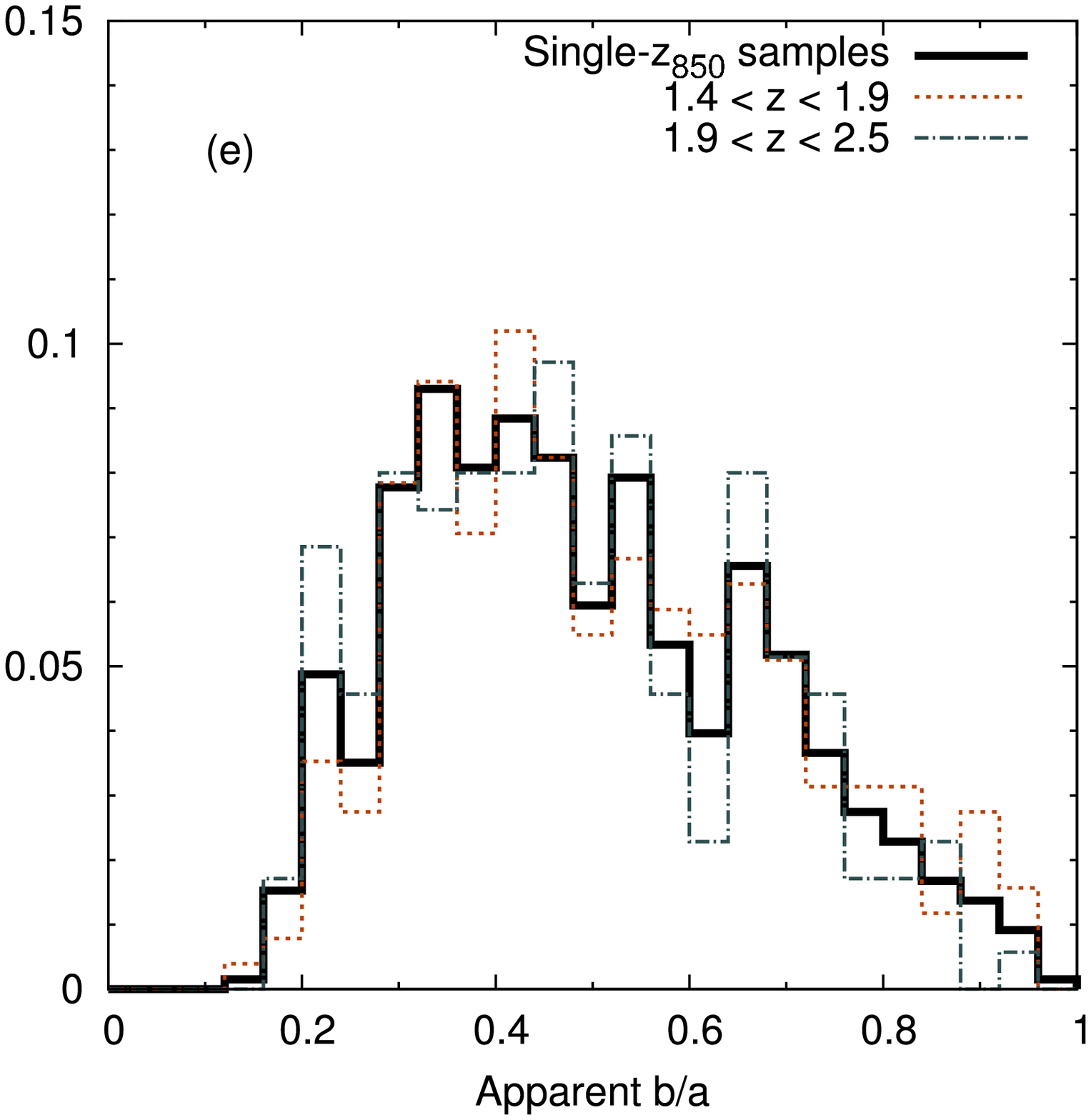} \\
\caption{Apparent $b/a$ histograms of the combined sample of single-\wz~sBzK subsamples in GOODS-S and GOODS-N. 
Solid histograms represent all single-\wz~sBzK galaxies with $0.5\leq n <2.5$.}
\label{sub_acsz}
\end{figure*}

\begin{deluxetable*}{c c cccc}
\tabletypesize{\footnotesize}
\tablewidth{0pt}
\tablecaption{Best-Fitting Parameters of Intrinsic Shape for sBzK Subsamples \label{tab:sub}}
\tablehead{
\colhead{Rest-Frame Wavelength} & \colhead{Subsamples} & \colhead{$\mu$} & \colhead{$\s$} & \colhead{$\mg$} & \colhead{$\sg$}\\
\colhead{(Observed Field)} & & & & &
}
\startdata
\multirow{11}{*}{UV (GOODS-S\&N)} & All single-\wz & $-1.05^{+0.10}_{-0.10}$ & $0.45^{+0.10}_{-0.10}$ & $0.26^{+0.02}_{-0.02}$ &  $0.055^{+0.015}_{-0.010}$ \\
			  & $\wz < 24.5$ mag  & $-1.00^{+0.30}_{-0.30}$ & $0.50^{+0.60}_{-0.30}$ & $0.30^{+0.04}_{-0.04}$ &  $0.030^{+0.060}_{-0.020}$ \\
			  \cmidrule{2-6}
			  & $0.5\leq n < 1.5$ & $-0.85^{+0.15}_{-0.30}$ & $0.50^{+0.45}_{-0.15}$ & $0.28^{+0.02}_{-0.04}$ & $0.050^{+0.020}_{-0.020}$\\
			  & $1.5\leq n < 2.5$ & $-1.15^{+0.30}_{-0.30}$ & $0.35^{+0.60}_{-0.15}$ & $0.24^{+0.08}_{-0.02}$ & $0.050^{+0.040}_{-0.020}$\\
			  & $2.5\leq n$	      & $-1.15^{+1.05}_{-0.45}$ & $0.80^{+1.05}_{-0.60}$ & $0.32^{+0.66}_{-0.08}$ & $0.110^{+0.240}_{-0.040}$\\
			  \cmidrule{2-6}
			  & $9.0 \leq\log (M^*/\Msun)<10.0$ & $-0.85^{+0.30}_{-0.00}$ & $0.50^{+0.30}_{-0.15}$ & $0.28^{+0.02}_{-0.04}$ & $0.030^{+0.020}_{-0.000}$\\
			  & $10.0 \leq\log(M^*/\Msun)$~~~~~~~~~ & $-1.15^{+1.05}_{-0.30}$ & $0.35^{+1.35}_{-0.15}$ & $0.26^{+0.44}_{-0.04}$ & $0.070^{+0.220}_{-0.020}$\\
			  \cmidrule{2-6}
			  & $B-K < 2.0$       & $-1.00^{+0.15}_{-0.15}$ & $0.50^{+0.30}_{-0.15}$ & $0.26^{+0.02}_{-0.04}$ & $0.050^{+0.020}_{-0.040}$\\
			  & $B-K \geq 2.0$  & $-1.00^{+0.90}_{-0.30}$ & $0.65^{+0.90}_{-0.30}$ & $0.30^{+0.04}_{-0.06}$ & $0.070^{+0.020}_{-0.020}$\\
			  \cmidrule{2-6}
			  & $1.4 \leq z < 1.9$& $-1.00^{+0.90}_{-0.30}$	& $0.50^{+1.20}_{-0.15}$ & $0.28^{+0.06}_{-0.04}$ & $0.050^{+0.040}_{-0.040}$\\
			  & $1.9 \leq z < 2.5$& $-0.85^{+0.75}_{-0.30}$ & $0.35^{+0.60}_{-0.15}$ & $0.24^{+0.20}_{-0.04}$ & $0.050^{+0.120}_{-0.040}$\\
\hline
\multirow{11}{*}{Optical (GOODS-S+SXDS)} & All single-\wh & $-1.30^{+0.10}_{-0.00}$ & $0.50^{+0.10}_{-0.20}$ & $0.27^{+0.01}_{-0.02}$ & $0.080^{+0.000}_{-0.010}$\\
                              & $H < 24.0$ mag  & $-1.30^{+0.00}_{-0.15}$ & $0.35^{+0.30}_{-0.00}$ & $0.26^{+0.04}_{-0.00}$ &  $0.070^{+0.020}_{-0.000}$ \\
			      \cmidrule{2-6}
			      & $0.5\leq n < 1.5$ & $-1.15^{+0.00}_{-0.15}$ & $0.50^{+0.15}_{-0.00}$ & $0.26^{+0.02}_{-0.02}$ & $0.070^{+0.000}_{-0.000}$\\
			      & $1.5\leq n < 2.5$ & $-1.30^{+0.30}_{-0.15}$ & $0.50^{+0.45}_{-0.30}$ & $0.30^{+0.08}_{-0.04}$ & $0.090^{+0.040}_{-0.020}$\\
			      & $2.5\leq n$       & $-1.60^{+1.20}_{-0.45}$ & $0.35^{+1.35}_{-0.15}$ & $0.34^{+0.64}_{-0.04}$ & $0.130^{+0.100}_{-0.100}$\\
			      \cmidrule{2-6}
			      & $9.0\leq \log (M^*/\Msun)<10.0$ & $-1.15^{+0.00}_{-0.15}$ & $0.50^{+0.15}_{-0.15}$ & $0.26^{+0.02}_{-0.02}$ & $0.070^{+0.000}_{-0.000}$\\
			      & $10.0\leq \log(M^*/\Msun)$~~~~~~~~~ & $-1.45^{+0.15}_{-0.15}$ & $0.35^{+0.30}_{-0.15}$ & $0.28^{+0.04}_{-0.00}$ & $0.090^{+0.020}_{-0.020}$\\
			      \cmidrule{2-6}
			      & $B-K < 2.0$       & $-1.15^{+0.15}_{-0.00}$ & $0.50^{+0.15}_{-0.15}$ & $0.26^{+0.02}_{-0.02}$ & $0.070^{+0.020}_{-0.000}$\\
			      & $B-K \geq 2.0$    & $-1.30^{+0.15}_{-0.15}$ & $0.50^{+0.30}_{-0.15}$ & $0.24^{+0.02}_{-0.02}$ & $0.090^{+0.000}_{-0.020}$\\
			      \cmidrule{2-6}
			      & $1.4 \leq z < 1.9$& $-1.30^{+0.15}_{-0.15}$ & $0.65^{+0.30}_{-0.15}$ & $0.30^{+0.02}_{-0.02}$ & $0.090^{+0.000}_{-0.020}$\\
			      & $1.9 \leq z < 2.5$& $-1.15^{+0.15}_{-0.15}$ & $0.20^{+0.30}_{-0.00}$ & $0.24^{+0.04}_{-0.02}$ & $0.070^{+0.040}_{-0.020}$\\
\enddata
\end{deluxetable*}

\begin{figure*}
\centering
\includegraphics[clip, angle=-90, width=0.3\textwidth]{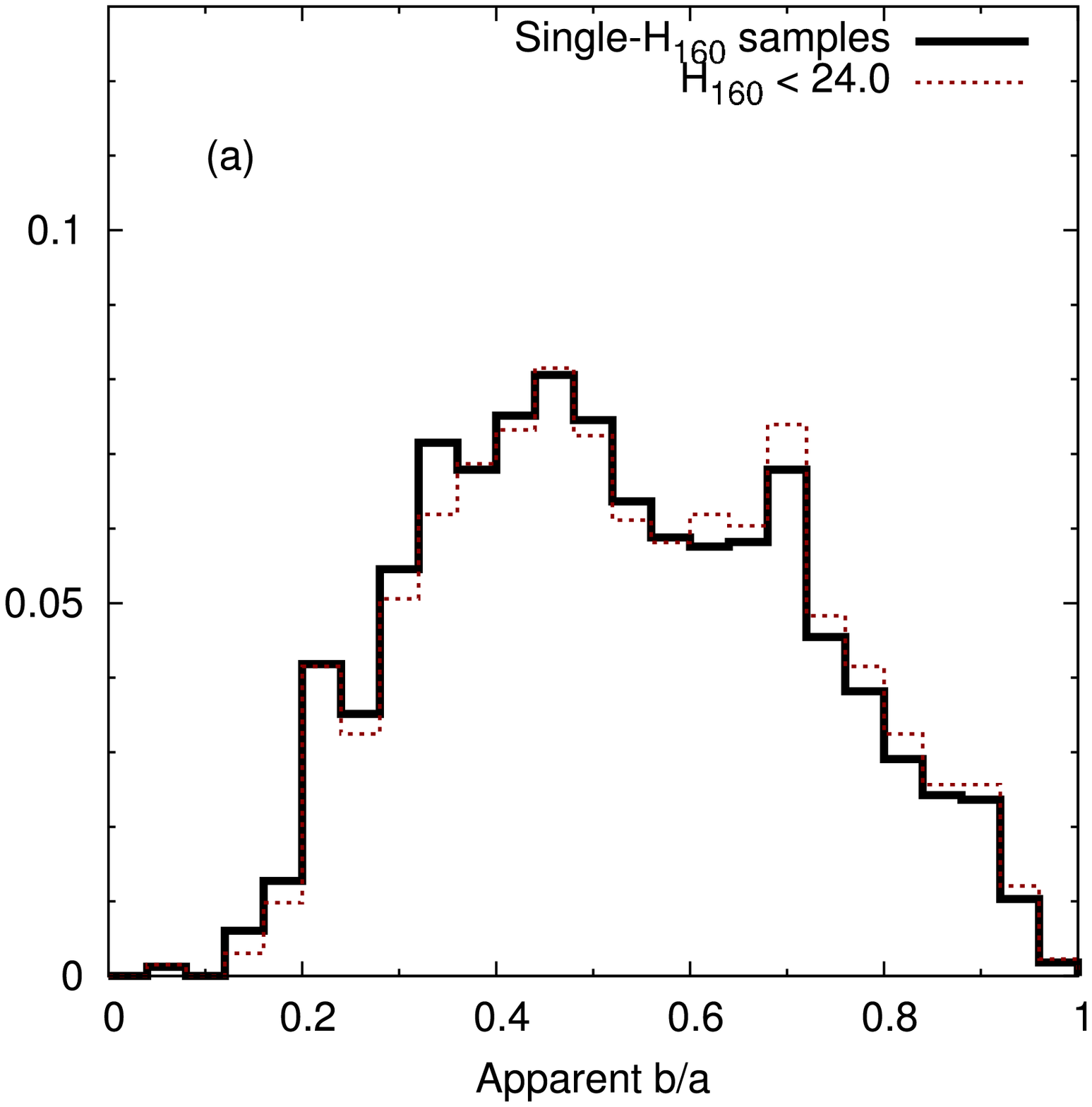} 
\includegraphics[clip, angle=-90, width=0.3\textwidth]{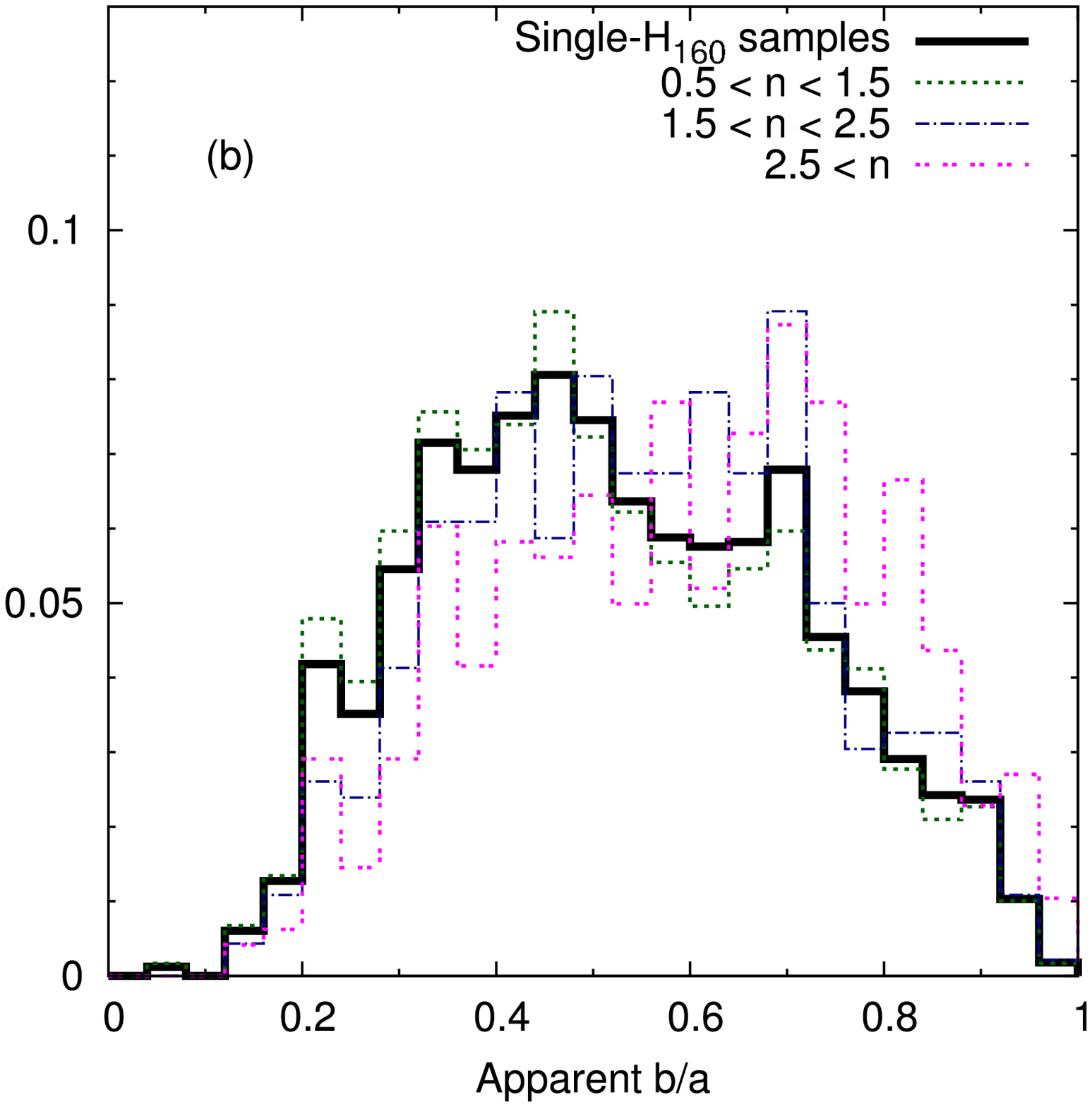} 
\includegraphics[clip, angle=-90, width=0.3\textwidth]{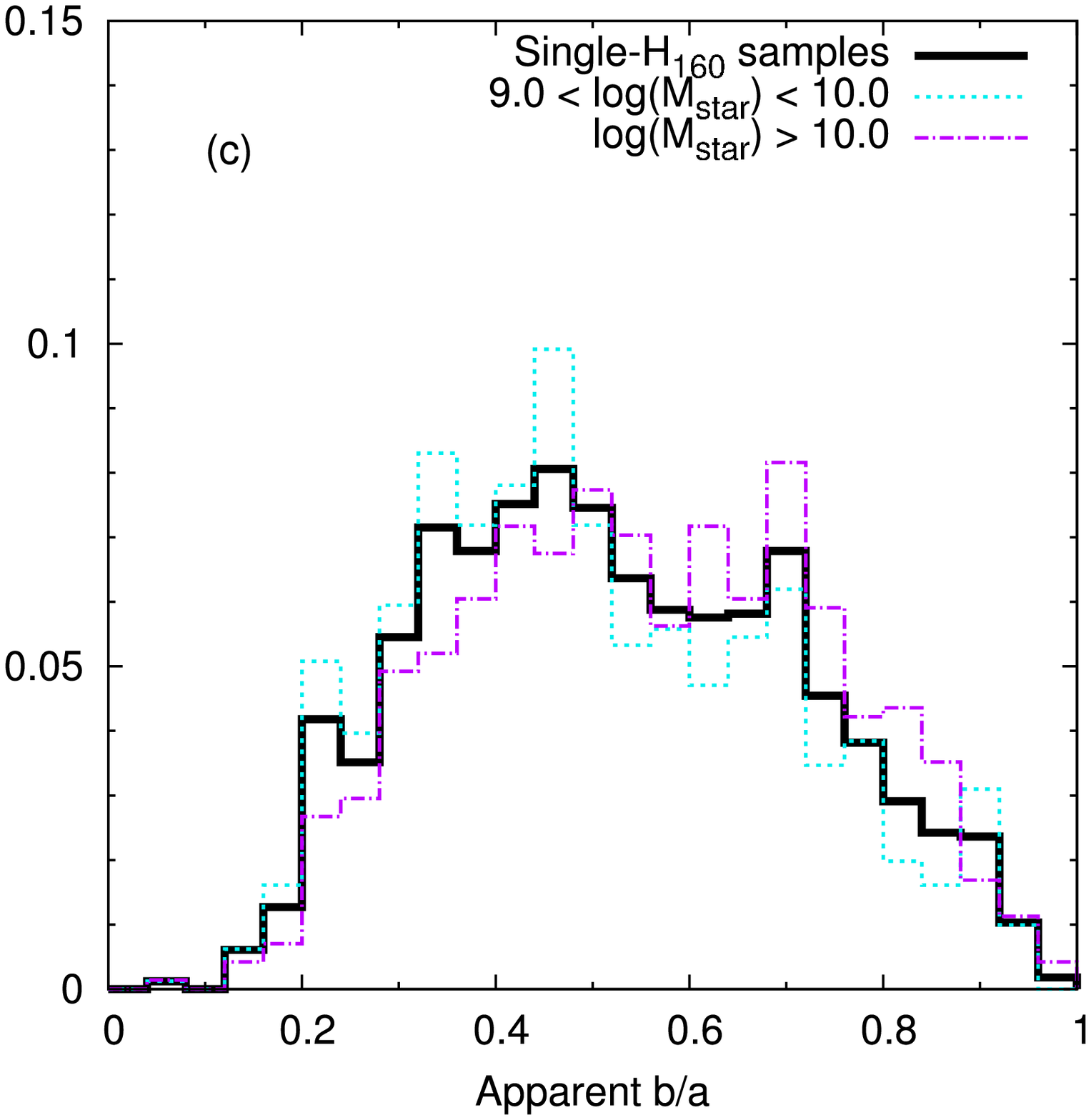} \\
\includegraphics[clip, angle=-90, width=0.3\textwidth]{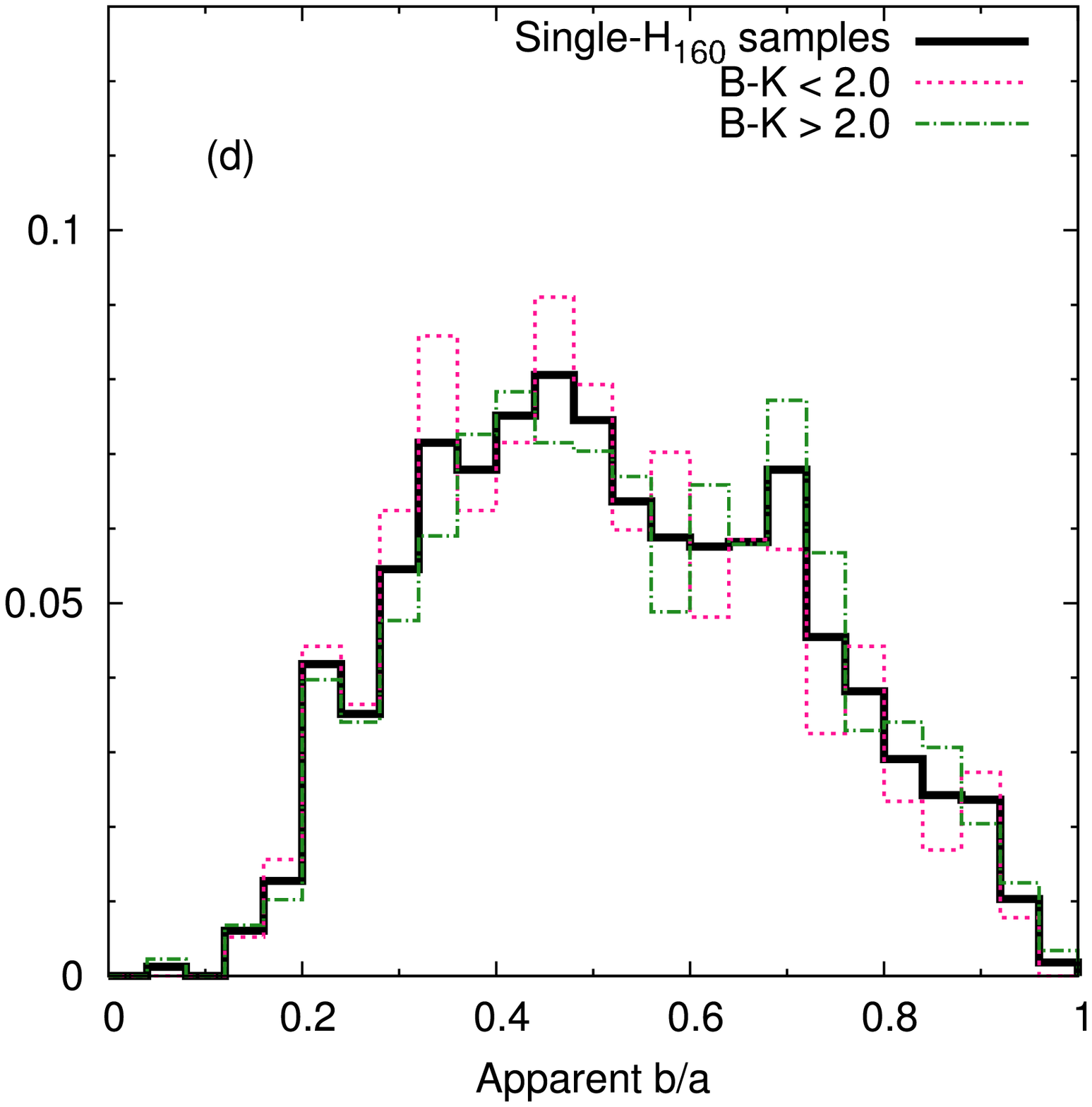} 
\includegraphics[clip, angle=-90, width=0.3\textwidth]{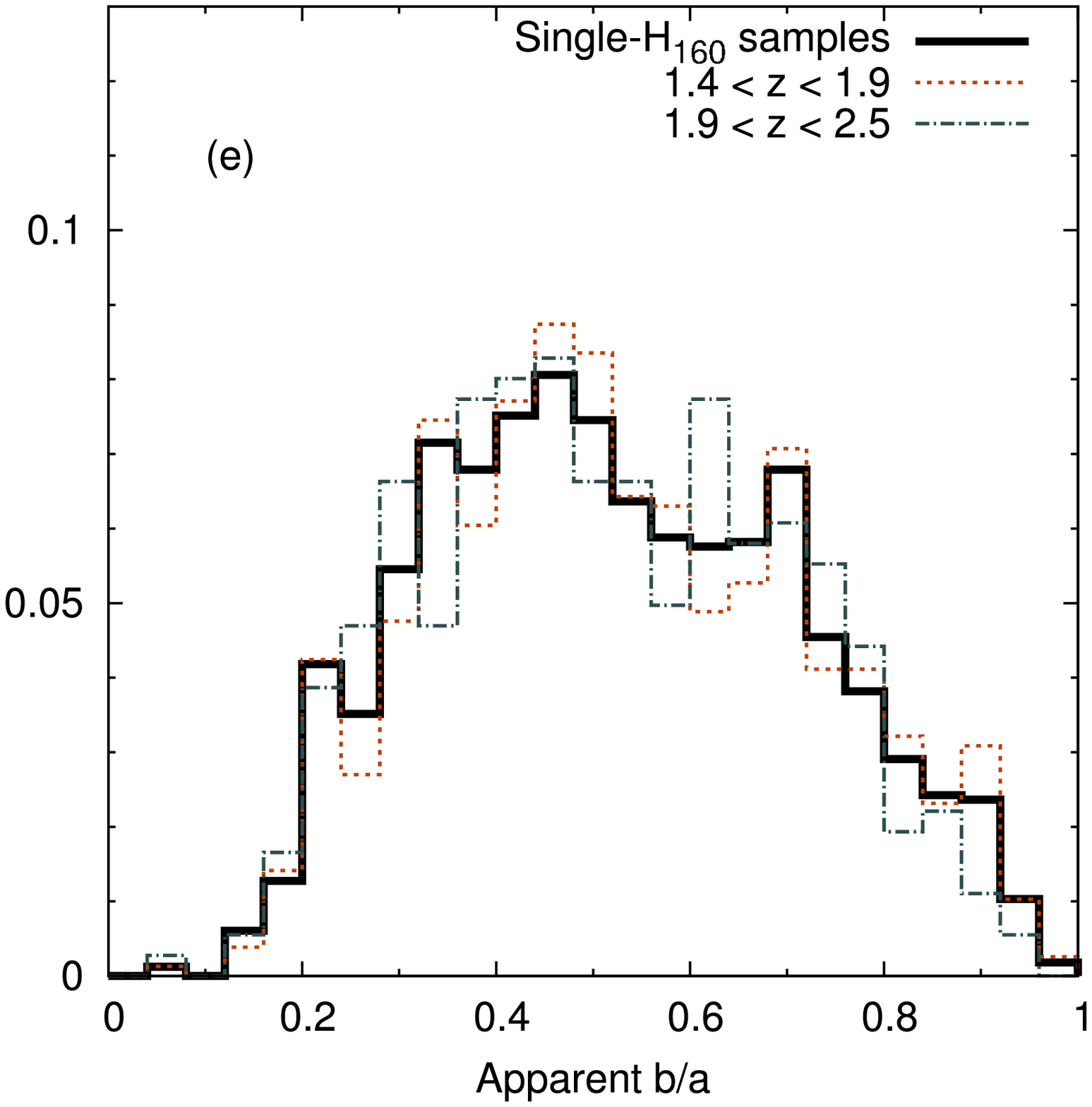} \\
\caption{Same as Figure \ref{sub_acsz} but for the combined sample of single-$\wh$~sBzK galaxies in GOODS-S and SXDS. }
\label{sub_wfc3h}
\end{figure*}

\subsection{Apparent $b/a$ Distribution and Intrinsic Shape of Subsamples}
\subsubsection{Rest-Frame UV in GOODS-S and GOODS-N}
In this section, we discuss the dependence of apparent $b/a$ distribution 
(if any) on various properties of the sBzK galaxies. 
As described in section \ref{subsec:restuvshape}, the $b/a$ distribution of the single-\wz~sample 
in both fields are drawn from the same distribution and their best-fitting shape parameters 
agree well within 68\% confidence level. Besides, the \wz~images in both fields have almost 
the same depth. Therefore, we combined together the sample from both fields to 
increase the statistics. 
Figure \ref{sub_acsz} shows $b/a$ histograms of subsamples of the single-\wz~sBzK 
galaxies in GOODS-S and GOODS-N. As described in section \ref{sec:accuracy} and also in paper I, 
the S\'ersic index can be derived accurately down to $\wz=24.5$ mag in the \wz~images, 
but we used the single-\wz~galaxies of which \wz~magnitudes reach $\wz=26.0$ mag in order 
to increase the number of samples. Thus we made a subsample of galaxies with 
$\wz<24.5$ mag to investigate the shape difference as compare to the whole sample. 
Figure \ref{sub_acsz}(a) shows the apparent $b/a$ distributions of both 
$\wz<24.5$ mag and the whole samples of single-\wz~galaxies. The distributions 
of both samples are close to each other in the way that the histogram of $\wz<24.5$ sample 
shows a peak at $b/a\sim0.3$ and gradually declines until $b/a=1.0$. 
We also derived the intrinsic shape parameters for subsamples and summarized in Table \ref{tab:sub}. 
The derived intrinsic shape for the brighter subsample agrees well with that of the whole 
sample within 68\% confidence level. Figure \ref{sub_acsz} (b) shows the very similar $b/a$ 
histograms between the whole sample and $0.5\leq n < 1.5$ subsample, while the histogram 
of the $1.5\leq n < 2.5$ subsample is little fuzzy. 
Their best-fitting parameters in Table \ref{tab:sub} are still in agreement within $1\sigma$ 
uncertainty, suggesting that diving the single-\wz~galaxies into subsamples of $0.5 \leq n < 1.5$ 
and $1.5\leq n < 2.5$ does not affect the statistical results of the intrinsic shape. 
It is interesting to note here that sBzK subsample with $n\geq2.5$ shows different $b/a$ distribution. 
The K-S test of two distributions between the whole sample 
($0.5 \leq n < 2.5$) and the $n\geq2.5$ subsample results in $P=0.02$, meaning 
that we can reject the null hypothesis that they are drawn from the same populations with 2\% significance level. 
Their $b/a$ histogram peaks at $b/a\sim0.6$ and has a lower fraction at $b/a < 0.5$, resulting 
in larger disk thickness $C/A$ but comparable face-on shape ($B/A$) as compared with $n<2.5$ sample. 

Figure \ref{sub_acsz} (c) shows the comparison of $b/a$ histograms for the lower-mass 
($9.0 \leq \log(M_{star}/\Msun) < 10.0$) and higher-mass ($\log(M_{star}/\Msun) \geq 10.0$) subsamples. 
The lower-mass subsample shows similar $b/a$ distribution as compared to the whole sample; 
the derived parameters of intrinsic shape also agree well with each other. 
The histogram of the higher-mass subsample is slightly different from that of the whole 
sample with $P=0.21$ from the K-S test; 
however, their intrinsic shape parameters are still in agreement within the 68\% uncertainty. 
We also divided the single-\wz~sBzK galaxies into subsamples 
according to their $B-K$ colors. 
Histogram of $B-K < 2.0$ subsample is well in agreement with that of the whole sample as seen 
in Figure \ref{sub_acsz} (d). Their derived intrinsic shapes are also consistent with each other. 
For the $B-K\geq2.0$ subsample, the $b/a$ histogram peaks at $b/a\sim0.40-0.45$, which is 
larger than the $b/a$ peak of the whole sample. This suggests a slightly higher $C/A$ thickness 
as also seen in Table \ref{tab:sub}, though an error is large. Figure \ref{sub_acsz} (e) shows the $b/a$ histograms 
of the $1.4\leq z <1.9$ and $1.9\leq z < 2.5$ subsamples. The redshifts are binned so that 
both subsamples have comparable number of galaxies. Histograms of both subsamples 
seem to look similar to those of the whole sample; their best-fitting parameters are 
also consistent within $1\sigma$ uncertainty. 

\subsubsection{Rest-Frame Optical in GOODS-S and SXDS}

Similar to the rest-frame UV wavelength, we combine the single-\wh~sBzK galaxies 
from GOODS-S and SXDS together in order to increase the statistical significance. 
Although the depth of the \wh~image in GOODS-S is currently shallower than that 
in SXDS, the resulting intrinsic shapes derived separately in both fields agree well with each other 
as shown in section \ref{subsec:restoptshape}. Thus combining the sample is unlikely 
to cause any false result or discussion. 

The single-\wh~sBzK galaxies were divided into subsamples according to their 
various properties and their $b/a$ histograms are shown in Figure \ref{sub_wfc3h}. 
Figure \ref{sub_wfc3h} (a) shows the $b/a$ histogram of the sBzK subsamples with $\wh<24.0$ mag, 
which is the magnitude limit for determining the accurate S\'ersic indices (section \ref{sec:accuracy}). 
Their $b/a$ histogram is remarkably identical to that of the whole sample and their derived intrinsic 
shape parameters are also the same (Table \ref{tab:sub}), indicating that using the 
whole sample to increase the statistic accuracy does not affect the results on intrinsic shape. 
Figure \ref{sub_wfc3h} (b) shows that the $b/a$ histograms of the $0.5\leq n <1.5$ and 
$1.5\leq n < 2.5$ subsample have no significant difference from those of the whole sBzK sample. 
The fitting results in Table \ref{tab:sub} also show an agreement within $1\sigma$ uncertainties. 
In contrast, the $n\geq 2.5$ subsample shows lower fraction at $b/a<0.5$ and 
larger fraction at $b/a>0.5$, similar to that seen in the rest-frame UV wavelength. 
According to Table \ref{tab:sub}, the peak $B/A$ (0.83) and peak $C/A$ (0.34) of the subsample 
are both different from those of the whole sample; the $n\geq2.5$ subsample statistically 
shows rounder and thicker shape than the $n<2.5$ sample. 
Although the $n\geq2.5$ subsample does not show the similar shape to 
the local elliptical galaxies \citep[peak $B/A=0.98$ and peak $C/A=0.43$;][]{padilla08}, 
the trend of being thicker in shape than the $n<2.5$ subsample 
is the same as that in local universe in the way that local elliptical galaxies show almost 
similar face-on ratio but thicker edge-on ratio than the local disk galaxies. 

The $b/a$ distributions of mass-divided subsamples are shown in Figure \ref{sub_wfc3h} (c). 
Similar to the single-\wz~galaxies, the lower-mass subsample with $9.0\leq \log (M_{star}/\Msun) < 10.0$ 
shows similar $b/a$ distribution to the whole sBzK sample, meanwhile 
the larger-mass subsample with $\log(M_{star}/\Msun) \geq 10.0$ shows slightly different distribution, 
fewer fraction at $b/a <0.5$ and larger fraction at $b/a > 0.5$. 
The derived parameters show that the larger-mass sample is slightly rounder and thicker than the 
lower-mass one; however, they are still within $1\sigma$ uncertainty. 
Figure \ref{sub_wfc3h} (d) illustrates the $b/a$ histograms of the $B-K$ subsamples. 
The $b/a$ histograms of both $B-K$ subsamples are similar to each other and similar 
to the whole sample; their derived parameters are also consistent within $1\sigma$ uncertainty. 
Likewise, both histograms of the redshift subsamples are not significantly different from those of 
the whole sample as seen in Figure \ref{sub_wfc3h} (e). 
Their derived intrinsic shapes are also in agreement with each other. 

\section{Discussion}\label{sec:discuss}
\subsection{Intrinsic Shape Comparison with Disk Galaxies in the Local Universe}

The apparent $b/a$ histogram of the local disk galaxies by \cite{padilla08} is shown 
in comparison purpose in the left panel of Figure \ref{qhistcombine}. 
As described in the previous section, the $b/a$ histogram of the local disks is 
significantly different from the sBzK galaxies both in rest-frame UV and optical wavelengths. 
The peak value of the derived $B/A$ distribution is 
$0.79\pm0.03$ for the sBzK galaxies (combined sample in the rest-frame optical), 
while it is $0.98\pm0.01$ for the local disk galaxies, indicating that the local disks are 
significantly rounder than the sBzK galaxies. 
The $\mg$ parameters of the local disks differs from that of the 
sBzK galaxies in both wavelengths, suggesting that the local disks are statistically flatter 
than the sBzK galaxies at $z\sim2$ (right panels of Figure \ref{qhistcombine}). 
It is implied that the population of galaxies showing round and flat disk seen in the local universe 
is unlikely to be exist yet at $z\sim2$. 

At this redshift, the outermost isophotes we can reach in this study are 
$\sim27-28$ and $\sim27$ mag arcsec$^{-2}$ for  $\mu_{z850}$ and $\mu_{H160}$, respectively. 
Due to the cosmological dimming effect, $\mu_{z850}$ of $\sim27-28$ mag arcsec$^{-2}$ 
at $z\sim2$ corresponds to $\mu_{B}\sim23-24$ mag arcsec$^{-2}$ at $z\sim0$ 
after simple $K$ correction by median $\wz-J$ color of the sample, while 
$\mu_{H160}\sim27$ mag arcsec$^{-2}$ corresponds to $\mu_{V}\sim 23.5$ mag arcsec$^{-2}$ 
at $z\sim0$. In contrast, the intrinsic shape study for local disk galaxies by \cite{padilla08} 
could reach surface brightness of 25 mag arcsec$^{-2}$. 
Although we could not reach the same surface brightness as observed in the local universe, 
it is possible that the surface brightness we see corresponds to $\sim25$ mag arcsec$^{-2}$ 
by considering $\sim1$ mag luminosity decrease from $z\sim1$ to $z\sim0$ \citep[e.g.,][]{scarlata07}. 
This surface brightness corresponds to far beyond bar ends seen in the local universe \citep[e.g.,][]{ohta90, kuchinski00, ohta07}. 
In fact, we trace the isophotes out to $\sim1$\ar~radius at $z\sim2$, which corresponds to 
$\sim8$ kpc at $z\sim0$. If the structural evolution was not drastic, we already trace 
the disk structure far enough. 

\subsection{Why Bar-Like?: Possible Origins of the Bar-Like/oval Structure}

As discussed in the previous section, the sBzK galaxies at $z\sim2$ show 
the similar light profile, effective radius, and stellar surface mass density to those 
of the disk galaxies in the local universe, suggesting they are likely to be disk galaxies. 
However, we found that the sBzK galaxies (at least 
those with a single component) are statistically in bar-like or oval shape instead 
of a round disk seen in the local disk galaxies. 
This result rises an interesting question of why or how the sBzK galaxies become bar-like. 
It is worth noting here that the bar-like structure seen in sBzK galaxies
is unlikely to be direct progenitors of bars in local barred galaxies,
because a fraction of barred galaxies decreases with increasing
redshifts and it is only 10-20\% at $z \sim 0.8$ \citep{sheth08}. 
Considering also the larger size as compared with a typical bar length in
local universe, we regard the bar-like shape seen in the sBzK galaxies 
as a global structure of galaxies. 

As discussed in paper I, one possibility is bar instability when the fraction of disk mass against 
halo mass within a disk radius exceeds the threshold \citep{ostriker73}. 
Using $N$-body simulations, \cite{am02} showed that different central 
concentration results in different morphology of the bars. 
Their massive disk model (MD model), where disk dominates the inner 
part of the galaxy, shows an agreement of the bar structure with the 
intrinsic shape of the sBzK galaxies. 

The bar structure can also be made by galaxy interaction 
\citep[e.g.,][]{noguchi87, gerin90, miwa98, berentzen04}. 
The galaxy is able to develop into a bar structure when perturbed by 
the tidal force of another galaxy. The galaxy loses its angular momentum 
due to tidal perturbation, resulting in elongating bar structure \citep{berentzen04}. 
Although this scenario seems to be less likely as the sBzK sample that we study 
is the single component, which shows no clear sign for strong close galaxy interaction, 
the bar-like structure generated through the galaxy interaction can last for at least a few Gyr \citep{noguchi96, berentzen04}. 
Thus it is still possible that the bar-like structure we see in the sBzK galaxies 
are probably the results of the galaxy interaction in the past.  
The pass-by galaxy interaction (galaxy harassment) and continuous minor merging 
process are also likely to be responsible for the bar-like shape of the sBzK 
galaxies. Studying galaxy shapes in cosmological simulations is desirable 
to answer which of the above scenarios is the most likely process for causing 
the bar-like structure of star-forming galaxies at high redshift. 

\subsection{Do We See A Thick Disk?}
The intrinsic thickness of $C/A \sim 0.3$ is close to a typical value for thick disks in 
the present-day disk galaxies \citep[e.g.,][]{yoachim06}.  
Age of the stellar population composing the thick disk is old; in the Milky Way galaxy, 
the age of the stellar population in the thick disk is older than $\sim 8$ Gyr and 
most of them are  between 10 and 13 Gyr \citep{reddy06}. 
Hence, there is a possibility that the structure we see is the progenitor of thick disks, 
though the radial scale length may be slightly smaller than that of present-day thick disk.  
The difference of peak thickness in the rest-frame UV and optical is very small, suggesting 
almost no age and/or metallicity gradient in the vertical structure.  
The thick disk may be heated up from a thin disk  \citep[e.g.,][]{quinn93}.  
In the redshift range we studied there seems to be a subtle evolution of thickness as seen 
in Table \ref{tab:sub}: $\mu_{\gamma} = 0.28$ in $1.4<z<1.9$ while it is 0.24 in $1.9<z<2.5$ in the 
rest-frame UV and 0.30 in $1.4<z<1.9$ and 0.24 in $1.9<z<2.5$ in the rest-frame optical wavelength. 
However, the difference is probably not significant when considering the large uncertainty. 

One problem for the thick disk interpretation may be the stellar mass of disks. 
\cite{yoachim06} showed $M_{\rm thick}/M_{\rm thin} = 
0.53 (V_c / 100~{\rm km\ s^{-1}})^{-2.3}$ for the local disk galaxies.  
Our sample galaxies show a typical stellar mass of $10^{10} M_{\odot}$ 
and a scale length of 2 kpc, giving a circular velocity of $\sim 150$ km s$^{-1}$ 
if the stellar mass dominates the total mass.   
Then the expected stellar mass ratio between thick and thin components at $z\sim0$ would be 
smaller than $\sim 0.2$, i.e., the stellar mass of the thin disk at $z\sim0$ should be larger 
than that of the sBzK galaxies at $z\sim2$ with at least factor of five. 
If this is the case, the data points for $z\sim2$ in Figure \ref{massden} should shift toward 
right by $\sim0.7$ dex or larger as the galaxies evolve, provided that the disk scale lengths 
do not evolve so much \citep{trujillo06}. Such a large shift can not reproduce the distribution 
of local disks in the figure. Thus the structure we see may partly include the thick disk, 
but would not totally be the thick disk. It should be worth noting that most of our 
sBzK galaxies are expected not to be evolving into present-day elliptical galaxies, 
because faint ($K \sim 23$ mag) sBzK galaxies are expected to reside in less massive 
dark halos and their counterparts at $z\sim 0$ are expected to be disk galaxies 
according to a clustering amplitude \citep{hayashi07}. 

\section{Summary}\label{sec:summary}

We study the intrinsic structure of star-forming galaxies at $z\sim2$ selected 
as sBzK based on their $B-z$ and $z-K$ colors. 1028 and 29835 galaxies 
were selected down to $K_s=24.0$ mag in GOODS-South and SXDS fields, 
respectively. For these objects, we made SEDs and derived the photometric redshifts 
and stellar masses. In the GOODS-S, the structure of the sBzK galaxies was studied in both 
\wz~and \wh~images obtained with $HST$, which correspond to the rest-frame UV ($\sim3000$\AA) 
and optical wavelengths ($\sim5300$\AA), respectively, while only the rest-frame 
optical structure was studied in SXDS due to the availability of the $HST$ images. 
The \wh~images cover only part of both GOODS-S and SXDS fields; therefore, 
only a fraction of sBzK galaxies (724 in GOODS-S and 2500 in SXDS) are 
used to study the rest-frame optical structure. 
Similar to paper I \citep{yuma11}, majority of the sBzK galaxies appear as a 
single component in high-resolution images. 57\% (583) of the sBzK galaxies were 
classified as a single component in \wz~images in GOODS-S, while 86\% (626) and 
82\% (2044) are classified as a single component in \wh~images in GOODS-S and SXDS, respectively. 
The larger fraction of single-component galaxies seen in \wh~images is partially 
due to their poorer resolution; however, more than half of the single-\wh~galaxies 
seen as multiple components in the \wz~images reveal the real morphological 
difference of galaxies between in the rest-frame UV and in the rest-frame optical wavelengths. 

Structural analysis was done for the single-component sBzK galaxies by fitting 2D light 
distributions of galaxies with a single S\'ersic profile by using GALFIT. 
Most of our samples show S\'ersic indices of 
$n=0.5-2.5$, suggesting a disk-like structure. Regardless of the observed fields, 
the effective radii are in the ranges of $1.0-3.0$ kpc and $1.5-4.0$ kpc in the \wz~and 
\wh~images, respectively. The median effective radius is 1.89 kpc for the single-\wz~galaxies 
in GOODS-S, while they are 2.52 kpc and 2.62 kpc for the single-\wh~galaxies 
in GOODS-S and SXDS, respectively. Ratio between the effective radii in \wz~and 
\wh~images ($r_{e,opt}/r_{e,UV}$) is 1.33, consistent with that of BM/BX or LBGs 
at similar redshift \citep{swinbank10}. Size-stellar mass relation of the sBzK galaxies with $0.5\leq n < 2.5$ 
indicates that most sBzK galaxies have comparable stellar surface mass density to 
disk galaxies at $z\sim0-1$, again suggesting they are disk galaxies. 

Although the S\'ersic index, effective radius, and stellar surface mass density of the sBzK galaxies 
are comparable to those of the local disk galaxies, it is important to study their intrinsic structure 
to see whether disk galaxies with similar shape as observed today already exist at $z\sim2$. 
Distribution of apparent axial ratios ($b/a$) of the galaxies was used to derived the intrinsic shape of galaxies. 
By assuming a tri-axial model with axes $A>B>C$, we found that the sBzK galaxies 
have peak face-on $B/A$ ratio of $0.70$ and peak edge-on $C/A$ ratio of $0.25-0.28$ 
in the rest-frame UV wavelength, where we also included the samples in the GOODS-N 
studied in paper I to improve statistical accuracy. For the rest-frame optical wavelength, 
the peak $B/A$ and $C/A$ ratios are $0.77-0.79$ and $0.26-0.27$, respectively. 
The results indicate that the sBzK galaxies have almost the same intrinsic 
shape in both rest-frame UV and optical wavelengths, though with slightly rounder 
in the rest-frame optical images. 

In order to examine the dependence of the intrinsic shape, 
we divided the sBzK galaxies into subsamples according to their properties. 
First, we determined the intrinsic shape of the brighter sample, whose 
structural parameters can be determined more reliably, and found that 
their derived intrinsic shape is consistent with the whole sample regardless 
of the rest-frame wavelength ranges. Likewise, dividing the sample into 
$0.5\leq n < 1.5$ and $1.5\leq n < 2.5$ does not affect the derived intrinsic shape. 
We, furthermore, divided the sBzK sample into subsamples by considering their 
stellar mass, $B-K$ color, and photometric redshifts. Except for the larger-mass 
subsample ($\log (M_{star}/\Msun) \geq 10.0$) that shows very slightly rounder and thicker shape, 
the intrinsic shapes of other subsamples are in agreement with those of the whole 
sBzK sample. 

Comparing the intrinsic shape of the sBzK galaxies at $z\sim2$ to those 
of local disk galaxies \citep[peak $B/A=0.95$, peak $C/A=0.21$;][]{padilla08}, we found that the sBzK 
galaxies are rather in bar-like or oval shape and slightly thicker than round and flat disks 
seen in the local universe. 
There are several possibilities explain the origin of the bar-like structure of the sBzK galaxies. 
First is the bar instability in a massive disk galaxy where disk mass dominate 
the central part of the galaxy. 
Interaction between galaxies is another possible process in which the galaxy can 
evolve into bar structure after influenced by the tidal perturbation from the pass-by galaxy. 
Even though no sign of interaction is seen in our single-component sample, 
we cannot reject the possibility of this galaxy interaction scenario as the resultant bar 
structure from the galaxy interaction can remain for a few more Gyr. 
Moreover, the continuous minor merge could also be the case.  
Study of structure evolution in cosmological simulations is desirable. 
As the intrinsic $C/A$ ratio of the sBzK galaxies are 
comparable to the typical value for thick disks in the local disk galaxies, 
we may see progenitors of thick disks. 
However, if all the stellar mass of the sBzK galaxies at $z\sim2$ is that of thick disk, 
the expected stellar mass of total disk at $z\sim0$ would be very large, 
suggesting they are not the direct progenitor of the present-day thick disks, 
though a part of them may make the thick disks. 
 
We are grateful to the anonymous referee for careful reading and valuable comments which improved clarity of the manuscript. 
This work is supported by the Grant-in-Aid for Scientific Research from Japan Society for the
Promotion of Science (24540230) and the Grant-in-Aid for Global COE program 
"The Next Generation of Physics, Spun from Universality and Emergence" 
from the Ministry of Education, Culture, Sports, Science, and Technology (MEXT) of Japan. 
This work is partially based on observations taken by the CANDELS Multi-Cycle Treasury Program 
with the NASA/ESA HST, which is operated by the Association of Universities for Research 
in Astronomy, Inc., under NASA contract NAS5-26555. 
Parts of the observations in GOODS-S have been carried out using the Very Large 
Telescope at the ESO Paranal Observatory under Program ID 168.A-0485. 
 
\bibliographystyle{apj}
\bibliography{yuma2012_sbzk}

\begin{thebibliography}{67}
\expandafter\ifx\csname natexlab\endcsname\relax\def\natexlab#1{#1}\fi

\bibitem[{{Akiyama} {et~al.}(2008){Akiyama}, {Minowa}, {Kobayashi},
  {et~al.}}]{akiyama08}
{Akiyama}, M., {Minowa}, Y., {Kobayashi}, N., {et~al.} 2008, \apjs, 175, 1

\bibitem[{{Athanassoula} \& {Misiriotis}(2002)}]{am02}
{Athanassoula}, E. \& {Misiriotis}, A. 2002, MNRAS, 330, 35

\bibitem[{{Balestra} {et~al.}(2010){Balestra}, {Mainieri}, {Popesso},
  {et~al.}}]{balestra10}
{Balestra}, I., {Mainieri}, V., {Popesso}, P., {et~al.} 2010, \aap, 512, A12+

\bibitem[{{Barden} {et~al.}(2005){Barden}, {Rix}, {Somerville},
  {et~al.}}]{barden05}
{Barden}, M., {Rix}, H., {Somerville}, R.~S., {et~al.} 2005, \apj, 635, 959

\bibitem[{{Berentzen} {et~al.}(2004){Berentzen}, {Athanassoula}, {Heller}, \&
  {Fricke}}]{berentzen04}
{Berentzen}, I., {Athanassoula}, E., {Heller}, C.~H., \& {Fricke}, K.~J. 2004,
  \mnras, 347, 220

\bibitem[{{Bertin} \& {Arnouts}(1996)}]{bertin96}
{Bertin}, E. \& {Arnouts}, S. 1996, A\&AS, 117, 393

\bibitem[{{Bolzonella} {et~al.}(2000){Bolzonella}, {Miralles}, \&
  {Pell{\'o}}}]{bolzonella00}
{Bolzonella}, M., {Miralles}, J., \& {Pell{\'o}}, R. 2000, A\&A, 363, 476

\bibitem[{{Bruzual} \& {Charlot}(1993)}]{bc93}
{Bruzual}, G. \& {Charlot}, S. 1993, \apj, 405, 538

\bibitem[{{Bruzual} \& {Charlot}(2003)}]{bc03}
---. 2003, \mnras, 344, 1000

\bibitem[{{Calzetti}(1997)}]{calzetti97}
{Calzetti}, D. 1997, \aj, 113, 162

\bibitem[{{Calzetti} {et~al.}(2000){Calzetti}, {Armus}, {Bohlin},
  {et~al.}}]{calzetti00}
{Calzetti}, D., {Armus}, L., {Bohlin}, R.~C., {et~al.} 2000, \apj, 533, 682

\bibitem[{{Cameron} {et~al.}(2011){Cameron}, {Carollo}, {Oesch},
  {et~al.}}]{cameron11}
{Cameron}, E., {Carollo}, C.~M., {Oesch}, P.~A., {et~al.} 2011, \apj, 743, 146

\bibitem[{{Conselice} {et~al.}(2011){Conselice}, {Bluck}, {Ravindranath},
  {et~al.}}]{conselice11}
{Conselice}, C.~J., {Bluck}, A.~F.~L., {Ravindranath}, S., {et~al.} 2011,
  \mnras, 417, 2770

\bibitem[{{Daddi} {et~al.}(2004){Daddi}, {Cimatti}, {Renzini},
  {et~al.}}]{daddi04}
{Daddi}, E., {Cimatti}, A., {Renzini}, A., {et~al.} 2004, \apj, 617, 746

\bibitem[{{F{\"o}rster Schreiber} {et~al.}(2009){F{\"o}rster Schreiber},
  {Genzel}, {Bouch{\'e}}, {et~al.}}]{forster09}
{F{\"o}rster Schreiber}, N.~M., {Genzel}, R., {Bouch{\'e}}, N., {et~al.} 2009,
  \apj, 706, 1364

\bibitem[{{Furusawa} {et~al.}(2008){Furusawa}, {Kosugi}, {Akiyama},
  {et~al.}}]{furusawa08}
{Furusawa}, H., {Kosugi}, G., {Akiyama}, M., {et~al.} 2008, \apjs, 176, 1

\bibitem[{{Gerin} {et~al.}(1990){Gerin}, {Combes}, \& {Athanassoula}}]{gerin90}
{Gerin}, M., {Combes}, F., \& {Athanassoula}, E. 1990, A\&A, 230, 37

\bibitem[{{Giavalisco} \& {Dickinson}(2001)}]{gm01}
{Giavalisco}, M. \& {Dickinson}, M. 2001, ApJ, 550, 177

\bibitem[{{Giavalisco} {et~al.}(2004){Giavalisco}, {Ferguson}, {Koekemoer},
  {et~al.}}]{giavalisco04}
{Giavalisco}, M., {Ferguson}, H.~C., {Koekemoer}, A.~M., {et~al.} 2004, \apj,
  600, L93

\bibitem[{{Grogin} {et~al.}(2011){Grogin}, {Kocevski}, {Faber},
  {et~al.}}]{candels2}
{Grogin}, N.~A., {Kocevski}, D.~D., {Faber}, S.~M., {et~al.} 2011, \apjs, 197,
  35

\bibitem[{{Hayashi} {et~al.}(2009){Hayashi}, {Motohara}, {Shimasaku},
  {et~al.}}]{hayashi09}
{Hayashi}, M., {Motohara}, K., {Shimasaku}, K., {et~al.} 2009, \apj, 691, 140

\bibitem[{{Hayashi} {et~al.}(2007){Hayashi}, {Shimasaku}, {Motohara},
  {et~al.}}]{hayashi07}
{Hayashi}, M., {Shimasaku}, K., {Motohara}, K., {et~al.} 2007, \apj, 660, 72

\bibitem[{{Koekemoer} {et~al.}(2011){Koekemoer}, {Faber}, {Ferguson},
  {et~al.}}]{candels1}
{Koekemoer}, A.~M., {Faber}, S.~M., {Ferguson}, {et~al.} 2011, \apjs, 197, 36

\bibitem[{{Kuchinski} {et~al.}(2000){Kuchinski}, {Freedman}, {Madore},
  {Trewhella}, {Bohlin}, {Cornett}, {Fanelli}, {Marcum}, {Neff}, {O'Connell},
  {Roberts}, {Smith}, {Stecher}, \& {Waller}}]{kuchinski00}
{Kuchinski}, L.~E., {Freedman}, W.~L., {Madore}, B.~F., {Trewhella}, M.,
  {Bohlin}, R.~C., {Cornett}, R.~H., {Fanelli}, M.~N., {Marcum}, P.~M., {Neff},
  S.~G., {O'Connell}, R.~W., {Roberts}, M.~S., {Smith}, A.~M., {Stecher},
  T.~P., \& {Waller}, W.~H. 2000, \apjs, 131, 441

\bibitem[{{Law} {et~al.}(2012){Law}, {Steidel}, {Shapley}, {et~al.}}]{law12}
{Law}, D.~R., {Steidel}, C.~C., {Shapley}, A.~E., {et~al.} 2012, \apj, 745, 85

\bibitem[{{Lawrence} {et~al.}(2007){Lawrence}, {Warren}, {Almaini},
  {et~al.}}]{lawrence07}
{Lawrence}, A., {Warren}, S.~J., {Almaini}, O., {et~al.} 2007, \mnras, 379,
  1599

\bibitem[{{Lilly} {et~al.}(1998){Lilly}, {Schade}, {Ellis}, {et~al.}}]{lilly98}
{Lilly}, S., {Schade}, D., {Ellis}, R., {et~al.} 1998, \apj, 500, 75

\bibitem[{{Miwa} \& {Noguchi}(1998)}]{miwa98}
{Miwa}, T. \& {Noguchi}, M. 1998, \apj, 499, 149

\bibitem[{{Mosleh} {et~al.}(2011){Mosleh}, {Williams}, {Franx}, \&
  {Kriek}}]{mosleh11}
{Mosleh}, M., {Williams}, R.~J., {Franx}, M., \& {Kriek}, M. 2011, \apj, 727, 5

\bibitem[{{Noguchi}(1987)}]{noguchi87}
{Noguchi}, M. 1987, MNRAS, 228, 635

\bibitem[{{Noguchi}(1996)}]{noguchi96}
---. 1996, ApJ, 469, 605

\bibitem[{{Nonino} {et~al.}(2009){Nonino}, {Dickinson}, {Rosati},
  {et~al.}}]{nonino09}
{Nonino}, M., {Dickinson}, M., {Rosati}, P., {et~al.} 2009, \apjs, 183, 244

\bibitem[{{Ohta} {et~al.}(2007){Ohta}, {Aoki}, {Kawaguchi}, \&
  {Kiuchi}}]{ohta07}
{Ohta}, K., {Aoki}, K., {Kawaguchi}, T., \& {Kiuchi}, G. 2007, \apjs, 169, 1

\bibitem[{{Ohta} {et~al.}(1990){Ohta}, {Hamabe}, \& {Wakamatsu}}]{ohta90}
{Ohta}, K., {Hamabe}, M., \& {Wakamatsu}, K. 1990, ApJ, 357, 71

\bibitem[{{Ostriker} \& {Peebles}(1973)}]{ostriker73}
{Ostriker}, J.~P. \& {Peebles}, P.~J.~E. 1973, ApJ, 186, 467

\bibitem[{{Ouchi} {et~al.}(2001){Ouchi}, {Shimasaku}, {Okamura},
  {et~al.}}]{ouchi01}
{Ouchi}, M., {Shimasaku}, K., {Okamura}, S., {et~al.} 2001, \apj, 558, L83

\bibitem[{{Overzier} {et~al.}(2010){Overzier}, {Heckman}, {Schiminovich},
  {et~al.}}]{overzier10}
{Overzier}, R.~A., {Heckman}, T.~M., {Schiminovich}, D., {et~al.} 2010, ApJ,
  710, 979

\bibitem[{{Padilla} \& {Strauss}(2008)}]{padilla08}
{Padilla}, N.~D. \& {Strauss}, M.~A. 2008, MNRAS, 388, 1321

\bibitem[{{Pannella} {et~al.}(2009){Pannella}, {Gabasch}, {Goranova},
  {et~al.}}]{pannella09}
{Pannella}, M., {Gabasch}, A., {Goranova}, Y., {et~al.} 2009, \apj, 701, 787

\bibitem[{{Peng} {et~al.}(2010){Peng}, {Ho}, {Impey}, \& {Rix}}]{peng10}
{Peng}, C.~Y., {Ho}, L.~C., {Impey}, C.~D., \& {Rix}, H.-W. 2010, \aj, 139,
  2097

\bibitem[{{Pickles}(1998)}]{pickles98}
{Pickles}, A.~J. 1998, PASJ, 110, 863

\bibitem[{{Popesso} {et~al.}(2009){Popesso}, {Dickinson}, {Nonino},
  {et~al.}}]{popesso09}
{Popesso}, P., {Dickinson}, M., {Nonino}, M., {et~al.} 2009, \aap, 494, 443

\bibitem[{{Quinn} {et~al.}(1993){Quinn}, {Hernquist}, \& {Fullagar}}]{quinn93}
{Quinn}, P.~J., {Hernquist}, L., \& {Fullagar}, D.~P. 1993, \apj, 403, 74

\bibitem[{{Reddy} {et~al.}(2006){Reddy}, {Lambert}, \& {Allende
  Prieto}}]{reddy06}
{Reddy}, B.~E., {Lambert}, D.~L., \& {Allende Prieto}, C. 2006, \mnras, 367,
  1329

\bibitem[{{Retzlaff} {et~al.}(2010){Retzlaff}, {Rosati}, {Dickinson},
  {et~al.}}]{retzlaff10}
{Retzlaff}, J., {Rosati}, P., {Dickinson}, M., {et~al.} 2010, \aap, 511, A50

\bibitem[{{Rudnick} {et~al.}(2003){Rudnick}, {Rix}, {Franx},
  {et~al.}}]{rudnick03}
{Rudnick}, G., {Rix}, H.-W., {Franx}, M., {et~al.} 2003, \apj, 599, 847

\bibitem[{{Ryden}(2004)}]{ryden04}
{Ryden}, B.~S. 2004, ApJ, 601, 214

\bibitem[{{Salpeter}(1955)}]{salpeter55}
{Salpeter}, E.~E. 1955, ApJ, 121, 161

\bibitem[{{Sargent} {et~al.}(2007){Sargent}, {Carollo}, {Lilly},
  {et~al.}}]{sargent07}
{Sargent}, M.~T., {Carollo}, C.~M., {Lilly}, S.~J., {et~al.} 2007, \apjs, 172,
  434

\bibitem[{{Sawicki}(2012)}]{sawicki11}
{Sawicki}, M. 2012, ArXiv e-prints; 1210.0285v1

\bibitem[{{Sawicki} \& {Yee}(1998)}]{sawicki98}
{Sawicki}, M. \& {Yee}, H.~K.~C. 1998, AJ, 115, 1329

\bibitem[{{Scarlata} {et~al.}(2007){Scarlata}, {Carollo}, {Lilly},
  {et~al.}}]{scarlata07}
{Scarlata}, C., {Carollo}, C.~M., {Lilly}, S., {et~al.} 2007, \apjs, 172, 406

\bibitem[{{S{\'e}rsic}(1963)}]{sersic63}
{S{\'e}rsic}, J.~L. 1963, Boletin de la Asociacion Argentina de Astronomia La
  Plata Argentina, 6, 41

\bibitem[{{Sersic}(1968)}]{sersic68}
{Sersic}, J.~L. 1968, {Atlas de galaxias australes}, ed. {Sersic, J.~L.}

\bibitem[{{Shen} {et~al.}(2003){Shen}, {Mo}, {White}, {et~al.}}]{shen03}
{Shen}, S., {Mo}, H.~J., {White}, S.~D.~M., {et~al.} 2003, MNRAS, 343, 978

\bibitem[{{Sheth} {et~al.}(2008){Sheth}, {Elmegreen}, {Elmegreen}, {Capak},
  {Abraham}, {Athanassoula}, {Ellis}, {Mobasher}, {et~al.}}]{sheth08}
{Sheth}, K., {Elmegreen}, D.~M., {Elmegreen}, B.~G., {Capak}, P., {Abraham},
  R.~G., {Athanassoula}, E., {Ellis}, R.~S., {Mobasher}, B., {et~al.} 2008,
  ApJ, 675, 1141

\bibitem[{{Smail} {et~al.}(2008){Smail}, {Sharp}, {Swinbank},
  {et~al.}}]{smail08}
{Smail}, I., {Sharp}, R., {Swinbank}, A.~M., {et~al.} 2008, \mnras, 389, 407

\bibitem[{Steidel {et~al.}(1996)Steidel, Giavalisco, Dickinson, \&
  Adelberger}]{steidel96}
Steidel, C., Giavalisco, M., Dickinson, M., \& Adelberger, K. 1996, AJ, 112,
  352

\bibitem[{{Steidel} {et~al.}(2004){Steidel}, {Shapley}, {Pettini},
  {et~al.}}]{steidel04}
{Steidel}, C.~C., {Shapley}, A.~E., {Pettini}, M., {et~al.} 2004, \apj, 604,
  534

\bibitem[{{Swinbank} {et~al.}(2010){Swinbank}, {Smail}, {Chapman},
  {et~al.}}]{swinbank10}
{Swinbank}, A.~M., {Smail}, I., {Chapman}, S.~C., {et~al.} 2010, MNRAS, 405,
  234

\bibitem[{{Trujillo} {et~al.}(2006){Trujillo}, {F{\"o}rster Schreiber},
  {Rudnick}, {Barden}, {Franx}, {Rix}, {Caldwell}, {McIntosh}, {Toft},
  {H{\"a}ussler}, {Zirm}, {van Dokkum}, {Labb{\'e}}, {Moorwood},
  {R{\"o}ttgering}, {van der Wel}, {van der Werf}, \& {van
  Starkenburg}}]{trujillo06}
{Trujillo}, I., {F{\"o}rster Schreiber}, N.~M., {Rudnick}, G., {Barden}, M.,
  {Franx}, M., {Rix}, H.-W., {Caldwell}, J.~A.~R., {McIntosh}, D.~H., {Toft},
  S., {H{\"a}ussler}, B., {Zirm}, A., {van Dokkum}, P.~G., {Labb{\'e}}, I.,
  {Moorwood}, A., {R{\"o}ttgering}, H., {van der Wel}, A., {van der Werf}, P.,
  \& {van Starkenburg}, L. 2006, \apj, 650, 18

\bibitem[{{Unterborn} \& {Ryden}(2008)}]{unterborn08}
{Unterborn}, C.~T. \& {Ryden}, B.~S. 2008, ApJ, 687, 976

\bibitem[{{Wuyts} {et~al.}(2012){Wuyts}, {Rigby}, {Sharon}, \&
  {Gladders}}]{wuyts12}
{Wuyts}, E., {Rigby}, J.~R., {Sharon}, K., \& {Gladders}, M.~D. 2012, ArXiv
  e-prints

\bibitem[{{Yabe} {et~al.}(2012){Yabe}, {Ohta}, {Iwamuro}, {et~al.}}]{yabe11}
{Yabe}, K., {Ohta}, K., {Iwamuro}, F., {et~al.} 2012, PASJ, 64, 60

\bibitem[{{Yoachim} \& {Dalcanton}(2006)}]{yoachim06}
{Yoachim}, P. \& {Dalcanton}, J.~J. 2006, \aj, 131, 226

\bibitem[{{Yoshikawa} {et~al.}(2010){Yoshikawa}, {Akiyama}, {Kajisawa},
  {et~al.}}]{yoshikawa10}
{Yoshikawa}, T., {Akiyama}, M., {Kajisawa}, M., {et~al.} 2010, \apj, 718, 112

\bibitem[{{Yuma} {et~al.}(2011){Yuma}, {Ohta}, {Yabe}, {Kajisawa}, \&
  {Ichikawa}}]{yuma11}
{Yuma}, S., {Ohta}, K., {Yabe}, K., {Kajisawa}, M., \& {Ichikawa}, T. 2011,
  \apj, 736, 92

\end{thebibliography}

\end{document}